\documentclass[11pt,a4paper]{article}
\usepackage[utf8]{inputenc}
\usepackage{amsmath}
\usepackage{amsfonts}
\usepackage{amssymb}
\usepackage{graphicx}
\usepackage{natbib}
\usepackage{subfig}
\usepackage[skip=-6pt]{caption}
\graphicspath{{Images/}}
\usepackage{mathtools}
\usepackage{authblk}
\usepackage{color}
\usepackage{subfig}
\usepackage[a4paper, left=1cm, right=1cm, top=2cm, bottom=2cm]{geometry}
\usepackage[toc]{appendix}
\usepackage{url}
\usepackage{hyperref}
\usepackage{xcolor}

\usepackage{fancyhdr}
\fancyhfoffset[L]{0pt}
\fancyhfoffset[R]{0pt}

\fancypagestyle{firstpage}{
  \fancyhf{}
  \fancyhead[C]{\textit{Non-peer-reviewed preprint submitted for consideration}}
  
}

\title{A divide--and--conquer strategy for fast elastodynamic simulation of earthquakes and aseismic slip on fault networks}
\author[1]{Federico Ciardo}
\author[2]{Pierre Romanet}
\affil[1]{Northwestern University, Department of Civil and Environmental Engineering, Evanston (IL), USA, email: \href{mailto:federico.ciardo@northwestern.edu}{federico.ciardo@northwestern.edu}}
\affil[2]{Université Côte d’Azur, CNRS, Observatoire de la Côte d’Azur, IRD, 
Geoazur, Sophia-Antipolis, France, email: \href{mailto:romanet@geoazur.unice.fr}{romanet@geoazur.unice.fr}}
\date{}

\begin{document}
\maketitle

\thispagestyle{firstpage}

% ABSTRACT
\begin{abstract}

Simulating long-term, fully dynamic sequences of earthquakes and aseismic slip (SEAS) on geometrically complex fault networks remains computationally demanding due to the cost of resolving elastodynamic fault interactions. While high-performance computing environments make such simulations more feasible, the associated expense remains substantial, particularly for multi-cycle simulations and long-term deformation evolution. Consequently, many studies adopt quasi-dynamic formulations that approximate elastodynamic wave-mediated stress transfer through radiation damping, enabling longer simulations at reduced computational cost.

Here we present an efficient numerical framework for fully elastodynamic SEAS simulations on complex fault networks. The method follows a divide--and--conquer strategy in which elastodynamic self-effects and fault-to-fault interaction effects are treated separately and evaluated using boundary integral formulations tailored to each interaction type. Self-interactions along planar fault segments are efficiently computed through a non-replicating spectral boundary integral formulation, eliminating spurious contributions from periodic fault images. Stress transfer across arbitrarily oriented faults is instead resolved using a fully dynamic space--time boundary integral representation accelerated by hierarchical matrices ($\mathcal{H}$-matrices).

A central methodological advance is the introduction of a selective $\mathcal{H}$-matrix compression strategy based on the fault-wise assembly of independent binary trees, which enables targeted low-rank approximation of long-range fault-to-fault interactions while preserving near-field accuracy and excluding self-interaction effects from the hierarchical structure. Additional computational gains arise from physics-informed truncation of elastodynamic histories, including mode-dependent time-window truncation for spectral self-effects and causality-based truncation for interaction kernels.

Benchmark multi-fault simulations validate the accuracy of the formulation against reference uncompressed space--time solutions. The method reduces the computational complexity of interaction calculations from $O(N^3)$ to $O(N^2 \log N)$, yielding up to three orders of magnitude reduction in computation time and approximately one order of magnitude reduction in memory usage for typical problem sizes ($\sim 3\cdot10^3$ degrees of freedom). These gains enable fully dynamic SEAS simulations on complex fault networks using workstation-class hardware.

\end{abstract}

\section{Introduction}

There is a growing need for accurate numerical models capable of simulating efficiently sequences of earthquakes and aseismic slip (SEAS) on realistic fault-zone structures. Such models are central to modern earthquake science, as they provide a physics-based framework for interpreting geophysical observations and for investigating the physical mechanisms that govern fault slip across a wide range of spatial and temporal scales.

Over the past two decades, major advances in seismology and geodesy have improved our ability to observe fault slip, rupture propagation, and seismic wave radiation. In parallel, the community has increasingly recognized the critical role of dense near-fault instrumentation. Near-fault measurements are essential to accurately resolve the spatio-temporal evolution of rupture fronts and slow slip processes, particularly in the immediate vicinity of faults where ground motion and stress transfer are most intense. Such observations also provide a unique opportunity to quantify the effects of heterogeneities in frictional properties and stress, as well as the role of fault geometry and segmentation, on near-fault rupture processes.

A great example of the impact of dense near-fault observations was provided by the 2023 Turkey--Syria earthquake doublet. Along the Amanos segment of the East Anatolian Fault system, a relatively dense array of strong-motion stations—approximately ten instruments located within a few kilometers of the surface trace—captured the rupture with unprecedented spatial and temporal resolution. These near-fault recordings enabled a detailed reconstruction of rupture evolution, revealing a large-amplitude, pulse-like rupture propagating at supershear speed \citep{jia_complex_2023,abdelmeguid_dynamics_2023,ren_supershear_2024}. The resulting velocity waveforms provided direct evidence of coherent rupture phases and their associated near-field ground-motion signatures, offering rare insight into the dynamics of supershear rupture in a natural earthquake \citep{yao_rupture_2025}.

In contrast, the recent devastating Myanmar (supershear) earthquake highlights the limitations imposed by sparse near-fault instrumentation \citep{bradley_mandalay_2025}. During the 2025 Mandalay earthquake, large sections of the Sagaing Fault were monitored by only a small number of strong motion stations (only two near-fault stations). As a consequence, the spatio-temporal evolution of rupture could not properly be resolved, leading to discrepancies across published studies. Not only inferred supershear rupture velocities vary between models, but the proposed initiation point of supershear rupture also differs by tens of kilometers along the fault (see e.g. \citep{wei_supershear_2025, xu_bimaterial_2025,latour_direct_2025,goldberg_ultralong_2025}).

As dense near-fault instrumentation continues to expand, the volume and resolution of available data are expected to
increase significantly. Fully exploiting these data requires accurate and efficient physics-based numerical models that can reproduce the observed spatio-temporal complexity of fault slip and that can be used to explore the physical conditions under which phenomena such as supershear rupture emerge.\\

%%%%%%%%%%%%%%%%%%%%%%%%%%%%%%%%%%%%%%%%v%%%%%%%%%%%%%%%%%%%%v%%%%%%%%%%%%%%%%%%%%v%%%%%%%%%%
A variety of numerical approaches have been developed to simulate SEAS, each characterized by different trade-offs between accuracy, geometrical flexibility, and computational cost. Volumetric methods, including finite-difference, finite-element, and continuous or discontinuous Galerkin formulations, explicitly resolve elastodynamic wave propagation and distributed deformation, and can accommodate complex and evolving fault geometries as well as material heterogeneities \citep{erickson_efficient_2014,thakur_effects_2020,galvez_multicycle_2021,dal_zilio_hydro-mechanical_2022,uphoff_discontinuous_2023}. 
However, their computational cost remains high for long-duration 3D SEAS simulations, particularly when inertial effects are fully included.

Boundary integral formulations offer an attractive alternative by restricting discretization to fault surfaces, hence reducing problem dimensionality. Within this class of methods, spectral boundary integral equation methods (SBIEM) are especially efficient, as convolution operators can be evaluated in the Fourier domain at relatively low computational cost. This efficiency has enabled long-duration simulations of earthquake sequences \citep{lapusta_elastodynamic_2000, lapusta_three-dimensional_2009, heimisson_spectral_2022}. Nevertheless, classical SBIEM formulations are restricted to particular (and planar) geometries \citep{barbot_spectral_2021} or mildly non-planar faults \citep{romanet_fully_2021}, limiting their applicability to realistic fault networks.

Space-time boundary integral formulations can overcome this geometrical limitation, as they naturally accommodate multiple interacting faults with arbitrary orientations. In their fully dynamic form, however, space–time BIEM entails a high computational cost due to the storage and evaluation of long elastodynamic interaction histories.

Both spectral and space–time boundary integral formulations can also be implemented under quasi-dynamic assumptions, with elastodynamic wave radiation being approximated through a radiation-damping term. These quasi-dynamic variants substantially reduce computational cost and have been widely employed to simulate earthquake cycles on complex fault systems \citep{ariyoshi_migration_2012, heimisson_crack_2020, cattania_precursory_2021, barbot_spectral_2021, barbot_motorcycle_2023, romanet_combined_2025, cheng_fastdash_2025}. However, because inertial effects and wave-mediated stress transfer are only approximated, quasi-dynamic formulations cannot fully capture dynamic rupture interactions, supershear transitions, or inertia-driven triggering between faults.

Bridging the gap between geometrical generality, computational efficiency, and the ability to fully resolve elastodynamic interactions therefore remains a central challenge in SEAS modeling. Hybrid numerical strategies have begun to emerge as a promising pathway to address this trade-off. For instance, \citet{abdelmeguid_novel_2019} proposed a coupled quasi-dynamic FEM–BIEM framework in which volumetric finite elements are combined with boundary integral representation to leverage the respective advantages of both methods. 

The strategy adopted in this work follows a similar hybrid philosophy, but operates entirely within the boundary integral framework while fully resolving elastodynamic interactions. Rather than coupling volumetric and boundary formulations, we combine complementary boundary integral operators operating in different representations. 
The formulation follows a \textit{divide--and--conquer} strategy in which elastodynamic self-effects and fault--to--fault interaction effects are treated separately and evaluated using the fast method best suited to each contribution. Self-interactions along individual fault segments are computed through a spectral boundary integral operator, leveraged where geometrical conditions permit efficient Fourier-domain evaluation, namely for planar fault segments. Interaction effects across geometrically complex and arbitrarily oriented faults are instead resolved using a fully dynamic space–time boundary integral operator. 
To alleviate the computational burden associated with long elastodynamic convolution histories, the space–time interaction operator is further accelerated through hierarchical matrices ($\mathcal{H}$-matrices).
 
This hybrid spectral / accelerated--space–time strategy combines the efficiency of spectral formulations for self-effects with the geometrical generality of space–time BIEM for interaction effects, thereby enabling fully dynamic SEAS simulations on complex fault networks without the need for large-scale supercomputing resources, at least in two-dimensional settings.\\

%A spectral boundary integral operator is leveraged where geometrical conditions permit efficient Fourier-domain evaluations (i.e., for elastodynamic self-interactions along individual planar fault segments), whereas a fully dynamic space–time boundary integral operator is employed to accommodate complex fault geometries and fault--to--fault elastodynamic interactions. The interaction operator is further accelerated through hierarchical matrix ($\mathcal{H}$-matrix) compression, enabling efficient storage and evaluation of long space–time convolution histories.
%This hybrid spectral / accelerated-space–time strategy combines the efficiency of spectral formulations with the geometrical generality of space–time BIEM, thereby enabling fully dynamic SEAS simulations on complex fault networks without the need for large-scale supercomputing resources, at least in two-dimensional settings.\\

This article is organized as follows. Section~2 presents the governing equations and numerical formulation of the proposed hybrid approach, including the spectral treatment of self-interactions, the space–time boundary integral representation of fault–to–fault interactions, and the hierarchical matrix compression strategy used to accelerate elastodynamic convolutions. Section~3 demonstrates the capabilities of the method through numerical examples of increasing geometrical complexity, culminating in large-scale simulations of earthquake sequences on a realistic fault network representative of the Gulf of Corinth rift system. Finally, Section~4 summarizes the main findings and outlines directions for future developments.

%------------------------------------------------------------------------------------------------------------------------------
\section{Modeling approach}
%------------------------------------------------------------------------------------------------------------------------------

\begin{figure}[t!]
\centering
\noindent\includegraphics[width=0.65\textwidth]{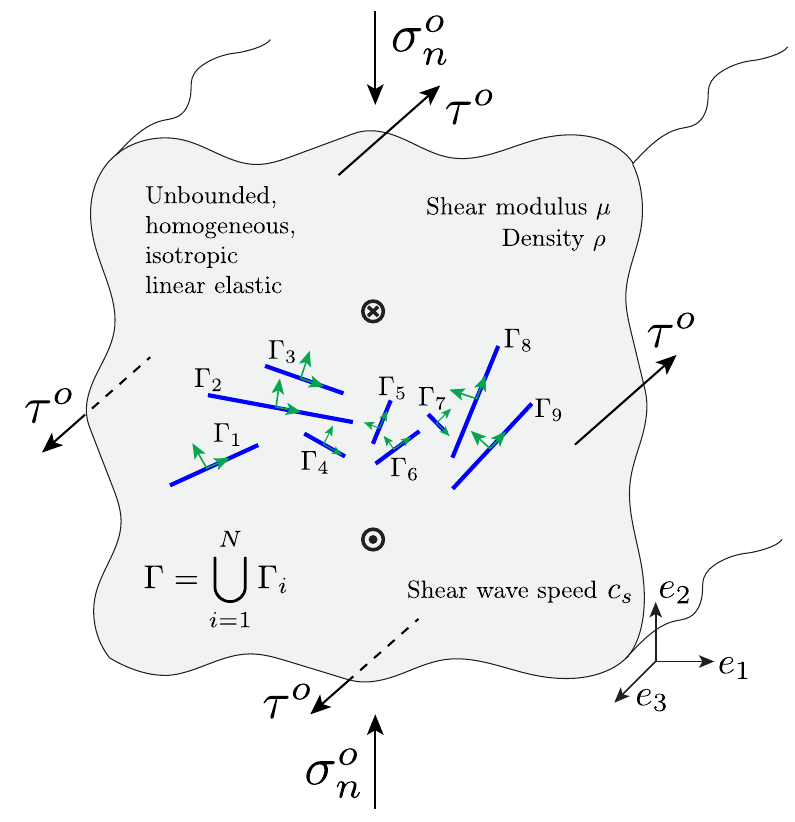}
\\[0.5cm]
\caption{Schematic representation of a fractured domain composed of pre-existing planar faults arbitrarily oriented in space. Fault segments $\Gamma_i$ may be spatially separated or intersect with one another and interact elastodynamically through stress transfer. In the present study, fault slip is governed by Rate-and-State friction and is modeled in an anti-plane (mode~III) setting, for which the problem is scalar and only shear tractions are involved.}
\label{fig:fault_network}
\end{figure}

\subsection{Equilibrium of forces}

We consider a system of pre-existing planar faults arbitrarily oriented within an unbounded linear elastic medium, as illustrated in Figure~1. Together, these faults form a fractured domain $\Gamma = \bigcup_i \Gamma_i$, where each $\Gamma_i$ denotes an individual fault segment. Faults may be spatially separated or intersect with one another.

In the present work, we focus on an anti-plane (mode~III) elastodynamic setting, for which fault slip and elastic fields are scalar and only shear tractions are involved (see also Appendix~\ref{app:Appendix1}). This choice allows us to introduce and validate the proposed modeling approach in the simplest fully dynamic configuration.

At any time $t$, equilibrium on fault $\Gamma_i$ is expressed as the balance between elastodynamic traction induced by elastic fault slip and external loading, and the frictional resistance of the fault:
\begin{equation}
\tau^{\text{el}}(\mathbf{x},t)
+\tau^{o}(\mathbf{x},t)
=
f_c(\mathbf{x},t) \times \left[\sigma_n^{\text{el}}(\mathbf{x},t)+\sigma_n^{o}(\mathbf{x},t)\right],
\label{eq:force_balance}
\end{equation}
where $\mathbf{x} \in \Gamma_i$, $\tau$ and $\sigma_n$ denote shear and normal traction respectively (with $^o$ denoting their background values which would have been created along each fault plane if that plane had been constrained against any slip), and $f_c(\mathbf{x},t)$ is the friction coefficient, here assumed to be rate- and state-dependent (see sub-section \ref{subsec:R&S}).
%The radiation damping term $\tau^{\text{rad}}$ accounts for energy loss due to seismic wave radiation and is proportional to slip rate \citep{rice_spatio-temporal_1993}.

A central computational challenge lies in evaluating the elastodynamic tractions generated by the history of fault slip or slip rate over the fractured domain $\Gamma$. For a given fault $\Gamma_i$, the elastic shear and normal traction can be naturally decomposed into a self-interaction term and contributions arising from all other faults in the system:
\begin{equation}
\begin{split}
\tau^{\text{el}}(\mathbf{x},t)
&=
\tau_{\Gamma_i \rightarrow\Gamma_i}^{\text{el}}(\mathbf{x},t)
+
\sum_{j\ne i}\tau^{\text{el}}_{\Gamma_j \rightarrow \Gamma_i}(\mathbf{x},t),\\
\sigma_n^{\text{el}}(\mathbf{x},t)
&=
\sigma^{\text{el}}_{n,\Gamma_i \rightarrow\Gamma_i}(\mathbf{x},t)
+
\sum_{j\ne i}\sigma^{\text{el}}_{n, \Gamma_j \rightarrow \Gamma_i}(\mathbf{x},t),
\end{split}
\label{eq:traction_decomposition}
\end{equation}
where the first term represents the elastodynamic response of fault $\Gamma_i$ to its own slip (or slip-rate) history, while the second term accounts for stress transfer from all other faults $\Gamma_j \subset \Gamma$.

In the anti-plane setting considered here, shear and normal degrees of freedom are not coupled. As a consequence, fault slip affects only the shear traction and does not modify the normal stress, so that the elastodynamic contribution to the normal traction vanishes, i.e.
$\sigma_n^{\text{el}}(\mathbf{x},t)=0$.

Both elastodynamic self-effects and fault--to--fault interactions are governed by the same boundary integral representation, which relates the shear traction at a point $\mathbf{x}$ on the fault to the history of slip $\delta(\mathbf{y},t)$ over the fractured domain $\Gamma$. In the anti-plane setting, this representation reads (see Appendix~\ref{app:Appendix1} for more details):
\begin{equation}
\tau^{\text{el}}(\mathbf{x},t)
=
-\mu^2\, n_l(\mathbf{x})
\sum_{i=1}^{N}
\int_{\Gamma_i}
\delta(\mathbf{y}(\xi),t)
*
n_j(\mathbf{y}(\xi))
\frac{\partial}{\partial x_l}
\frac{\partial}{\partial x_j}
U_3^3\!\left(\mathbf{x}(\xi),t,\mathbf{y}(\xi)\right)
\,\mathrm{d}\xi ,
\label{eq:representation_antiplane}
\end{equation}
where $*$ denotes convolution product in time (see Eq. \ref{eq:convolution_product}) and $U_3^3$ is the anti-plane elastodynamic Green’s function \eqref{eq:GreenFunction}.

\subsection{Constitutive law for fault strength}
\label{subsec:R&S}

To close the governing equations on each fault plane, the shear traction obtained from the elastodynamic boundary integral representation must be balanced by the frictional resistance of the fault. In this work, fault strength is described using a rate-and-state friction law, which captures the dependence of friction on slip rate and the evolving state of the fault surface \citep{dieterich_modeling_1979,ruina_slip_1983}.
Specifically, the local shear strength of the fault is prescribed as
\begin{equation}
\tau(\mathbf{x},t)
=
f_c(\mathbf{x},t) \times \sigma_{n}^o(\mathbf{x},t)
= \left[
f_* 
+ a\, \ln\!\left(\frac{V(\mathbf{x},t)}{V_*}\right)
+ b\, \ln\!\left(\frac{V(\mathbf{x},t)\,\theta(\mathbf{x},t)}{D_c}\right)
\right]
\, \times \sigma_n^o (\mathbf{x},t),
\quad \mathbf{x}\in\Gamma,
\label{eq:R&S_friction}
\end{equation}
where $V = \partial \delta/\partial t$ is the slip rate, $a$ and $b$ are the direct effect and state evolution parameters respectively, $D_c$ is a characteristic slip distance, and $\sigma_n^o$ is the background normal stress acting on the fault. The constants $f_*$ and $V_*$ denote the friction coefficient and slip rate at a reference steady state, for which the state variable satisfies $\theta = D_c/V_*$.

The evolution of the state variable $\theta$ is governed by an intermediate rate-and-state evolution law \citep{ruina_slip_1983,ciardo_nonlinear_2025}, given by
\begin{equation}
\frac{\partial \theta(\mathbf{x},t)}{\partial t}
=
\frac{1}{\varepsilon}
\left[
\left(\frac{V(\mathbf{x},t)\,\theta(\mathbf{x},t)}{D_c}\right)^{-\varepsilon}
-1
\right]
\frac{V(\mathbf{x},t)\,\theta(\mathbf{x},t)}{D_c},
\label{eq:state_evolution_law}
\end{equation}
where $\varepsilon$ is a dimensionless parameter taking values in the interval $(0,1]$. For $\varepsilon=1$, Eq.~\eqref{eq:state_evolution_law} reduces to the classical Dieterich--Ruina aging law, while in the limit $\varepsilon \to 0$ it recovers the Ruina--Dieterich slip law. All numerical results presented in this manuscript are obtained using $\varepsilon=1$.

\subsection{Computation of fault elastodynamic response}
\label{sec:elastodynamic_response_fault}

The equilibrium equation~\eqref{eq:force_balance} involves the evaluation of the elastodynamic shear traction $\tau^{\text{el}}(\mathbf{x},t)$ generated by the history of fault slip over the fractured domain $\Gamma$. Starting from the boundary integral representation given in Eq.~\eqref{eq:representation_antiplane}, the total elastic traction acting on the fault can be expressed, after regularization due to the hyper--singular kernel, in a compact form as the sum of an instantaneous contribution and a history-dependent term \citep{cochard_dynamic_1994,tada_non-hypersingular_1997}:
\begin{equation}
\tau^{\text{el}}(\mathbf{x},t)
=
\tau^{\text{rad}}(\mathbf{x},t)
+
f(\mathbf{x},t),
\label{eq:superimposition2}
\end{equation}
where $\tau^{\text{rad}}$ is the radiation damping contribution, which accounts for energy loss due to seismic wave radiation and is proportional to slip rate \citep{rice_spatio-temporal_1993}. In the anti-plane setting, it is given by
$\tau^{\text{rad}}(\mathbf{x},t) = -\dfrac{\mu}{2c_s}\,V(\mathbf{x},t)$,
with $\mu$ denoting the shear modulus and $c_s$ the shear-wave speed. The second term, $f(\mathbf{x},t)$, is a linear elastic functional of the full history of fault slip (or slip rate), typically in the form of a boundary integral operator. It depends on $\delta(\mathbf{y}(\xi),\Theta)$ for all source points $\mathbf{y}(\xi)$ and times $\Theta$ lying within the elastodynamic wave cone influencing the observation point $(\mathbf{x},t)$, and can be evaluated through a double space--time convolution operator acting on the slip history (like in Eq. \eqref{eq:representation_antiplane}).

The central idea of this work is to evaluate this integral operator using distinct numerical representations, depending on whether the observation point $\mathbf{x}$ lies on the same fault as the source point $\mathbf{y}$ (self-effects) or on a different fault (fault--to--fault interactions).

Elastodynamic self-interactions are computed using a spectral boundary integral representation. In this formulation, the elastodynamic linear operator becomes diagonal in the spectral domain, reducing the space--time convolution to a set of independent time convolutions for each spectral mode. To eliminate spurious elastic interactions associated with periodic Fourier representations, a non-replicating spectral formulation is adopted following \citet{cochard_spectral_1997}, in which each physical fault is embedded in an auxiliary (padding) domain where slip is constrained to vanish outside the rupture.

In contrast, fault--to--fault interactions generally involve arbitrarily oriented and potentially intersecting faults, for which a spectral representation is no longer applicable. These interactions are therefore evaluated in the space--time domain. Because fault--to--fault interactions are long-range when faults are sufficiently separated, as well as sufficiently smooth far from elastodynamic wave fronts, the resulting operators are well suited for hierarchical low-rank approximation, enabling efficient acceleration through $\mathcal{H}$-matrix techniques.

This hybrid strategy provides the foundation for the numerical framework developed in this study, which is presented in detail in the following subsections.

\subsubsection{Computation of self-effects via spectral-time representation without replication}
\label{subsec:self-effects}

Focusing on a generic plane $\Gamma_i$, with local along-fault coordinate $\xi$, and expressing the slip as a discrete superimposition of $n$ spectral modes $\delta(\xi,t) = \sum_{n=-\infty}^{+\infty}D_n(t) e^{(2 i \pi n \xi/\lambda)}$, periodic with spatial replication period $\lambda$, the linear functional $f(\xi,t)$ describing the slip-history contribution in (\ref{eq:superimposition2}) can be written as \citep{perrin_self-healing_1995}
\begin{equation}
f(\xi,t) = \sum_{n=-\infty}^{\infty} F_n(t) e^{i k_n \xi}, \quad F_n(t) = - \frac{\mu \left|k_n\right|}{2} \left( \frac{J_1(\left|k_n\right| c_s t)}{t} * D_n(t) \right), \quad k_n = \frac{2\pi n}{\lambda}
\label{eq:functional_replication}
\end{equation}
where $J_1()$ is the Bessel function of the first kind. 
%Although \eqref{eq:functional_replication} is valid for each fault $\Gamma_i$, the Fourier representation implicitly assumes a spatially periodic slip distribution.
%Consequently, the resulting elastic response corresponds to an infinite periodic repetition of the fault, and using it directly would introduce spurious interactions between the physical fault and all of its periodic images.
The above spectral–time formulation applies independently to each fault $\Gamma_i$ composing $\Gamma$ (with $i=1,\dots,N$). 
However, because the Fourier expansion is periodic, it implicitly represents an infinite periodic repetition of the fault along the $\xi$-direction.
The direct use of this formulation would therefore introduce spurious elastic interactions on $\Gamma_i$ due to elastic waves generated from replicating ruptures.
To eliminate these nonphysical effects, we adopt the non-replicating spectral method introduced by \cite{cochard_spectral_1997}.
The key idea is to embed the physical fault of length $L_i$ inside a larger integration domain of length $2L_i$	and to extend the slip field such that i) $\delta(\xi,t)$ is non-zero only on the physical rupture $\left[ -L_i/2,L_i/2\right]$, ii) $\delta(\xi,t) = 0$ everywhere else on the extended domain.
This extended interval is therefore a virtual (padding) domain used solely for the Fourier representation. A Fourier series with period $2L_i$ is then constructed over this domain. Because the slip is constrained to vanish outside the physical fault, the periodic extension sums to zero outside the rupture, thereby eliminating all contributions from periodic replicas.
With this modification, the functional form in antiplane settings becomes \citep{cochard_spectral_1997}
\begin{equation}
f(\xi,t) = \sum_{n=-\infty}^{\infty} F_n(t) e^{i k_n \xi}, \quad F_n(t) = -\frac{\mu c_s k_n^2}{2} \left( K(n,t) * D_n(t)\right), \quad k_n = \frac{\pi n}{L_i}
\label{eq:functional_no_replication}  
\end{equation}
where
%\begin{equation}
%    K(n,t) = \begin{cases}
%\frac{1}{2 \pi}\int_{0}^{2\pi} \cos(\psi)^2 \cos(\frac{n \pi c_s t}{L_i} \sin \psi) \, \text{d}\psi, \quad c_s t/L_i < 1 \\[0.5cm]
%\frac{2}{\pi}\int_{0}^{\arcsin{(L_i/(c_s t))}} \cos(\psi)^2 \cos(\frac{n \pi c_s t}{L_i} \sin \psi) \, \text{d}\psi, \quad c_s t/L_i > 1
%\end{cases}
%\label{eq:non_replicant_kernel}
%\end{equation}
\begin{subequations}
\label{eq:non_replicant_kernel}
\begin{align}
K(n,t) &= \frac{1}{2 \pi}\int_{0}^{2\pi} \cos^{2}\!\psi \,
\cos\!\left(\frac{n \pi c_s t}{L_i}\sin\psi\right)\,\mathrm{d}\psi,
\qquad \frac{c_s t}{L_i} < 1,
\label{eq:non_replicant_kernel_a}\\[0.3cm]
K(n,t) &= \frac{2}{\pi}\int_{0}^{\arcsin\!\left(\frac{L_i}{c_s t}\right)} \cos^{2}\!\psi \,
\cos\!\left(\frac{n \pi c_s t}{L_i}\sin\psi\right)\,\mathrm{d}\psi,
\qquad \frac{c_s t}{L_i} > 1.
\label{eq:non_replicant_kernel_b}
\end{align}
\end{subequations}
%Note that, more recently, \cite{noda_dynamic_2021} proposed an alternative approach to remove spatial replication that generalizes naturally to all deformation modes.
Note that, more recently, \cite{noda_dynamic_2021} proposed an alternative approach to remove spatial replications by separating the kernel into a static part and a dynamic part (velocity formulation) and by cutting appropriately in time the dynamic kernel before the waves coming from the replicated fault are coming to the considered fault. 

From Equation (\ref{eq:functional_replication}) and (\ref{eq:functional_no_replication}) we can readily deduce that:
\begin{itemize}
    \item [-] In the spectral–time domain, the linear boundary integral operator becomes spatially uncoupled: the double space–time convolution appearing in the classical space-time formulation (\ref{eq:representation_antiplane}) reduces to a pure time convolution for each Fourier (spectral) mode.
    \item [-] The non-replicating spectral formulation yields accurate results with a finite number of modes and avoids the large replication distances required by the classical (replicating) spectral method. Indeed, \cite{noda_dynamic_2021} showed that the replicating formulation gives similar results than the non-replicating one only when the replication distance $\lambda$ is at least four times the rupture length $L_i$, which is often impractically large. 
    \item [-] Although the convolution kernel in the non-replicating formulation is more complex than that of the classical spectral method and requires a numerical integration for the evaluation of (\ref{eq:non_replicant_kernel_b}), it possesses an important property: for each spectral mode, the kernel $K(n,t)$ decays more rapidly and monotonically compared to the replicating kernel at sufficiently large convolving time (see Figure \ref{fig:Kernels_decay} in Appendix \ref{app:Appendix4}). This enables an efficient mode-dependent truncation of the elastodynamic time window (as described below), significantly accelerating the evaluation of the time convolution.
\end{itemize}
For these reasons, the spectral–time representation provides one of the most efficient numerical framework for computing the elastodynamic self-effects on each pre-existing fault $\Gamma_i$.

\paragraph{Discretization \& truncation of elastodynamic time-window}
To evaluate $f(\xi,t)$ in \eqref{eq:functional_no_replication}, we approximate the slip field using a finite number of spectral modes. The Fourier summation is truncated to the interval $\left[ -N_p/2, N_p/2\right]$, where $N_p$ denotes the total number of retained modes. In practice, we choose $N_p$ equal to the number of straight, constant-slip elements of length $\Delta x_i$ used to discretize the fault plane $\Gamma_i$. The elements divide the rupture interval $\left[ -L_i/2, L_i/2\right]$ into $N_p$ segments of uniform size $\Delta x_i = L_i/N_p$. Slip $\delta(\xi,t)$ is sampled at the midpoint of each element, i.e. at $\xi_j = (j+1/2)\Delta x_i$ with $j=0,1,\dots,N_p-1$, and the corresponding spectral coefficients $D_n(t)$ are obtained through a Fast Fourier Transform (FFT). This choice ensures that the Fourier representation of slip is fully consistent with the spatial discretization adopted for the physical fault, the same that will also be used for the evaluation of the fault--to--fault interactions (see sub-section \ref{subsubsec:fault-to-fault_interactions}).

The time convolution is discretized using a constant time step $\Delta t_{\text{min}}$, chosen as a fixed fraction of the time required for a shear wave to traverse one spatial element. Specifically, $\Delta t_{\text{min}} = \beta \Delta x_i/c_s$, where $\beta \leq 1/2$ in an integer factor. This time step is used to discretized the elastodynamic time-window over which the history of slip spectral coefficients $D_n(t)$ is convolved with the mode-dependent kernel $K(n,t)$. The length of such a window depends on the spectral mode $n$: 
since $K(n,t)$ decays fast and monotonically for sufficiently large values of the nondimensional ratio $\pi n c_s t/L_i$, the convolution for higher modes can be truncated with shorter times. Vice-versa, lower modes retain long memory and require a longer convolution history.     

Here we adopt the same strategy proposed by \cite{lapusta_elastodynamic_2000} for out-of-plane and write the mode-dependent truncation time as
\begin{equation}
    T_{\text{end}}(n) = T_{\text{end}}(1) \cdot \mathcal{L}(n)
\end{equation}
where $T_{\text{end}}(1) = \eta_w L_i/c_s$ is the time window of the lowest mode (with $1\leq\eta_w\leq4$) defined as the time required by a shear wave to travel at least the entire fault rupture $\Gamma_i$, and $\mathcal{L}(n)$ is a dimensionless function that depends on the spectral mode $n$ as 
\begin{equation}
    \mathcal{L}(n) = \frac{1}{\left| n\right|} + \frac{q_w -1}{N_p/2-1} \left(1 - \frac{1}{\left| n\right|} \right),
    \label{eq:mode_dependent_truncation}
\end{equation}
In Eq. (\ref{eq:mode_dependent_truncation}), $1\leq q_w\leq N_p/2$ is an integer that modulates the variation of truncation time with spectral mode number $n$. If $q_w = N_p/2$, $\mathcal{L}(n) = 1$ hence the truncation time is equal to value associated with the lowest mode ($T_{\text{end}}(1)$) for all the spectral modes. Large computational advantages arise when $q_w$ approaches the lower bound value $1$. \cite{lapusta_elastodynamic_2000} suggest acceptable values of $q_w$ can be as low as 4 for some problem resolved with the replicant version of the spectral boundary integral method. With the non-replicant version used here, a more efficient truncation can be used and $q_w$ can approach the lower bound 1, for which $T_{\text{end}}(n) = T_{\text{end}}(1)/\left| n \right|$. 

Using the spatial-temporal discretization and the mode-dependent truncation strategy described above, the elastic functional (\ref{eq:functional_no_replication}) in the spectral-time domain reduces to
\begin{equation}
    F_n^t = -\frac{\pi \mu c_s n}{2 L_i} \sum_{T=0}^{T_{\text{end}}(n)} K^{T}_n D_n^{t-T},
    \label{eq:funcational_discretized_form}
\end{equation}
where $D_n^{t-T}$ is the spectral slip coefficient for mode $n$ in the self-effect elastodynamic time window $\left[ t-T_{\text{end}}(n), t\right]$. Inverse FFT of (\ref{eq:funcational_discretized_form}) provides the elastic functional for the fault self-effects in the space-time domain.

To perform the time convolution (\ref{eq:funcational_discretized_form}), the history of spectral coefficients $D_n^{t-T}$ for all modes are stored, and the kernel $K^{T}_n$ pre-integrated over the (truncated) elastodynamic time window. It is worth noticing that if a mode-dependent truncation of such a window is not used and the kernel is pre-calculated over the $T_{\text{end}}(1)$-long time period dictated by the lowest mode, the memory complexity to store the kernel would be $\mathcal{O}(n\cdot (T_{\text{end}}(1)/\Delta t_{\text{min}}))$. For long faults and/or time step $\Delta t_{\text{min}}$ very small, the complexity increases considerably, becoming larger than the computational complexity $\mathcal{O}(n \cdot \log(n))$ associated with FFT to move from spatial to spectral domain. Using the mode-dependent truncation presented above, instead, such a disadvantage disappear and a sub-quadratic complexity is always guaranteed.  

\subsubsection{Computation of fault--to--fault interactions via space-time representation accelerated using $\mathcal{H}$-matrices}
\label{subsubsec:fault-to-fault_interactions}
We now describe the computation of fault--to--fault interactions in the space–time domain. 
These interactions are evaluated \textit{pairwise}: for each receiver fault $\Gamma_{i^\prime}$, we compute the shear-stress changes induced by the slip history on a distinct source fault $\Gamma_{j^\prime}$, with $i^\prime \neq j^\prime$. A loop over all the pairs $(i^\prime, j^\prime)$ therefore yields the complete set of interactions among the faults in the system.
%In this pairwise approach, the observation point $\mathbf{x}$ lies on $\Gamma_{i^\prime}$ while the source $\mathbf{y}$ lies on $\Gamma_{j^\prime}$.

We use the same piecewise‐constant interpolation of slip used for the self-effects and the same constant time step $\Delta t_{\min}$ to discretize the fault--to--fault elastodynamic time window.
The stress change at node $i$ and time $t$ on fault $\Gamma_{i^\prime}$ due to history of slip or slip rate on all the other faults $\Gamma_{j^\prime}$ can be written in the discrete form as \citep{tada_non-hypersingular_1997,ando_efficient_2007}
\begin{equation}
    %\tau_i^{\text{el},t} = -\frac{\mu}{2 c_s} \left[ V_i^t + n_{l}^i \cdot \sum_{j} \sum_{\mathcal{T} = \mathcal{T}_{\text{init}}^{i^\prime j^\prime}}^{\mathcal{T}_{\text{end}}^{i^\prime j^\prime}} \mathcal{K}_{\sigma_{3l}}^{i,j,\mathcal{T}} V_j^{t-\mathcal{T}}\right],
    f_i^t = -\frac{\mu}{2 c_s} \cdot n_{l}^i \cdot \sum_{j} \sum_{\mathcal{T} = \mathcal{T}_{\text{init}}^{i^\prime j^\prime}}^{\mathcal{T}_{end}^{i^\prime j^\prime}} \mathcal{K}_{\sigma_{3l}}^{i,j,\mathcal{T}} V_j^{t-\mathcal{T}},
    \label{eq:fault-to-fault_interactions_discretized}
\end{equation}
where $l=1,2$ is a dummy index (summed over shear components), $V_j^t$ is the slip rate at node $j$ and time $t$, $\mathbf{n}^i = (n_1^i,n_2^i)$ is the unit normal at node $i$, and $\mathcal{K}_{\sigma_{3l}}^{i,j,\mathcal{T}}$ is the discrete elastodynamic kernel associated with stress component $\sigma_{3l}$ generated by a unit slip rate on node $j$ applied over the interval $\left[ 0, \Delta t_{\text{min}}\right]$. Equation \eqref{eq:fault-to-fault_interactions_discretized} thus represents 
slip-history contribution in \eqref{eq:superimposition2} due to interaction effects.

Following \cite{ando_efficient_2007}, the kernel $\mathcal{K}_{\sigma_{3l}}^{i,j,\mathcal{T}}$ may be expressed as a spatial finite-difference of the primitive functions $I_{3l}$:
\begin{equation}
\begin{split}
    \mathcal{K}_{\sigma_{3l}}(r_1^\prime,r_2^\prime,t) & = I_{3l} \left( r_1^\prime + \Delta s/2, r_2^\prime,t+\Delta t_{\text{min}}\right) - I_{3l} \left( r_1^\prime - \Delta s/2, r_2^\prime,t+\Delta t_{\text{min}}\right) \\ & - I_{3l} \left( r_1^\prime + \Delta s/2, r_2^\prime,t\right) + I_{3l} \left( r_1^\prime - \Delta s/2, r_2^\prime,t\right),
    \label{eq:kernel_interactions}
\end{split}
\end{equation}
where $\Delta s$ is the size of the source element on $\Gamma_{j^\prime}$, and $(r_1^\prime,r_2^\prime)$ are the coordinates of the observation point in the local coordinate system $\mathcal{R}^\prime = (\mathcal{O}^\prime,e_1^\prime, e_2^\prime)$ centered on the source element. In (\ref{eq:kernel_interactions}), the index $l$ is not summed but selects the stress component ($l=1,2$).

The anti-plane primitives $I_{31}$ and $I_{32}$ admit the following closed-form expressions \citep{ando_efficient_2007}
\begin{equation}
\begin{split}
I_{31}(r_1,r_2,t) &= 
-\,\frac{r_2}{\pi r}\,
H\!\left( t-\frac{r}{c_s}\right)
\sqrt{\left(\frac{c_s t}{r}\right)^2 - 1} \\
I_{32}(r_1,r_2,t) &= 
H(r_1)\,
H\!\left( t-\frac{|r_2|}{c_s}\right) + \\
&\quad + \frac{1}{\pi}\,\text{sgn}(r_1)\,
H\!\left(t-\frac{r}{c_s}\right)
\left[
\frac{|r_1|}{r}\sqrt{\left(\frac{c_s t}{r}\right)^2 - 1}
- \arccos\!\left( 
\frac{|r_1|}{\sqrt{(c_s t)^2 - r_2^2}}
\right)
\right]
\end{split}
\label{eq:primitives}
\end{equation}
where again $r = \left\Vert \mathbf{x} - \mathbf{y}\right\Vert$ and $H(\cdot)$ is the Heaviside step function.

\paragraph{Truncation of elastodynamic time-window}

From \eqref{eq:primitives}, we can observe that the primitives become nonzero only after specific arrival times.  
In particular,
\begin{equation}
I_{31}(r_1,r_2,t)\neq 0 \;\Leftrightarrow\; t>\frac{r}{c_s},
\end{equation}
while
\begin{equation}
I_{32}(r_1,r_2,t)\neq 0 \;\Leftrightarrow\;
\left[ \big(r_1>0 \,\, \& \,\, t>|r_2|/c_s\big)
\, \lor \, t>\frac{r}{c_s}\right]
\end{equation}
Hence the elastodynamic kernel $\mathcal{K}_{\sigma_{3l}}$ is identically zero until the earliest of these activation times.

For a given source--receiver fault pair $\left(\Gamma_{i^\prime},\Gamma_{j^\prime}\right)$, we therefore compute, for each receiver node and each source node:
\[
\mathcal{T}_{\mathrm{arr}}(i,j) =
\begin{cases}
\displaystyle
\min\!\left(\frac{r_{ij}}{c_s},\,\frac{|r_{2,ij}'|}{c_s}\right),
& r_{1,ij}'>0,\\[0.35cm]
\displaystyle
\frac{r_{ij}}{c_s},
& r_{1,ij}'\le 0,
\end{cases}
\]
where $(r_{1,ij}',r_{2,ij}')$ is the receiver–source offset expressed in the local coordinate system of the source element and $r_{ij}=\|\mathbf{x}_i-\mathbf{y}_j\|$.

The earliest time for interaction between the two faults is then given by
\begin{equation}
\mathcal{T}_{\mathrm{earl}}^{\,i'j'}=\min_{i,j} \mathcal{T}_{\mathrm{arr}}(i,j)
\label{eq:early_time}
\end{equation}
To remain on the conservative side, we introduce a safety factor $\eta_{\mathcal{T}_{\min}}$ and begin the convolution slightly earlier than the time defined in (\ref{eq:early_time}):
\begin{equation}
\mathcal{T}_{\mathrm{init}}^{\,i'j'} = (1-\eta_{\mathcal{T}_{\min}})\, \mathcal{T}_{\mathrm{earl}}^{\,i'j'}.
\end{equation}
Thus, the convolution time window is expanded slightly toward earlier times, guaranteeing that the evaluation of the elastodynamic kernel starts just before the arrival of the shear wave. Values of $\eta_{\mathcal{T}_{\min}}$ may range between $10^{-6}$ and $10^{-1}$. In this contribution, $\eta_{\mathcal{T}_{\min}}=0.01$.

The final convolution time $\mathcal{T}_{\text{end}}^{\,i'j'}$ is defined so that the farthest receiver element falls sufficiently inside the causality cone $(r,t)$ generated by the shear wave \citep{ando_efficient_2007}.
More specifically, for each source–receiver pair, we compute the distance between every source element on $\Gamma_{j'}$ and the endpoints of the receiver fault $\Gamma_{i'}$. The largest of these distances, divided by the shear-wave speed, defines the arrival time of the shear-wave front at the farthest receiver point. We denote this time as $\mathcal{T}_{\max}^{\,i'j'}$. The maximum truncation time is then taken to be proportional to this arrival time through a factor $\eta_{\mathcal{T}{\max}}>0$, i.e.,
\begin{equation}
\mathcal{T}_{\text{end}}^{\,i'j'} = (1+\eta_{\mathcal{T}_{\max}}) \mathcal{T}_{\text{max}}^{\,i'j'}.
\end{equation}
Thus, $\mathcal{T}_{\text{init}}^{\,i'j'}$ and $\mathcal{T}_{\text{end}}^{\,i'j'}$ define the time interval over which the space–time convolution must be evaluated for the fault pair $(\Gamma_{i'},\Gamma_{j'})$. 

To evaluate the spatial--temporal convolution (\ref{eq:fault-to-fault_interactions_discretized}), we follow the same approach used for the self-effects and pre-integrate the kernels over the truncated elastodynamic time window. For a single interaction pair, the memory required to store such pre–integrated kernels scales as $\mathcal{O}(p \cdot q \cdot n_t)$, where $p$ is the number of receiver elements on $\Gamma_{i'}$, $q$ is the number of source elements on $\Gamma_{j'}$, and  
\(n_t = (\mathcal{T}_{\mathrm{end}}^{\,i'j'} - 
       \mathcal{T}_{\mathrm{init}}^{\,i'j'}) / \Delta t_{\min}\)
is the number of convolution time steps in the interaction window.  
The subsequent evaluation of the space–time convolution for this \emph{single} pair also requires $\mathcal{O}(n_t \cdot p \cdot q)$ operations. Here denoting by $n$ the characteristic number of degrees of freedom along each fault (so that $p,q=\mathcal{O}(n)$), this corresponds to $\mathcal{O}(n_t \cdot n^2)$. When many fault pairs are present, the total cost grows proportionally to the number of such pairs, making this term the dominant computational cost of the interaction calculation.
In the next subsection, we show how this bottleneck is substantially reduced through the use of hierarchical matrices.

\begin{figure}[t!]
\centering
\noindent\includegraphics[width=\textwidth]{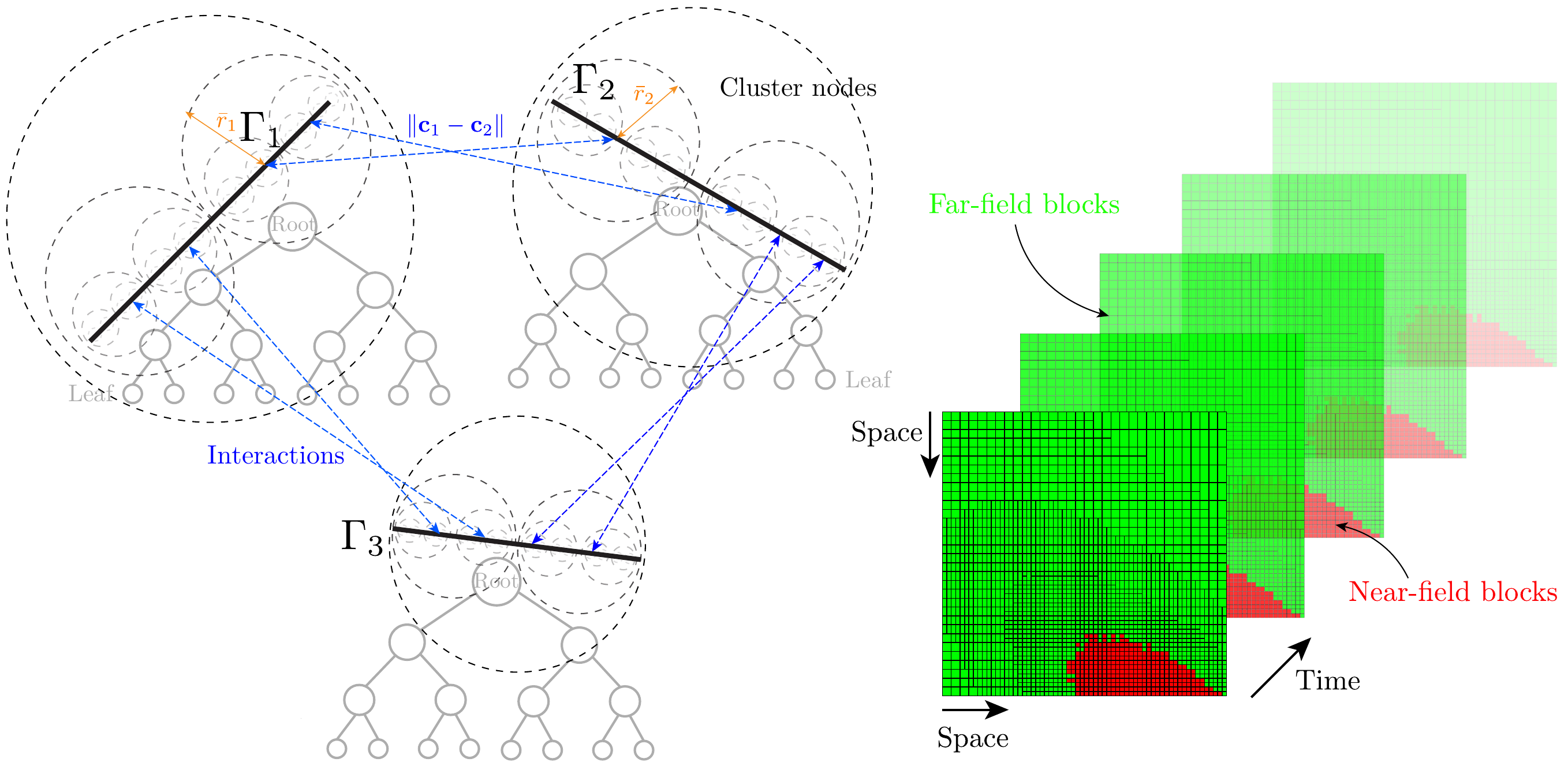}
\\[0.5cm]
\caption{Illustration of the hierarchical matrix (\(\mathcal{H}\)-matrix) construction used to accelerate fault--to--fault interaction calculations. 
Left: each fault is decomposed into a binary tree of spatial clusters down to leaf nodes. Interactions between well-separated clusters are identified as admissible and represented using low-rank approximations, while near-field interactions are stored in full.
Right: schematic representation of the resulting stack of \(\mathcal{H}\)-matrices associated with the discretized space--time convolution, with one matrix per convolution time lag. Green blocks denote admissible interactions stored in low-rank form, whereas red blocks correspond to near-field interactions stored as full, dense matrices.}
\label{fig:h-mat}
\end{figure}
%\paragraph{Acceleration of space--time convolution using hierarchical matrices}
%Hierarchical matrices ($\mathcal{H}$-matrices) provide an efficient representation for large, dense matrices that exhibit data sparsity and admit low-rank structure \cite{hackbusch_sparse_1999,borm_introduction_2003}.
%For a single fault--to--fault interaction pair, the pre-integration of the discretized elastodynamic kernel in (\ref{eq:fault-to-fault_interactions_discretized}) produces a fully populated matrix for each convolution time step.
%The entries of these matrices decay with the source--receiver distance as $\sim r^{-2}$, which suggests that interactions between sufficiently separated subsets of degrees of freedom can be approximated, to an user-defined accuracy, using low-rank representations.
%$\mathcal{H}$-matrices exploit this property to significantly reduce both memory usage and computational cost.
%The central idea is to partition each dense matrix into a hierarchy of sub-blocks and determine, for each block, whether it can be accurately represented in low rank.
%A geometric admissibility condition based on the relative sizes of the source and receiver clusters and their mutual separation identifies (i) admissible (far-field) blocks, which exhibit rapid decay of singular values and are therefore compressed via truncated singular value decomposition (SVD), and (ii) inadmissible (near-field) blocks, which are stored in full.
%In the following, we describe how this selective, hierarchical approximation is constructed for a single convolution slice.
\paragraph{Acceleration of space--time convolution using hierarchical matrices}
Hierarchical matrices ($\mathcal{H}$-matrices) provide an efficient representation for large, dense matrices that exhibit data sparsity and admit low-rank structure \cite{hackbusch_sparse_1999,borm_introduction_2003}. 
For a single fault--to--fault interaction pair, the pre-integration of the discretized elastodynamic kernel in (\ref{eq:fault-to-fault_interactions_discretized}) produces a fully populated matrix for each convolution time step. The entries of these matrices decay with the source--receiver distance as $\sim r^{-2}$, which suggests that interactions between sufficiently separated subsets of degrees of freedom can be approximated, to a user-defined accuracy, using low-rank representations. $\mathcal{H}$-matrices exploit this property to significantly reduce both memory usage and computational cost.

In the present formulation, however, the hierarchical construction departs from classical $\mathcal{H}$-matrix implementations in two important ways, both motivated by the divide--and--conquer strategy adopted in this work. First, matrix blocks associated with elastodynamic self-effects—i.e., the diagonal contributions arising from interactions within the same fault—are entirely excluded from the hierarchical structure. These terms are instead evaluated through the spectral boundary integral formulation described in section~\ref{subsec:self-effects}. As a result, one of the most computationally demanding components of the space--time operator is removed a priori from the hierarchical compression stage.

Second, rather than constructing a single global hierarchical tree over the entire discretized domain, we adopt a fault-wise dual-tree decomposition in which each fault is associated with its own binary cluster tree. This design prevents the hierarchical partitioning from mixing self- and interaction contributions and is particularly well suited to fractured geometries composed of multiple disconnected or intersecting fault segments. It also enables a more targeted identification of admissible far-field blocks for fault--to--fault stress transfer.

The central idea then follows the classical $\mathcal{H}$-matrix philosophy: each dense interaction matrix is partitioned into a hierarchy of sub-blocks, and for each block we determine whether it can be accurately represented in low rank. A geometric admissibility condition based on the relative sizes of the source and receiver clusters and their mutual separation identifies (i) admissible (far-field) blocks, which exhibit rapid decay of singular values and are therefore compressed via truncated singular value decomposition (SVD), and (ii) inadmissible (near-field) blocks, which are stored in full. In the following, we describe how this selective, hierarchical approximation is constructed for a single convolution slice.

\paragraph{Fault-wise hierarchical decomposition, admissibility, and block extraction}
%\comment{I didn't really realised it before reading it, but this is really brilliant way of doing the tree! And it is also much more efficient probably for planar faults.}
Classical $\mathcal{H}$-matrix approaches in slip modeling of pre-existing faults typically construct a single quadtree for the entire computational mesh, where each node tree corresponds to a spatial subdomain (e.g., \cite{ciardo_fast_2020,ozawa_large-scale_2023,cheng_fastdash_2025,romanet_combined_2025}).
Such a global decomposition is not suitable here because the self-effects must not be taken into account as they are evaluated using the spectral--time representation introduced above.
A ``global" tree would therefore mix elements belonging to the same fault and would inevitably generate blocks associated with self-interactions, which must not enter the $\mathcal{H}$-matrix structure.

To avoid this, we construct one hierarchical binary tree per fault. Each node in a fault-wise tree represents a spatial subdomain (or cluster) of that fault, obtained through recursive subdivision of its geometry until it contains no more than $\eta_{\text{leaf}}$ elements. In this case, the node is labeled as a leaf (see Figure~\ref{fig:h-mat}).

Each spatial subdomain of a fault is enclosed in a bounding circle. For a cluster containing $m$ element centroids $\{\mathbf{x}_1,\dots,\mathbf{x}_m\}$, the center of the circle is defined as
\begin{equation}
\mathbf{c} = \frac{1}{m} \sum_{i=1}^m \mathbf{x}_i,
\end{equation}
and its radius is chosen to fully enclose all the elements
\begin{equation}
\bar{r} = \max_{1\leq i \leq m} \left( \|\mathbf{x}_i - \mathbf{c}\| + \Delta s/2 \right),
\end{equation}
where $\Delta s$ is the uniform element size used to discretize the fault.

The bounding circles enter directly into the admissibility criterion. Two clusters, one from the receiver tree and one from the source tree, are considered well separated whenever
\begin{equation}
\|\mathbf{c}_{\mathrm{obs}} - \mathbf{c}_{\mathrm{src}}\|
    > \eta_{\mathcal{H}} \left( \bar{r}_{\mathrm{obs}} + \bar{r}_{\mathrm{src}} \right),
\label{eq:admissibility_condition}
\end{equation}
where $\eta_{\mathcal H}$ is a safety factor (typically $\eta_{\mathcal H}=1.5$).
Satisfying this condition indicates that the corresponding matrix block is suitable for low-rank approximation; otherwise the block must be refined or treated as near-field.

The admissible and near-field blocks are then identified through a dual-tree traversal \citep{yokota_fmm_2012}.  
Starting from the roots of the two trees associated with the interaction pair $(\Gamma_{i'},\Gamma_{j'})$ (with $i' \neq j'$), each cluster pair is subjected to the above criterion.  
If the clusters are well separated, the block is recorded as admissible; if both clusters are leaves and the criterion fails, the block is labeled near-field.  
Otherwise, further refinement is required. The cluster with the larger bounding radius is subdivided—unless it is already a leaf, in which case the other cluster is split. When both clusters have same size, the receiver (observer) cluster is preferentially subdivided. Each resulting child cluster is then paired with the other cluster and tested recursively. This hierarchical procedure yields a geometric partition of the dense interaction matrix into sub-blocks of varying sizes, each uniquely tagged as admissible or near-field.

\paragraph{SVD-based low-rank compression}
Once the admissible blocks have been identified by the dual-tree traversal, each such block is compressed via a truncated singular value decomposition (SVD).
Given an admissible block $B \in \mathbb{R}^{p^\prime\times q^\prime}$, we compute its SVD and retain only the singular values exceeding a prescribed tolerance $\varepsilon_{\text{SVD}}$.
If $k$ singular values are retained, the block is approximated as
\begin{equation}
    B \approx U_k \Sigma_k V_k^{\top},
\end{equation}
where $U_k \in \mathbb{R}^{p^\prime \times k}$, $V_k \in \mathbb{R}^{q^\prime \times k}$, and $\Sigma_k \in \mathbb{R}^{k \times k}$ is diagonal.
This representation reduces the storage requirement from $\mathcal{O}(p^\prime q^\prime)$ to $\mathcal{O}(k (p^\prime+q^\prime))$, with $k \ll \min(p^\prime,q^\prime)$ for well-separated clusters. 
To quantify the efficiency of this compression, we define the \textit{compression ratio} as the percentage of memory required to store the low-rank representation relative to the memory required for the corresponding full dense block.
Note that if $k$ is not significantly smaller than $\min(p^\prime,q^\prime)$, the SVD-based representation may in fact increase memory usage, as two matrices of comparable size must be stored instead of a single dense block.

Near-field blocks, which do not satisfy the admissibility criterion, are stored in full to preserve accuracy.
Because the space--time convolution involves many time steps, the compression of individual convolution slices is embarrassingly parallel, with all low-rank approximations performed independently for each time level in the elastodynamic window.

\paragraph{Computational complexity}
Once the admissible blocks have been compressed, the matrix--vector products required to evaluate a  \emph{single spatial convolution} (i.e., one time slice of the kernel) can be performed in $\mathcal{O}(k \cdot n \cdot \log n)$ time, where $n$ here is the characteristic number of degrees of freedom involved in the interaction and $k$ is the rank of the admissible blocks. This represents a substantial improvement over the dense cost $\mathcal{O}(n^2)$ for the same operation. Because $k$ remains small and essentially independent of $n$, the effective scaling for a single spatial convolution is close to $\sim \mathcal{O}(n \cdot \log n)$\footnote{The same computational complexity is obtained when elastic self-effects are also included in the $\mathcal{H}$-matrix compression (see, e.g., \citep{cheng_fastdash_2025}). Because self-effects are not incorporated in the hierarchical compression considered here, the reported log-linear scaling should be interpreted as an upper-bound estimate. For faults that are widely separated, the higher compression ratios can further reduce the cost, yielding an approximately linear computational complexity.}

For the full spatial--temporal convolution, each fault-to-fault interaction involves $n_t$ time steps within its elastodynamic interaction window. Since each time slice can be evaluated independently, the overall computational cost for the full space--time convolution becomes $\sim \mathcal{O}\!\left( n_t \cdot  n \cdot \log n \right)$, in contrast to the dense cost $\mathcal{O}\!\left( n_t \cdot n^2 \right)$ for the same operation.

%%%%%%%%%%%%%%%%%%%%%%%%%%%%%%%%%%%%%%%%%%%%%%%%%%%%%%%%%%%%%%%%%%%%%%%%%%%%%%%%%%%%%%%%%%%%%%%%%%%%%%%%%

\subsection{Time integration and updating scheme}
\subsubsection{One--time--step integration}

We use a second--order predictor--corrector explicit scheme with variable time stepping to integrate the anti--plane governing equations (\ref{eq:force_balance}) and (\ref{eq:R&S_friction}--\ref{eq:state_evolution_law}). This scheme follows the approach of \cite{lapusta_elastodynamic_2000} and is widely used in earthquake--cycle simulations (e.g., \cite{romanet_fully_2021}). 

In the following, all field variables are understood as nodal (centroidal) quantities defined on fault $\Gamma_i$, i.e.\ functions of $\mathbf{x}\in\Gamma_i$, and evaluated pointwise unless stated otherwise.

The main steps of the time integration and updating scheme are summarized below:

\begin{enumerate}
    \item \textbf{Time--step prediction}.  
    The variable time step is chosen as a multiple of the minimum time step $\Delta t_{\min}$ used to discretize both the self-effect and interaction elastodynamic time--window. Following \cite{lapusta_elastodynamic_2000}, $\Delta t = \max \{ \Delta t_{\min},\ \gamma\,\Delta t_{\min} \},$ where $\gamma = \mathrm{ceil}(\Delta t_{\text{ev}}/\Delta t_{\min})$ and $\Delta t_{\text{ev}} = \min_i \left(\xi_i\, D_c / V_i\right)$.
    Here $\xi_i$ is a parameter that depends on both the stiffness of element $i$ and the rate--and--state parameters (see Eq.~(15) of \cite{lapusta_elastodynamic_2000}).  
    This adaptive scheme ensures that the time step never drops below $\Delta t_{\min}$ during fast (seismic) slip, while increasing significantly during slow (interseismic) deformation. It is worth noticing that this adaptive time stepping scheme does not control the error made at each timestep. Hence, in some problems involving fluid-injection for instance, it may lead to large errors. A solution that would overcome this would be a Runge-Kutta inspired algorithm that produces an estimation of the error made at each time step \citep{press_adaptive_1992}. For example, this error may be estimated by computing in parallel one full time step $\Delta t$, and two successive time step $\Delta t /2$ \citep{romanet_fully_2021}. In this contribution, however, we do not follow this route and stick with the physics-inspired adaptive time stepping scheme described above.

    \item \textbf{Slip-- and state--rate prediction}. Given the history of slip rate on fault $\Gamma_i$ in the truncated interaction time window, $\mathbf{V}^{\,t-\mathcal{T}} = \mathbf{V}(t-\mathcal{T})$, the history of slip spectral coefficients, $\mathbf{D}^{\,t-T}_n = \mathbf{D}_n(t-T)$, in the truncated self-effect time window, and the current state variable $\boldsymbol{\theta}(t)$ and its rate $\dot{\boldsymbol{\theta}}(t)$, the predicted variables (denoted with superscript $^*$) at time $t+\Delta t$ are computed as \begin{equation} 
    \begin{split} 
    \boldsymbol{\delta}^*(t+\Delta t) &\leftarrow \boldsymbol{\delta}(t) + \Delta t \cdot \mathbf{V}(t), \\ \boldsymbol{\theta}^*(t+\Delta t) &\leftarrow \boldsymbol{\theta}(t) + \Delta t \cdot \dot{\boldsymbol{\theta}}(t), \\ \mathbf{D}_n^*(t+\Delta t) &\leftarrow \mathbf{D}_n(t) + \Delta t \cdot \dot{\mathbf{D}}_n(t). 
    \end{split} 
    \end{equation} Here $\dot{\mathbf{D}}_n(t)$ is obtained via FFT from the current slip rate $\mathbf{V}(t)$, corresponding to the portion of the elastodynamic window with initial convolving time $\mathcal{T} = 0$.

    Once $\mathbf{D}_n^*(t+\Delta t)$ is known, we compute the functional coefficients $\mathbf{F}_n^*(t+\Delta t)$ using (\ref{eq:funcational_discretized_form}) and obtain, via inverse FFT, the elastic shear-stress change from self-effects, $\tau_{\Gamma_i \rightarrow \Gamma_i}^{\mathrm{el}}(t+\Delta t)$.

    The predicted slip velocity at time $t+\Delta t$ is obtained by solving the nonlinear equilibrium equation
    \begin{equation}
        \tau_{\Gamma_i \rightarrow \Gamma_i}^{\mathrm{el}}(t+\Delta t)
        + \sum_{j \ne i} \tau^{\mathrm{el}}_{\Gamma_j \rightarrow \Gamma_i}(t+\Delta t)
        + \tau^{o}_{\Gamma_i}(t+\Delta t)
        = f\!\big(V(t+\Delta t), \theta(t+\Delta t)\big)\, \times \sigma^o_n,
    \end{equation}
    where the interaction effects from $\Gamma_j$ (with $j\ne i$) are evaluated using (\ref{eq:fault-to-fault_interactions_discretized}) (making sure that the sources $\mathbf{y}$ do not lie on $\Gamma_i$).  
    We solve this equation using a Newton--Raphson method with a tolerance of $10^{-8}$.  
    Finally, given $\mathbf{V}^*(t+\Delta t)$, the predicted state rate $\dot{\boldsymbol{\theta}}^*(t+\Delta t)$ is evaluated using (\ref{eq:state_evolution_law}).

    \item \textbf{Slip-- and state--rate correction}.  
    The prediction step is repeated after replacing
    \begin{equation}
    \begin{split}
        \mathbf{V}(t) &\rightarrow \frac{\mathbf{V}(t) + \mathbf{V}^*(t+\Delta t)}{2}, \\
        \dot{\boldsymbol{\theta}}(t) &\rightarrow 
            \frac{\dot{\boldsymbol{\theta}}(t) + \dot{\boldsymbol{\theta}}^*(t+\Delta t)}{2}.
    \end{split}
    \end{equation}
    This yields the corrected values $\mathbf{D}_n^{**}(t+\Delta t)$, $\mathbf{V}^{**}(t+\Delta t)$, and $\dot{\boldsymbol{\theta}}^{**}(t+\Delta t)$.  
    The first two are then appended chronologically to the truncated self-effect and interaction elastodynamic time windows, respectively.
\end{enumerate}

Steps~1--3 are applied independently to each pre-existing fault $\Gamma_i$ in the network $\Gamma$, after which the simulation proceeds to the next time step.\\

The entire algorithm is implemented in \cite{Mathematica}. In the next section we present several numerical examples that illustrate the capabilities of our approach and demonstrate its computational advantages. 
All simulations shown in this work were performed on a laptop equipped with an Apple M3 Max processor (16 cores) and 48~GB of memory.
\section{Numerical examples}
\subsection{Two--fault problem}
\label{subsec:two_fault_problem}

\begin{figure}[t!]
\centering
\noindent\includegraphics[width=0.75\textwidth]{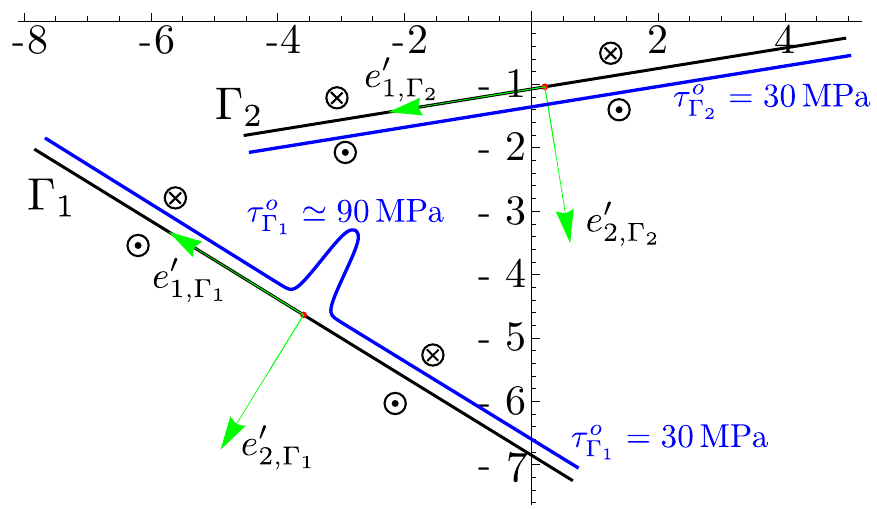}
\\[0.5cm]
\caption{Geometry of the two-fault problem, with all the dimensions expressed in kilometers. Two disconnected planar faults, $\Gamma_1$ and $\Gamma_2$, with arbitrary orientations are embedded in an unbounded elastic medium. Local fault reference systems are defined consistently so that signed slip, slip rate, and shear traction follow the same convention on both faults. Blue profiles indicate the imposed initial shear stress distributions: uniform on fault $\Gamma_2$ and locally peaked on fault $\Gamma_1$ to artificially trigger a dynamic rupture.}
\label{fig:Two_fault_problem}
\end{figure}
We first consider a simple benchmark consisting of two disconnected, planar faults with arbitrary orientations embedded in an unbounded, homogeneous, isotropic elastic medium (Figure~\ref{fig:Two_fault_problem}). Fault centroids are located at $(-3591.98 \, \text{m},-4635.56 \, \text{m})$ for $\Gamma_1$ and $(214.108 \, \text{m},-1037.84 \, \text{m})$ for $\Gamma_2$ in a global Cartesian reference frame with origin at $(0,0)$. The fault lengths are $L_1 = 9927.57\,\mathrm{m}$ and $L_2 = 9578.55\,\mathrm{m}$. Each fault is represented by a straight line segment under a plane-strain assumption with anti-plane deformation, and its orientation is specified by a unit normal. Specifically, on each fault we introduce a local orthonormal reference system $(\mathbf e'_1,\mathbf e'_2,\mathbf e'_3)$. The out-of-plane unit vector is common to all faults and coincides with the global basis direction, $\mathbf e'_3=\mathbf e_3$. The in-plane unit normal to the fault is defined as $\mathbf e'_2=\mathbf n$, and the in-plane tangential direction is given by $\mathbf e'_1=\mathbf e'_3\times\mathbf e'_2$, ensuring a right-handed basis. For fault $\Gamma_1$, $\mathbf{e^\prime_2} = -0.524322 e_1 - 0.85152 e_2$, and for fault $\Gamma_2$, $\mathbf{e^\prime_2} = 0.159658 e_1 -0.987172 e_2$.  
In the anti-plane setting considered here, slip and slip rate are scalar quantities defined as the jump of the out-of-plane displacement component $u_3$ across the fault. Their sign is determined by the orientation of the unit normal vector, which defines the positive and negative sides of the fault.

Variables are expressed in the local fault reference system using a rotation matrix $R(\mathbf n,\theta)$, defined as a rotation by a signed angle $\theta\in(-\pi,\pi]$ about the unit axis $\mathbf n$, following the right-hand rule. The rotation matrix is given by Rodrigues’ formula
\begin{equation}
R(\mathbf n,\theta)
=
\cos\theta\,\mathbf I
+
(1-\cos\theta)\,\mathbf n\otimes\mathbf n
+
\sin\theta
\begin{bmatrix}
0 & -n_3 & n_2\\
n_3 & 0 & -n_1\\
-n_2 & n_1 & 0
\end{bmatrix},
\label{eq:Rodrigues}
\end{equation}
so that the components of a vector $\mathbf v$ expressed in the local fault reference system are obtained as $\mathbf v'=\mathbf v\,R(\mathbf n,\theta)$. Note that the rotation matrix serves only to transform vectors between reference frames and does not, by itself, define the sign convention.

The shear traction acting on a fault is defined as the out-of-plane component of the traction vector,
$\tau = \sigma_{13} n_1 + \sigma_{23} n_2$,
and is reported with a sign consistent with the same choice of unit normal. Reversing the orientation of $\mathbf n$ reverses the associated local fault reference system, which in turn leads to a corresponding change in the rotation matrix $R(\mathbf n,\theta)$ and a sign reversal of the shear traction, slip, and slip rate. %This change affects only the sign of these quantities, while their magnitudes remain unchanged. 
In the configuration shown in Figure~\ref{fig:Two_fault_problem}, the local reference systems on $\Gamma_1$ and $\Gamma_2$ are chosen such that the reported signed quantities have the same sign on both faults.\\

%\begin{figure}%
 %   \centering
 %   \subfloat[\centering $\Gamma_1 \to \Gamma_2$]{{\includegraphics[width=0.4\textwidth]{Images/Compression12.pdf} }}
%    \,\,
%    \subfloat[\centering $\Gamma_2 \to \Gamma_1$]{{\includegraphics[width=0.4\textwidth]{Images/Compression21.pdf} }}\\[0.5cm]
%\caption{$\mathcal{H}$-matrix block structure and compression ratios for the first temporal convolution slice of the fault--to--fault interaction kernel. (a) Interaction from fault $\Gamma_1$ to fault $\Gamma_2$. (b) Reciprocal interaction from fault $\Gamma_2$ to fault $\Gamma_1$. Green blocks denote admissible interactions represented in low-rank form, while the numerical values indicate the corresponding local compression ratios.}
 %   \label{fig:compression_problem1}
%\end{figure}

We have devised this problem such to serve as a benchmark for elastodynamic numerical codes incorporating fault slip and rate-and-state friction. All material, frictional, and geometric parameters are summarized in Table~\ref{tab:Table1}. Each fault is discretized using $10^3$ equally sized straight segments, which is sufficient to resolve the cohesive-zone length scale \citep{perfettini_dynamics_2008,lapusta_three-dimensional_2009}
\begin{equation}
\Lambda = \frac{9\pi}{32} L_b,
\qquad
L_b = \frac{\mu D_c}{b\,\sigma_n} = 450 \, \text{m}
\label{eq:cohesive_zone_length_scale}
\end{equation}
on each fault with more than 40 elements.

\begin{table}[t!]
  \centering
  \begin{tabular}{lcl}
    \hline
    Name & symbol & Value \\
    \hline
    Reference friction coefficient      & $f_*$     & 0.6 \\
    Reference velocity                  & $V_*$     & $10^{-9}\,\mathrm{m/s}$ \\
    Critical slip distance   (uniform)           & $D_c$     & $10^{-2}\,\mathrm{m}$ \\
    Rate and state parameter (uniform)           & $b$       & $(1.2/0.9) \cdot 10^{-2}$ \\
    Rate and state parameter  (uniform)          & $a$       & $1.2 \cdot 10^{-2}$ \\
    State evolution law parameter               & $\varepsilon$   & 1 (Aging law)\\
    \hline
    Shear modulus                       & $\mu$      & $30\,\mathrm{GPa}$ \\
    Shear velocity                      & $c_s$      & $3464\,\mathrm{m/s}$ \\
    Initial normal stress     & $\sigma_n^o$ & $50\,\mathrm{MPa}$ \\
    \hline
    Tolerance Singular Value Decomposition          & $\varepsilon_{\text{SVD}}$        & $10^{-6}$ \\
    Minimum number of elements in a leaf node  & $\eta_{\text{leaf}}$ & $50$ \\
    Safety factor $\mathcal{H}$-matrix admissibility condition & $\eta_{\mathcal{H}}$      & $1.5$ \\
    Simulation time               & $T_{\text{end}}$       & $2\,\mathrm{s}$ \\
    Minimum time step               & $\Delta t_{\text{min}}$       & $\sim 1.38 \times 10^{-3}\,\mathrm{s}$ \\
    Number of elements on $\Gamma_1$               & $N_{\text{elts},\Gamma_1}$       & $10^3$ \\
    Number of elements on $\Gamma_2$               & $N_{\text{elts},\Gamma_2}$       & $10^3$ \\
    Total length of fault $\Gamma_1$               & $L_{1}$       & $9927.57\,\mathrm{m}$ \\
    Total length of fault $\Gamma_2$               & $L_{2}$       & $9578.55\,\mathrm{m}$ \\
    Truncation factor for interaction histories & $\eta_{\mathcal{T}_{\min}}$       & $10^{-2}$ \\
    Truncation factor for interaction histories   & $\eta_{\mathcal{T}_{\max}}$       & $10^{-2}$ \\
    Truncation factor for self-effect histories & $q_w$       & $200$ \\
    Truncation factor for self-effect histories   & $\eta_w$       & $1.5$ \\
    \hline
  \end{tabular}\\[0.5cm]
\caption{Input parameters for two-fault problem.}
\label{tab:Table1}
\end{table}

Initial conditions are specified as follows. On fault $\Gamma_2$, the system is initially in steady-state equilibrium. At time $t=0$, the slip rate is uniform and equal to the reference creeping rate $V_{\mathrm{init}} = V_* = 10^{-9}\,\mathrm{m/s}$, and the associated state variable is $\theta_{\mathrm{init}} = D_c / V_{\mathrm{init}}$. Assuming no prior slip and slip-rate history for $t<0$, the corresponding uniform background shear stress that satisfies equilibrium is $\tau^o_{eq} = f_* \sigma_n + \dfrac{\mu}{2c_s} V_* = 30\,\mathrm{MPa}$. Hence we set $\tau^o(\mathbf{x}, t=0) = \tau^o_{eq} = 30\,\mathrm{MPa}$ for $\mathbf{x}\in \Gamma_2$.
The same initial conditions are prescribed on fault $\Gamma_1$, except that the background shear stress is perturbed to violate equilibrium and trigger dynamic rupture. Specifically, we superimpose on the equilibrium value $\tau^o_{eq} = 30 \, \text{MPa}$ a smooth, localized shear-stress perturbation,
\begin{equation}
\tau^o(x^\prime,t=0) = \tau^o_{eq} + \Delta\tau\, B(x^\prime),
\end{equation}
where $x^\prime$ is here the along-fault coordinate in the local orthonormal basis $(\mathbf{e}_1^\prime,\mathbf{e}_2^\prime)$ centered on $\Gamma_1$ and $B(\xi)$ is a dimensionless function defined as
\begin{equation}
B(\xi)=
\begin{cases}
\dfrac{1}{2}\!\left[\tanh\!\left(\kappa\!\left(\dfrac{\xi}{w_-}+1\right)\right)+1\right], & \xi \le 0,\\[8pt]
\dfrac{1}{2}\!\left[1-\tanh\!\left(\kappa\!\left(\dfrac{\xi}{w_+}-1\right)\right)\right], & x \ge 0.
\end{cases}
\end{equation}
Here $w_- = w_+ = 500\,\mathrm{m}$ and $\kappa = 2.5$ control the width and smoothness of the perturbation. We take $\Delta\tau = 2\tau_{eq}^o$, yielding a peak shear stress of approximately $3\tau_{eq}^o$ at the fault center (see Figure~\ref{fig:Two_fault_problem}). This localized overstress nucleates an earthquake that propagates bilaterally along $\Gamma_1$ until reaching the fault extremities, at which point the simulation is terminated. The corresponding final simulation time is $T_{\mathrm{end}} = 2\,\mathrm{s}$.

Since fast (seismic) slip nucleates immediately at $t=0^+$, the adaptive time step remains at its minimum value throughout the simulation:
\[
\Delta t_{\min} = \beta \cdot \min\!\left(\frac{\Delta x_{\Gamma_1}}{c_s},\,\frac{\Delta x_{\Gamma_2}}{c_s}\right) \simeq 1.38\times10^{-3}\,\mathrm{s},
\]
where $\beta = 1/2$, and $\Delta x_{\Gamma_1} \simeq 9.93 \, \text{m}$ and $\Delta x_{\Gamma_2} \simeq 9.58 \, \text{m}$ denote the mesh cell sizes on faults $\Gamma_1$ and $\Gamma_2$, respectively. 
This (minimum) time step is also used to pre-integrate the elastodynamic kernels associated with both self-effects and fault--to--fault interaction effects over the truncated time window.
In this particular example, the truncation strategy described above for interaction effects yields a maximum truncation time of $\mathcal{T}_{\mathrm{end}}^{\,i'j'} \simeq 3.76\,\mathrm{s}$, which exceeds the total simulation time. 
Consequently, slip-rate histories are continuously inserted into the truncated elastodynamic time window throughout the simulation, without fully populating it. The ``effective" upper bound is therefore given by $\mathcal{T}_{\mathrm{end}}^{,i’j’} = T_{\mathrm{end}} = 2\, \mathrm{s}$. These stored histories are then used to evaluate the space–time convolution in Eq.~\eqref{eq:fault-to-fault_interactions_discretized} at each time step. For the $\mathcal{H}$-matrix parameters reported in Table~\ref{tab:Table1},  the resulting discretized space--time convolution operator associated with  fault--to--fault interactions admits an efficient hierarchical low-rank  representation, with an  overall compression ratio of $\sim24.7\%$. This results in a substantial  reduction in memory storage and convolution cost.\\

\begin{figure}[t!]
  \centering
  \subfloat[a][Fault $\Gamma_1$]{\includegraphics[width=\textwidth]{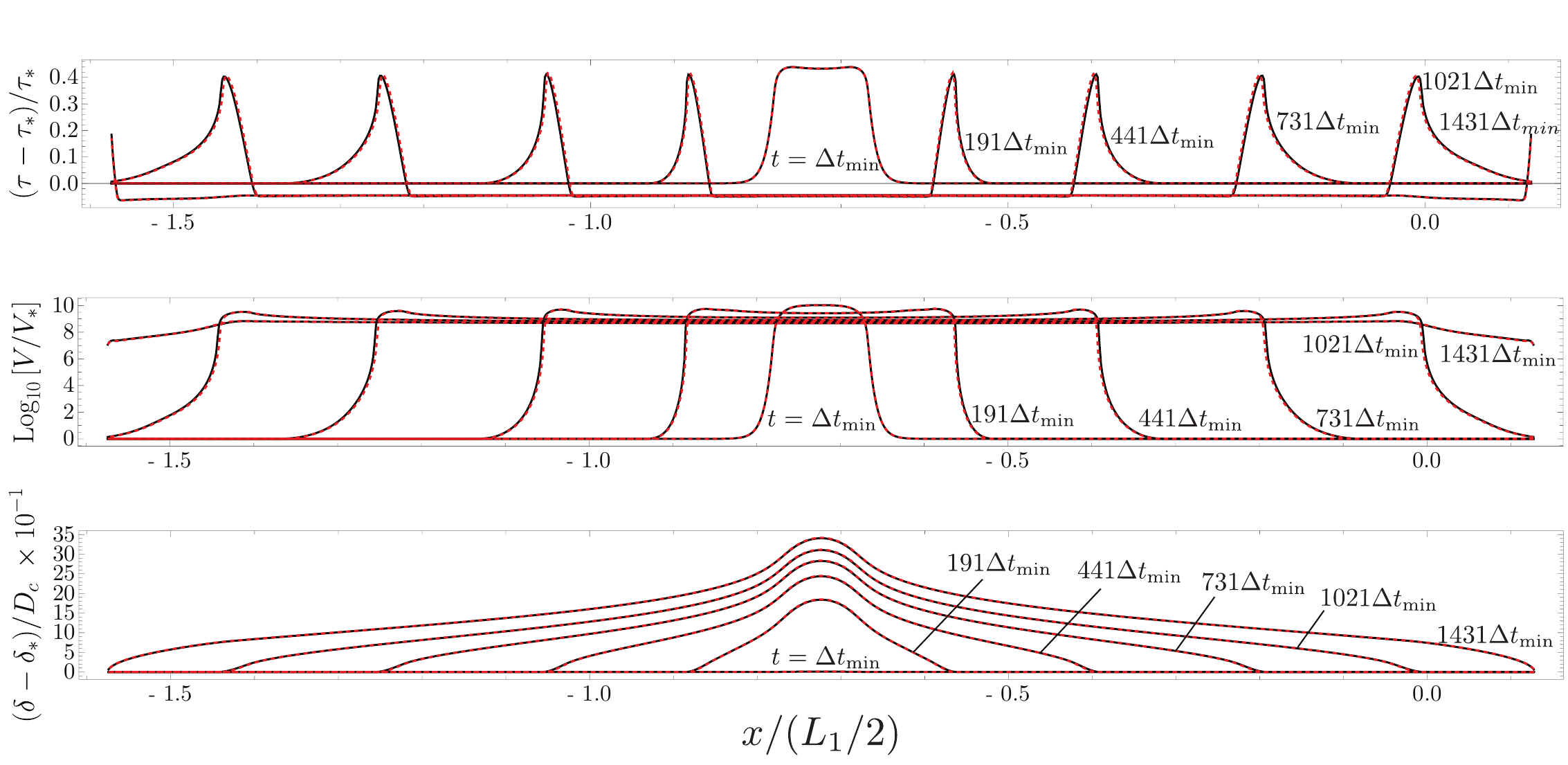}} \\
  \subfloat[b][Fault $\Gamma_2$]{\includegraphics[width=\textwidth]{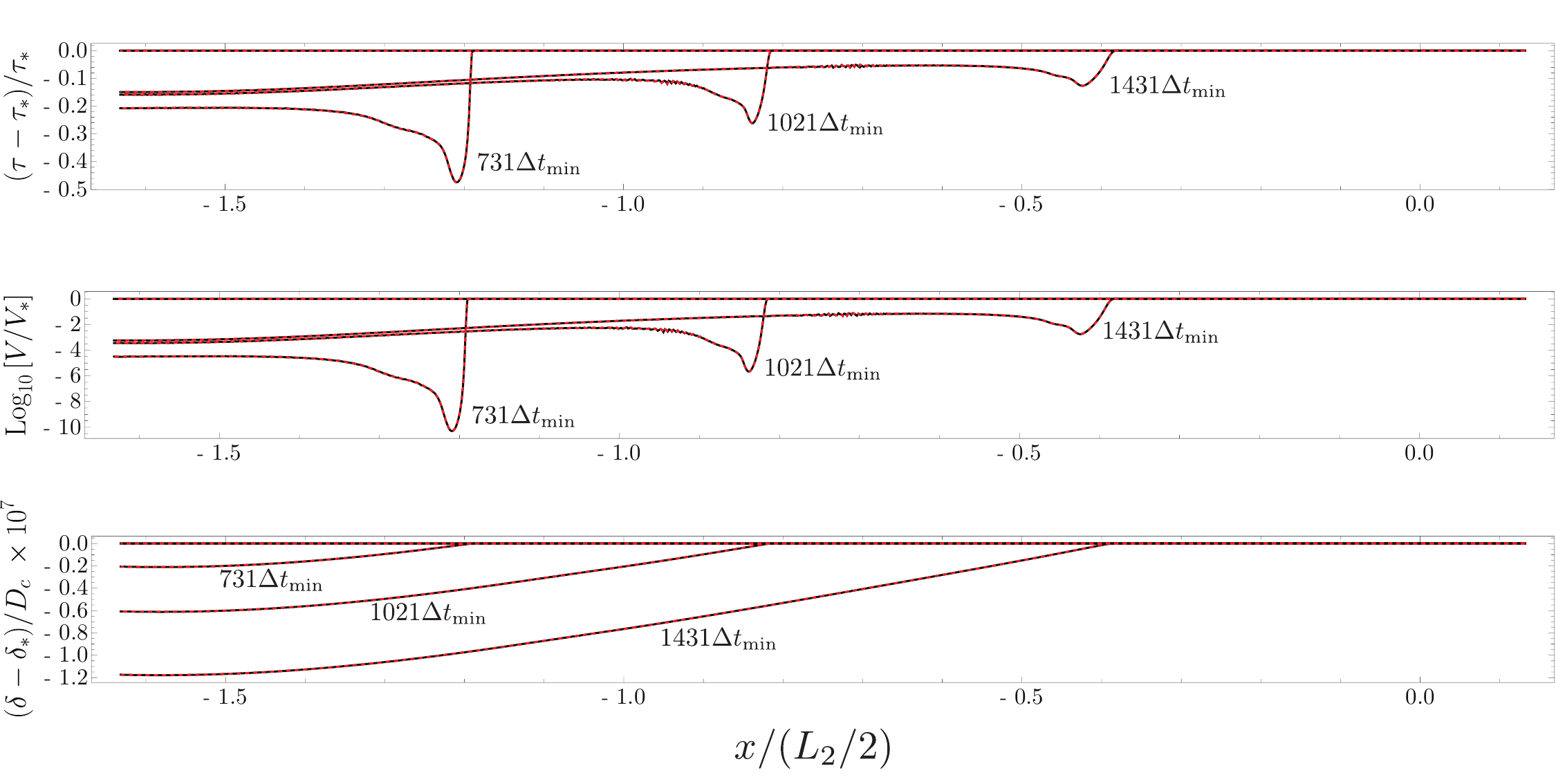}}\\[0.5cm] 
  \caption{Comparison between the proposed hybrid spectral / space--time formulation with $\mathcal{H}$-matrix acceleration (black solid lines) and the classical uncompressed space--time BIEM (red dashed lines). Results are shown for shear traction, slip rate, and slip at selected time snapshots on (a) fault $\Gamma_1$ and (b) fault $\Gamma_2$. The two solutions are in excellent agreement, indicating no observable loss of accuracy due to hierarchical compression.}
  \label{fig:Results_twofault_problem}
\end{figure}

%We first examine the structure and effectiveness of the $\mathcal{H}$-matrix compression applied to the fault--to--fault interaction kernels. Figure~4 shows the block partitioning and local compression ratios for the first temporal convolution slice of the discretized space--time convolution operator (\ref{eq:fault-to-fault_interactions_discretized}). 
%The left panel corresponds to the interaction from fault $\Gamma_1$ to fault $\Gamma_2$, while the right panel shows the reciprocal interaction from $\Gamma_2$ to $\Gamma_1$. For the $\mathcal{H}$-matrix parameters listed in Table~\ref{tab:Table1}, all sub-blocks satisfy the admissibility condition and are therefore represented in low-rank form, as indicated by the uniformly green coloring.
%The reported compression ratios demonstrate that the interaction blocks are highly compressible, resulting in a total compression ratio (obtained by summing the compression ratios of all temporal convolution slices) of $\sim 24.7\%$, and hence in a considerable computational speed-up, as shown below.

We first assess the accuracy of the proposed hybrid approach. Figure~\ref{fig:Results_twofault_problem} compares the accelerated formulation (black solid lines) with the classical space--time BIEM without compression (red dashed lines), the latter being assumed to be the reference numerical solution. The two solutions are in excellent agreement for all reported scaled quantities, including shear traction, slip rate, and slip, on both faults and at all time snapshots shown. This confirms that the combination of the hybrid formulation and $\mathcal{H}$-matrix compression introduces no observable loss of accuracy, at least with the input parameter chosen in this simulation.\\

\begin{figure}[t!]
\centering
\subfloat[]{{\includegraphics[width=0.65\textwidth]{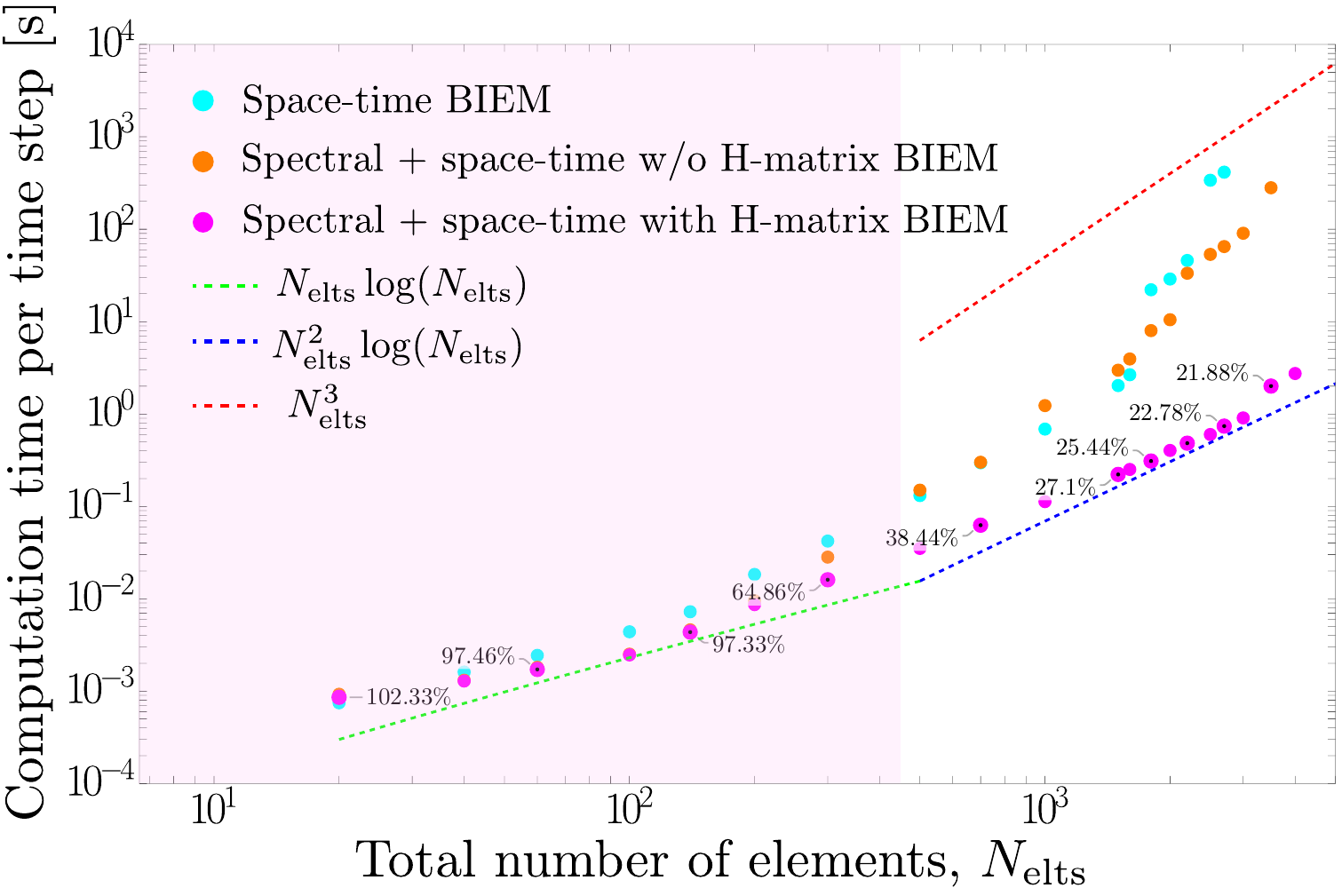} }}
\,\,
\subfloat[]{{\includegraphics[width=0.65\textwidth]{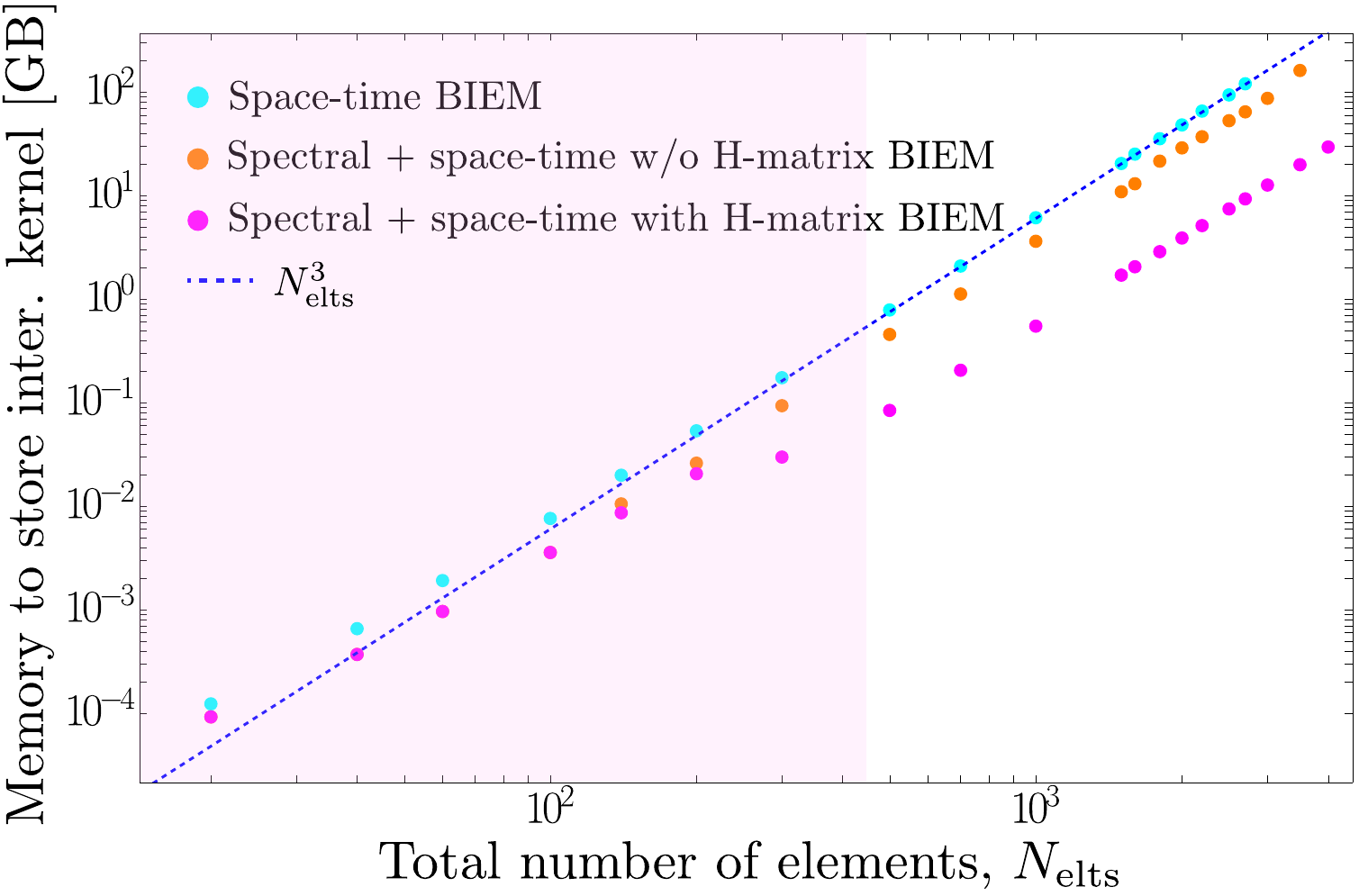} }}\\[0.5cm] 
\caption{Computational performance comparison for the two--fault problem as a function of the total number of fault elements $N_{\mathrm{elts}} = N_{\mathrm{elts},\Gamma_1} + N_{\mathrm{elts},\Gamma_2}$. 
(a) Computation time per time step (in seconds). 
(b) Memory used to store pre-integrated interaction kernel. 
For sufficiently large $N_{\mathrm{elts}}$, the hybrid spectral / space--time formulation with \(\mathcal{H}\)-matrix acceleration achieves approximately three orders of magnitude reduction in computation time and one order of magnitude reduction in memory usage compared to the classical space--time BIEM. Pink regions denote the simulation results obtained with cohesive zone length-scales resolved with less than 10 elements.}
\label{fig:performance}
\end{figure}

Finally, Figure~\ref{fig:performance} quantifies the computational benefits of the proposed approach. Panel (a) shows the computation time per time step as a function of the total number of fault elements $N_{\mathrm{elts}}$, while panel (b) reports the corresponding memory requirements associated with the storage of interaction kernels.

Focusing first on Figure~\ref{fig:performance}a, the magenta symbols correspond to the new, hybrid spectral / space--time formulation with $\mathcal{H}$-matrix compression. For relatively small values of total number of elements $N_{\mathrm{elts}} = N_{\mathrm{elts},{\Gamma_1}}+N_{\mathrm{elts},{\Gamma_2}}$ (with $N_{\mathrm{elts},{\Gamma_1}}= N_{\mathrm{elts},{\Gamma_2}}$), the computational cost is dominated by the spatial convolution performed at each time step. Self-effects are evaluated using the spectral formulation, which has $\mathcal{O}(N_{\mathrm{elts}}\log N_{\mathrm{elts}})$ computational complexity (thanks to FFT), while fault--to--fault interaction effects are calculated via $\mathcal{H}$-matrix--vector multiplications, which exhibit the same $\mathcal{O}(N_{\mathrm{elts}}\log (N_{\mathrm{elts}}))$ scaling. As a result, the overall cost follows an $N_{\mathrm{elts}}\log (N_{\mathrm{elts}})$ trend for small problem sizes.

As the number of fault elements increases, the number of temporal convolution steps increases accordingly. The spatial discretization is $\Delta x \sim L_{\text{fault}}/N_{\mathrm{elts}}$, and the minimum stable time step therefore scales as $\Delta t_{\min}\propto \Delta x/c_s$. For a fixed simulation duration $T_{\mathrm{end}}$, this implies that the number of time steps required for the space--time convolution grows proportionally to $N_{\mathrm{elts}}$. Consequently, repeating an $\mathcal{O}(N_{\mathrm{elts}}\log N_{\mathrm{elts}})$ spatial operation over $\mathcal{O}(N_{\mathrm{elts}})$ time steps yields the observed $\mathcal{O}(N_{\mathrm{elts}}^2\log N_{\mathrm{elts}})$ scaling for the $\mathcal{H}$-accelerated hybrid formulation at larger problem sizes.

The results reported in Figure~\ref{fig:performance} are obtained using fixed values of the \(\mathcal{H}\)-matrix admissibility and compression parameters (see Table \ref{tab:Table1}). A systematic study of the impact of different compression levels, in particular of the admissibility parameter $\eta_{\mathcal{H}}$, on computational performance and solution accuracy is presented in Appendix~\ref{app:Appendix3}.

The orange symbols in Figure~\ref{fig:performance} correspond to the hybrid spectral / space--time formulation without $\mathcal{H}$-matrix compression, while the cyan symbols represent the classical space--time BIEM. For both uncompressed approaches, the spatial application of interaction effects involves dense matrix--vector products with $\mathcal{O}(N_{\mathrm{elts}}^2)$ complexity per time step. Accordingly, for small $N_{\mathrm{elts}}$, when the number of time steps remains limited, the total cost scales approximately quadratically. As $N_{\mathrm{elts}}$ increases and the number of time steps grows proportionally to $N_{\mathrm{elts}}$, the overall scaling transitions to $\mathcal{O}(N_{\mathrm{elts}}^3)$ for both formulations. The orange curve lies below the cyan curve due to the spectral treatment of self-effects, which reduces the prefactor of the computational cost even though the interaction effects remain uncompressed.

Taken together, these results demonstrate a reduction of approximately three orders of magnitude in computation time per time step compared to the classical, uncompressed space--time BIEM, for a sufficiently large number of elements. In practical terms, a simulation that would require on the order of one year of wall-clock time using the traditional space--time formulation can be completed in approximately 10 hours with the proposed hybrid spectral / space--time approach.\\

%The pink-shaded regions in Figure~6(a) denote simulations that are not sufficiently resolved. In these cases, the cohesive-zone length scale is discretized with fewer than the minimum required number of elements (here taken as more than 10).

Figure~\ref{fig:performance}b shows the memory requirements as a function of the total number of fault elements $N_{\mathrm{elts}}$. For all formulations, the memory usage ultimately scales as $\mathcal{O}(N_{\mathrm{elts}}^3)$. This cubic scaling arises because the dominant memory cost is associated with the discretized fault-to-fault interaction kernel, which can be interpreted as a three-dimensional space--time object (or ``cube''): two spatial dimensions corresponding to the discretizations of the source and receiver faults, and one temporal dimension associated with the discretized interaction history. Since the number of time steps grows proportionally to $N_{\mathrm{elts}}$, the total number of stored kernel entries scales as $\mathcal{O}(N_{\mathrm{elts}}^2 \times N_{\mathrm{elts}})=\mathcal{O}(N_{\mathrm{elts}}^3)$.

\begin{figure}[t!]
\centering
\noindent\includegraphics[width=0.95\textwidth]{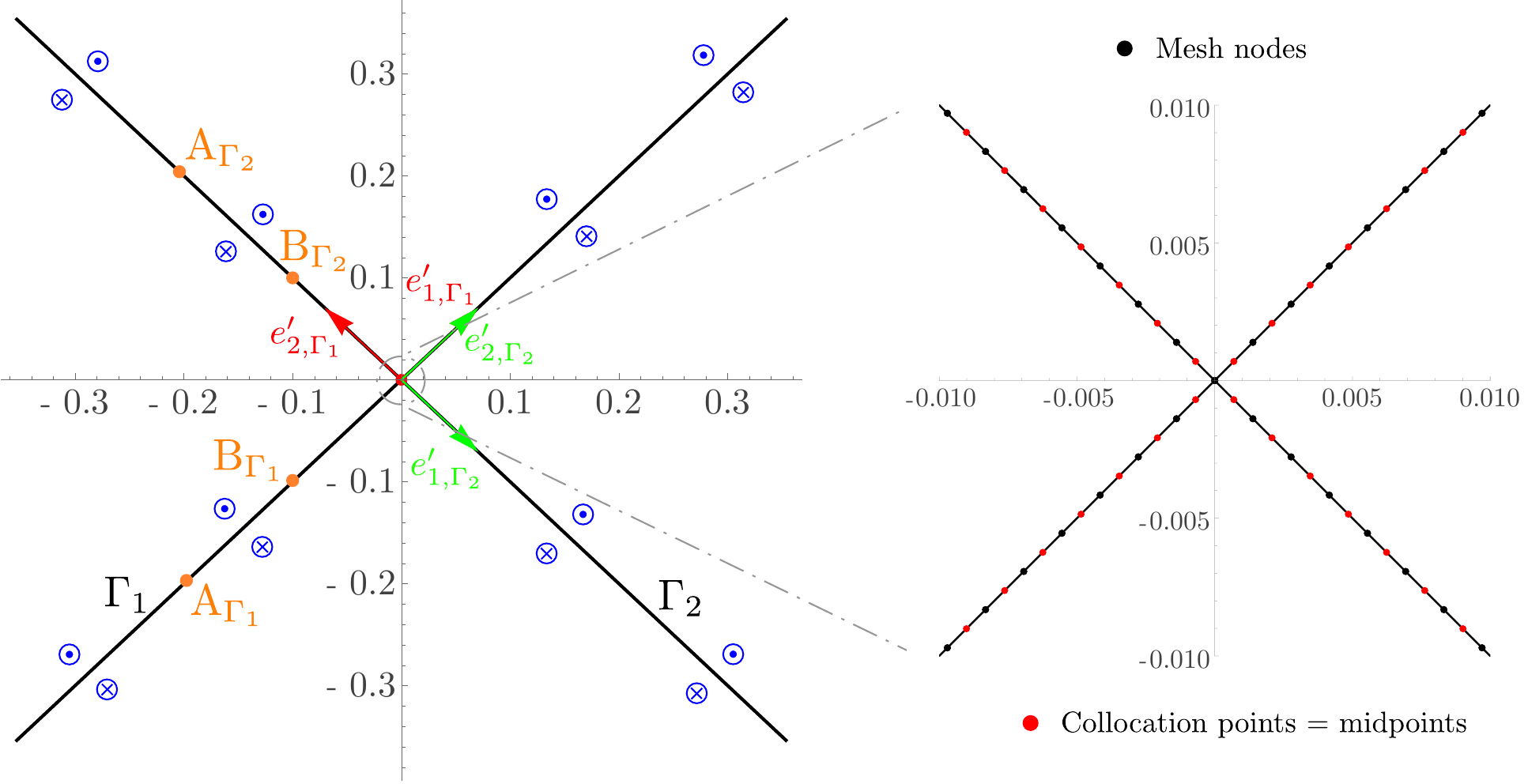}
\\[0.5cm]
\caption{Left: Geometry of the X-shaped fault problem. Two planar faults of equal length intersect at their midpoints $(0,0)$, forming a symmetric X-shaped configuration. Axes of the global reference system are expressed in kilometers. Orange points denote the observation points for slip accumulation plot in Fig. \ref{fig:Xfault_results}-right. Right: Zoom at intersecting point. Red points denote the collocation points for elasticity which correspond to mesh midpoints. Black points instead denote the mesh nodes.}
\label{fig:X_fault_problem}
\end{figure}

Although the asymptotic scaling is cubic for all methods, the prefactor differs substantially between formulations. In particular, the $\mathcal{H}$-accelerated hybrid spectral / space--time formulation exhibits a significantly smaller prefactor due to the low-rank representation of admissible blocks within the space--time interaction kernel, resulting in an approximately one order of magnitude reduction in memory usage compared to the classical, uncompressed space--time BIEM. The hybrid spectral / space--time formulation without $\mathcal{H}$-matrix compression (orange symbols) lies slightly below the classical space--time BIEM (cyan symbols), reflecting the spectral evaluation of self-effects, which reduces the memory footprint of the self-interaction contribution even though the fault-to-fault interaction kernels remain dense. As a result, while the asymptotic scaling remains $\mathcal{O}(N_{\mathrm{elts}}^3)$, the proposed $\mathcal{H}$-accelerated approach enables simulations on personal-computer hardware at problem sizes that are impractical for uncompressed space--time formulations.

\subsection{X-shaped fault problem}
\label{subsec:X-fault_problem}

We next consider a problem consisting of an X-shaped fault system, formed by two pre-existing planar faults intersecting at their midpoints (see Figure~\ref{fig:X_fault_problem}). Each fault has equal length of 1\,Km and symmetric orientation.

The first motivation for this configuration is to demonstrate that the proposed hybrid spectral / accelerated--space--time BIEM can naturally accommodate intersecting faults/fractures. In the present discretization, the fault intersection coincides with a mesh node; however, since all boundary integral equations are collocated at element centroids (or midpoints--see Zoom in Fig. \ref{fig:X_fault_problem}-right), no singular collocation points arise and no special numerical treatment is required at the intersection.

\begin{table}[t!]
  \centering
  \begin{tabular}{lcl}
    \hline
    Name & symbol & Value \\
    \hline
    Reference friction coefficient      & $f_*$     & 0.6 \\
    Reference velocity                  & $V_*$     & $10^{-9}\,\mathrm{m/s}$ \\
    Critical slip distance   (uniform)           & $D_c$     & $10^{-3}\,\mathrm{m}$ \\
    Rate and state parameter (uniform)           & $b$       & $(1.2/0.9) \cdot 10^{-2}$ \\
    Rate and state parameter  (uniform)          & $a$       & $1.2 \cdot 10^{-2}$ \\
    State evolution law parameter               & $\varepsilon$   & 1 (Aging law)\\
    \hline
    Shear modulus                       & $\mu$      & $30\,\mathrm{GPa}$ \\
    Shear velocity                      & $c_s$      & $3464\,\mathrm{m/s}$ \\
    Initial normal stress     & $\sigma_n^o$ & $100\,\mathrm{MPa}$ \\
    Background stressing rate     & $\dot{\tau^o}$ & $0.01\,\mathrm{Pa/s}$ \\
    \hline
    Tolerance Singular Value Decomposition          & $\varepsilon_{\text{SVD}}$        & $10^{-4}$ \\
    Minimum number of elements in a leaf node  & $\eta_{\text{leaf}}$ & $50$ \\
    Safety factor $\mathcal{H}$-matrix admissibility condition & $\eta_{\mathcal{H}}$      & $1.5$ \\
    Simulation time               & $T_{\text{end}}$       & $\sim 100\,\mathrm{years}$ \\
    Minimum time step               & $\Delta t_{\text{min}}$       & $\sim 2.83 \times 10^{-4}\,\mathrm{s}$ \\
    Number of elements on $\Gamma_1$               & $N_{\text{elts},\Gamma_1}$       & $510$ \\
    Number of elements on $\Gamma_2$               & $N_{\text{elts},\Gamma_2}$       & $510$ \\
    Total length of fault $\Gamma_1$               & $L_{1}$       & $1\,\mathrm{Km}$ \\
    Total length of fault $\Gamma_2$               & $L_{2}$       & $1\,\mathrm{Km}$ \\
    Truncation factor for interaction histories & $\eta_{\mathcal{T}_{\min}}$       & $10^{-2}$ \\
    Truncation factor for interaction histories   & $\eta_{\mathcal{T}_{\max}}$       & $10^{-2};10^{-1};1;2$ \\
    Truncation factor for self-effect histories & $q_w$       & $102$ \\
    Truncation factor for self-effect histories   & $\eta_w$       & $1.5$ \\
    \hline
  \end{tabular}\\[0.5cm]
\caption{Input parameters for X-shape fault problem.}
\label{tab:Table2}
\end{table}

The second motivation is to provide a test of the truncation strategy adopted for the double space--time convolution of fault--to--fault interaction effects. In contrast to the two--fault configuration discussed in Section~\ref{subsec:two_fault_problem}, the X--shaped geometry is characterized by very short interaction distances between fault branches, leading to strong interaction effects due to elastic waves that are activated at early times. Under plane--strain conditions, the elastodynamic interaction kernels associated with anti--plane slip \eqref{eq:kernel_interactions} decay only slowly in time and approach their static limits only asymptotically as $t \to \infty$.
As a consequence, truncating the interaction history over a finite time window introduces an approximation whose accuracy depends on the chosen truncation parameters. The X--shaped fault system thus represents a near ``worst--case scenario'' for the adopted truncation strategy, since the interacting fault segments are in close proximity and the corresponding interaction kernels remain far from negligible over the (truncated) elastodynamic time window.

In this example, fault slip is driven by a uniform tectonic stressing rate, $\dot{\tau}^o = 0.01\,\mathrm{Pa/s}$, acting on all fault branches. This loading provides the sole driving mechanism for dynamic slip nucleation and evolution in the X-shaped system.

Each fault is discretized using 510 uniformly sized elements, ensuring that the cohesive-zone length scale $\Lambda_0$ is resolved with approximately ten elements. The adopted faults length (1~km) also exceeds the theoretical nucleation length required for the onset of dynamic slip. For the aging law ($\varepsilon = 1$) and $a/b = 0.9$ considered here, this nucleation length may be estimated as \citep{rubin_earthquake_2005,viesca_self-similar_2016}
\begin{equation}
L_{\mathrm{nuc}} = \frac{L_b}{\pi (1 - a/b)^2} \approx 716~\text{m},
\end{equation}
where the elasto-frictional length-scale $L_b$ is defined in Eq.~\eqref{eq:cohesive_zone_length_scale} and can be evaluated from the input parameters summarized in Table~\ref{tab:Table2}.

As reported in Table~\ref{tab:Table2}, we consider four different truncation times obtained by varying the parameter $\eta_{\mathcal{T}_{\max}}$. Specifically, the final convolution time $\mathcal{T}_{\text{end}}^{\,i'j'}$ for the interaction effects is extended beyond the shear-wave arrival time at the farthest receiver point by 1\%, 10\%, 100\%, and 200\%, corresponding to $\eta_{\mathcal{T}_{\max}} = 10^{-2},\,10^{-1},\,1,$ and $2$, respectively.

Initial conditions are prescribed such that the two faults start from different frictional states. On fault~$\Gamma_1$, the system is initialized at steady state, with the state variable satisfying $\theta_{\text{init}} = D_c / V_{*}$, so that the nondimensional quantity $\Omega = V\theta_{\text{init}}/D_c = 1$. On fault~$\Gamma_2$, the initial state variable is increased to place the fault above steady state, with $\Omega = 1.5$. 

\begin{figure}
\centering
\subfloat[]{{\includegraphics[width=0.48\textwidth]{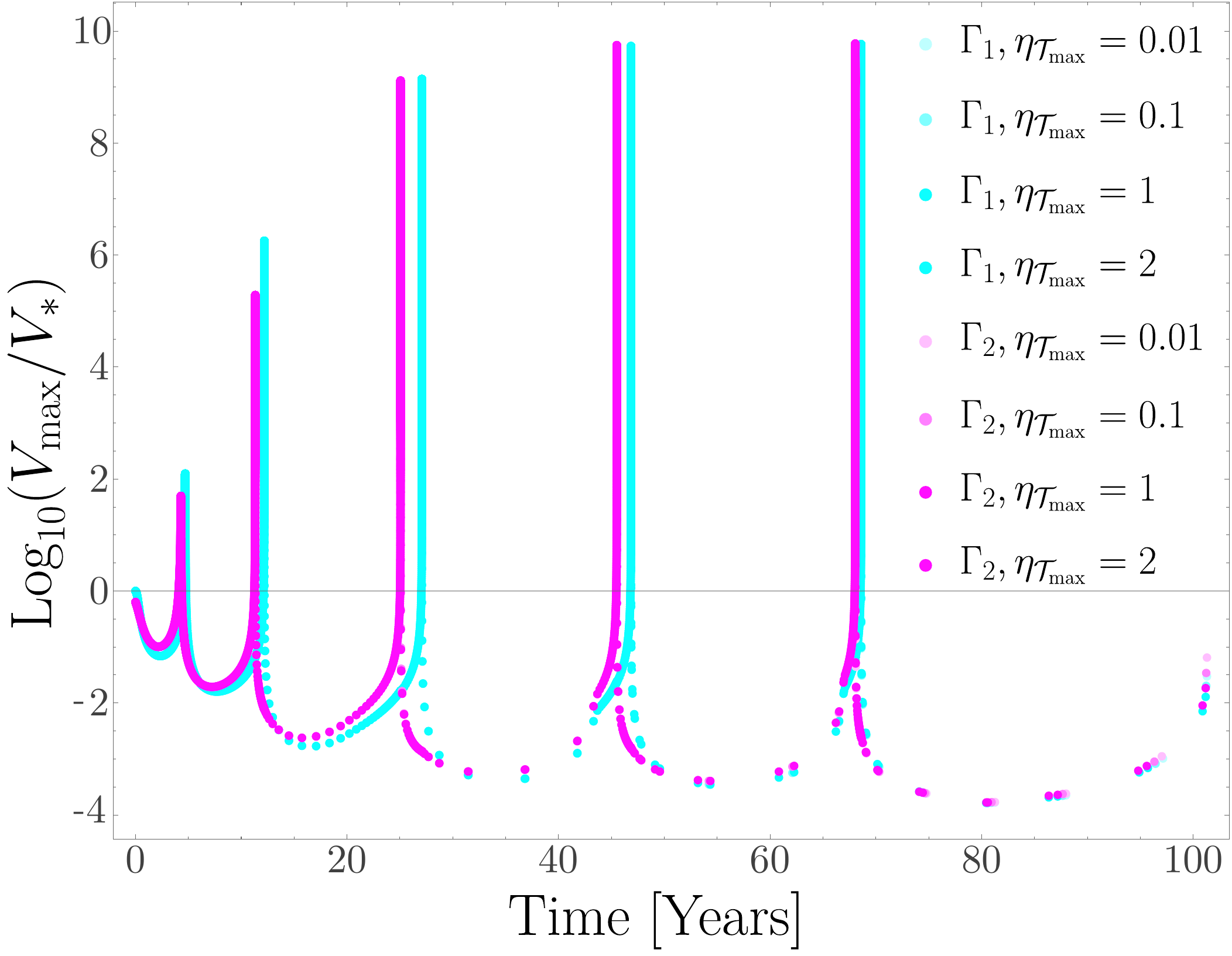}}}
\,\,
\subfloat[]{{\includegraphics[width=0.50\textwidth]{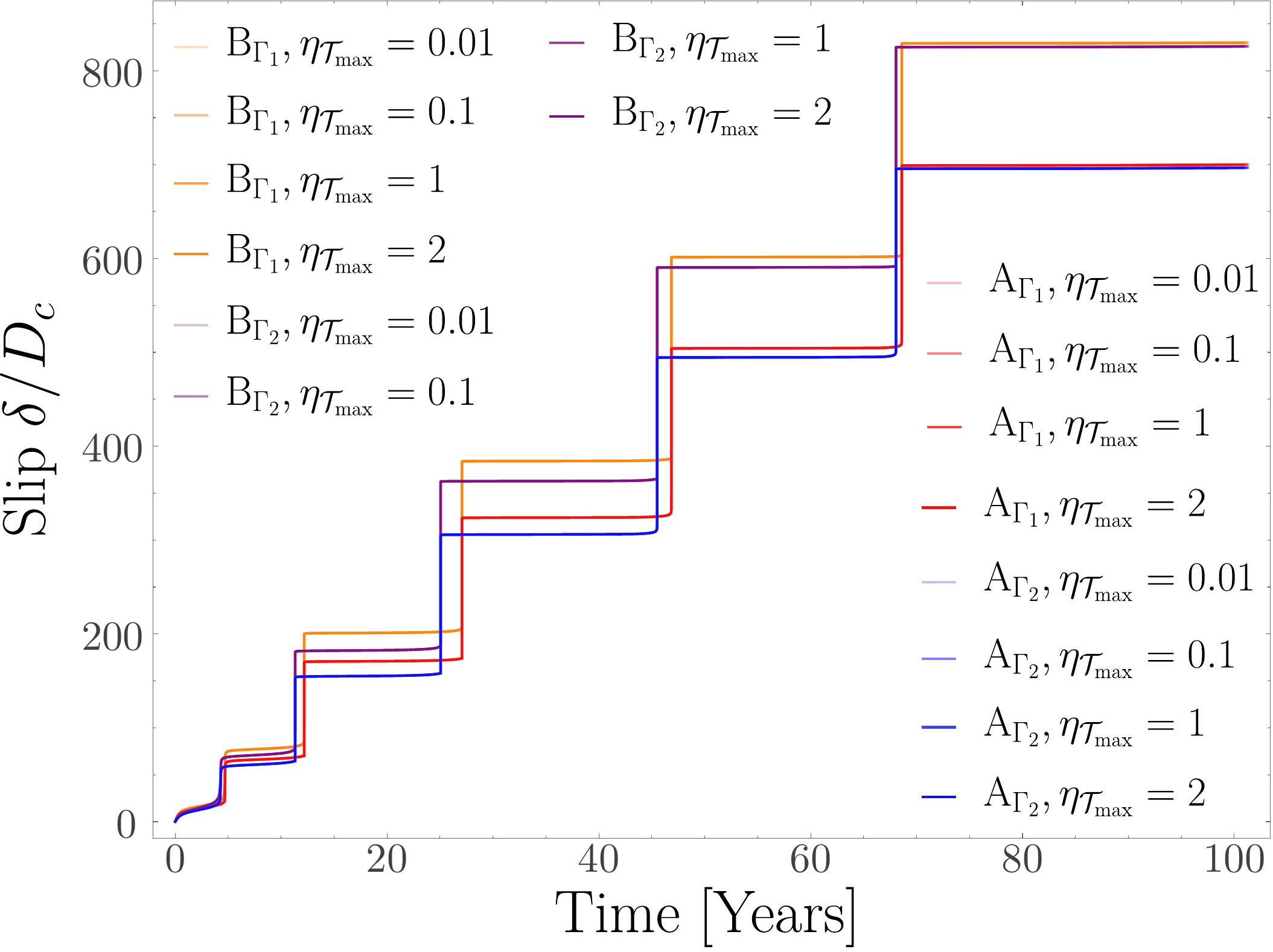}}}\\[0.5cm] 
\caption{Slip evolution in the X-shaped fault system for different truncation levels of the elastodynamic interaction history. 
(a) Maximum slip rate attained on faults $\Gamma_1$ (cyan) and $\Gamma_2$ (magenta) within the simulated time window $[0,100\,\mathrm{years}]$, shown for varying values of the truncation parameter $\eta_{T_{\max}}$. Different shades of the same color denote different truncation levels. 
(b) Temporal accumulation of slip at observation points A and B (see Figure~\ref{fig:X_fault_problem}), located on the two fault branches. Different shades denote results obtained with different truncation levels at each observation point. For a given observation point, the corresponding slip histories are basically indistinguishable.}
\label{fig:Xfault_results}
\end{figure}

Figure~\ref{fig:Xfault_results} summarizes the resulting slip evolution for the different truncation levels. Panel~(a) reports the maximum slip rate achieved on each fault over the course of the simulation for all values of the truncation parameter $\eta_{T_{\max}}$. Results for faults $\Gamma_1$ and $\Gamma_2$ are distinguished by color, with cyan and magenta symbols, respectively, while different shades denote simulations performed with different truncation levels.

Firstly we note, as expected, that the maximum slip rate on fault $\Gamma_2$ systematically occurs earlier than on fault $\Gamma_1$ (within the simulated time window $[0,100\,\text{years}]$). This temporal offset reflects the different initial conditions prescribed on the two faults: fault $\Gamma_2$ is initialized above steady state, which promotes earlier slip acceleration relative to fault $\Gamma_1$, initialized at steady state. Despite this difference in timing, both faults reach seismic slip rates as the system evolves.

We also observe that, for a given fault, the maximum slip rate is essentially insensitive to the truncation of the elastodynamic interaction history. Points corresponding to different values of $\eta_{T_{\max}}$ plot nearly on top of one another, indicating that the peak slip rate attained during both fast and slow slip episodes is not affected by the adopted truncation window.

A similar lack of sensitivity is observed for the cumulative slip. Panel~(b) shows the temporal evolution of slip accumulation at two representative observation points, A and B (see Figure~\ref{fig:X_fault_problem}), located on the two fault branches. The slip histories at these points are essentially identical for all values of $\eta_{T_{\max}}$, demonstrating that the long-term slip evolution is likewise unaffected by the truncation level considered here.

\begin{table}[t!]
\centering
\begin{tabular}{ccc}
\hline
$\eta_{T_{\max}}$ & Total compression ratio (\%) & Runtime (hours) \\
\hline
0.01 & 51.49 & $\sim$0.88 \\
0.1  & 44.75 & $\sim$1.02 \\
1    & 28.24 & $\sim$2.80 \\
2    & 25.02 & $\sim$4.53 \\
\hline
\end{tabular}
\\[0.5cm]
\caption{Compression ratios and total runtimes for the X-shaped fault simulations at different truncation levels.}
\label{tab:Table3}
\end{table}

Finally, we examine the associated computational performance. Table \ref{tab:Table3} summarizes the compression ratios achieved for the interaction kernels together with the corresponding total simulation runtimes, obtained using identical $\mathcal{H}$-matrix input parameters.
%As expected, increasing the truncation window (i.e., larger $\eta_{T_{\max}}$) leads to a substantial increase in total runtime, reflecting the larger interaction history retained in the space--time convolution.
As expected, increasing the truncation window (i.e., larger $\eta_{T_{\max}}$) leads to a substantial increase in total runtime, reflecting the larger interaction history retained in the space--time convolution for fault--to--fault interaction effects. This trend is accompanied by a reduction in compression ratio: although it decreases from 51.49\% for $\eta_{T_{\max}}=0.01$ to 25.02\% for $\eta_{T_{\max}}=2$, the corresponding $\sim 200\%$ extension of the elastodynamic time window results in an increase in total runtime by more than a factor of five.

\subsection{Slip dynamics in the extensional Corinth rift, Greece}

Finally, to illustrate the performance of the proposed approach on a large--scale fault system with realistic geometrical complexity, we consider a fault network representative of the active extensional regime of the Gulf of Corinth, Greece. The Corinth rift is one of the most seismically active continental rifts in Europe and is characterized by a network of normal faults accommodating N--S extension also in the form of shallow slow-slip events \citep{mildon_transient_2024}. 

The fault geometry used in this example is derived from the mapping reported in \citep{nixon_increasing_2024}. Specifically, we consider $44$ major normal faults from their Fig.~1A that belong to the active offshore and onshore rift system. Only the major active faults were retained, while secondary splays and minor structures were neglected. Each mapped fault trace was approximated as planar. Free-surface effects are neglected.

\begin{figure}[t!]
\centering
\noindent\includegraphics[width=\textwidth]{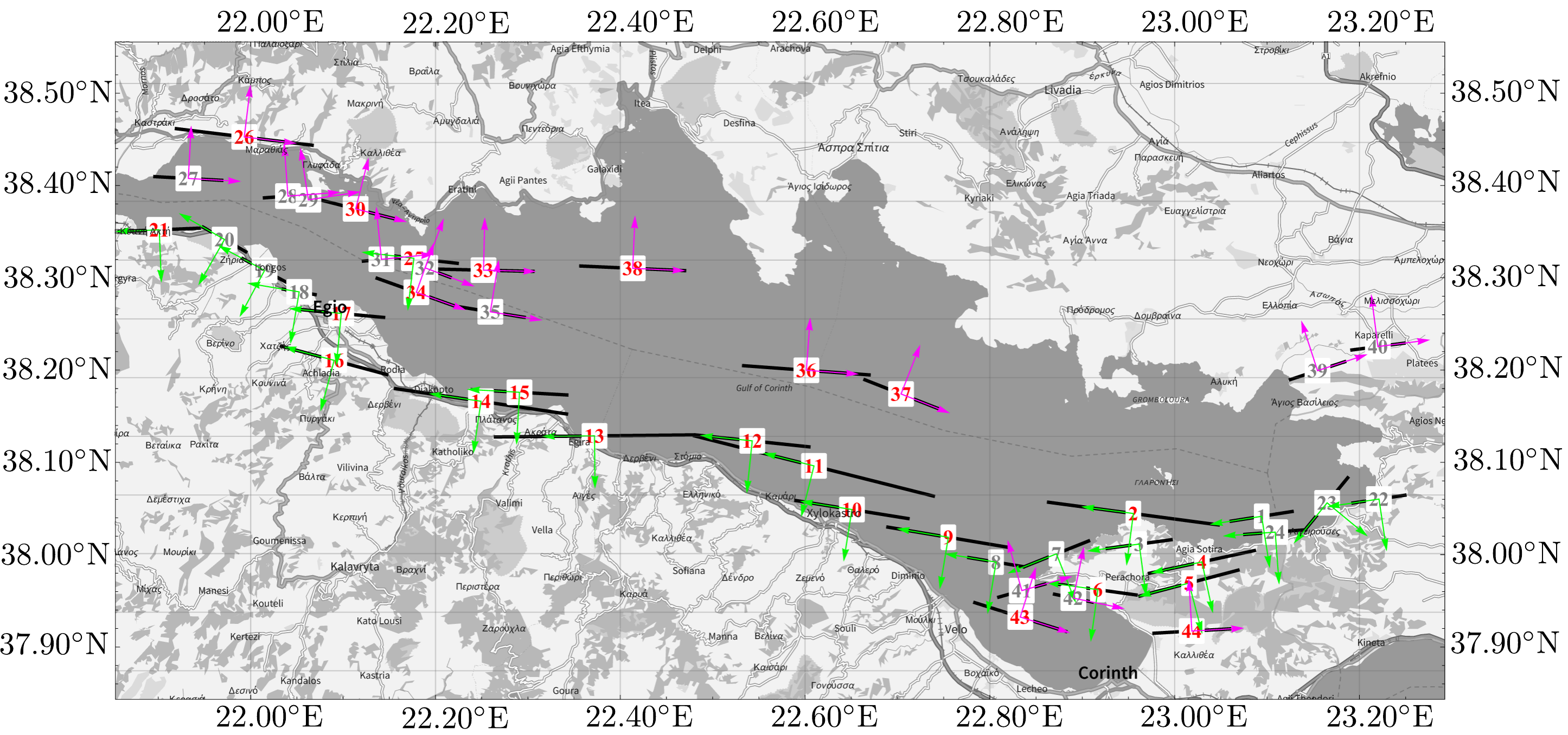}
\\[0.5cm]
\caption{
Fault geometry adopted for the Gulf of Corinth large--scale problem. The network is constructed by mapping 44 major normal faults from
Fig.~1A of \citet{nixon_increasing_2024}. Only the major active faults are retained, while secondary splays are neglected. Each fault trace is approximated as
planar lines (see black, thick lines). Faults with magenta local reference systems are dipping South, whereas faults with green reference system are dipping North. Faults numbered with red numbers have lengths exceeding the theoretical nucleation length $L_{\text{nuc}}\approx 7162$~m, while those labeled in gray are shorter than the corresponding nucleation size.}
\label{fig:Golf_Corinth_problem}
\end{figure}
The resulting geometry is shown in Fig.~\ref{fig:Golf_Corinth_problem}. Because the Corinth rift includes both north--dipping and south--dipping normal faults, we differentiate the slip sign convention according to the dip direction. North-dipping faults are associated with one orientation of the local coordinate system (in particular one orientation of the normal vector), whereas south-dipping faults adopt the opposite orientation. In Fig.~\ref{fig:Golf_Corinth_problem}, these reference systems are illustrated using different colors: green for north dip direction and magenta for south dip direction.\\ 

It is important to stress that the objective of this example is not to reproduce or match observed seismicity, slip distributions, or geodetic measurements from rift. Instead, this test case is designed to evaluate the scalability the proposed divide--and--conquer strategy when applied to a large-scale fault network. Accordingly, the frictional and loading parameters adopted in the simulations are not tuned to the Corinth rift and should not be interpreted as site--specific. %Uniform rate--and--state friction properties together with simplified background stressing conditions are prescribed across all fault segments.
Indeed, the input frictional, material, and loading parameters are essentially the same as those reported in Table~\ref{tab:Table2}, with the exception that the critical slip distance is set here to $D_c = 10^{-2}$~m. This choice leads to a characteristic length-scale $L_b = 225$~m and a theoretical nucleation length $L_{\text{nuc}} = \dfrac{L_b}{\pi (1-a/b)^2} \approx 7162~\text{m}.$

Many of the pre-existing faults, denoted with red integer labels in Fig.~\ref{fig:Golf_Corinth_problem}, have total lengths exceeding this nucleation length and are therefore potentially seismogenic. In total, $24$ out of the $44$ modeled faults satisfy $L > L_c$. The remaining faults, labelled with a gray number, are shorter than the nucleation size and can only host aseismic slip transients.

Initial conditions are prescribed such that $\Omega = \dfrac{V\theta_{\text{int}}}{D_c} = 1.5$ (with $\theta_{\text{init}} = D_c/V_*$)
on all faults, meaning that the system initially lies above steady state. 
All faults are discretized using a sufficiently fine mesh of straight elements to resolve the cohesive-zone length scale $\Lambda = 9\pi L_b/32$ with at least four elements.
As far as $\mathcal{H}$-matrix compression is concerned, the parameters adopted are $q_w = 78$, $\eta_w = 1.5$,  $\eta_{\text{leaf}} = 50$, $\eta_H = 1.5$.
Regarding the truncation of elastodynamic interaction histories, we set $\eta_{t_{\min}} = \eta_{t_{\max}} = 10^{-2}$, consistent with the X-fault problem reported in section \ref{subsec:X-fault_problem}, where truncation was shown not to introduce additional error.
The total simulated duration is approximately $32$~years, during which each fault is loaded by a uniform background tectonic stressing rate of $0.01$~Pa/s.\\

\begin{figure}
\centering
\subfloat[]{{\includegraphics[width=0.48\textwidth]{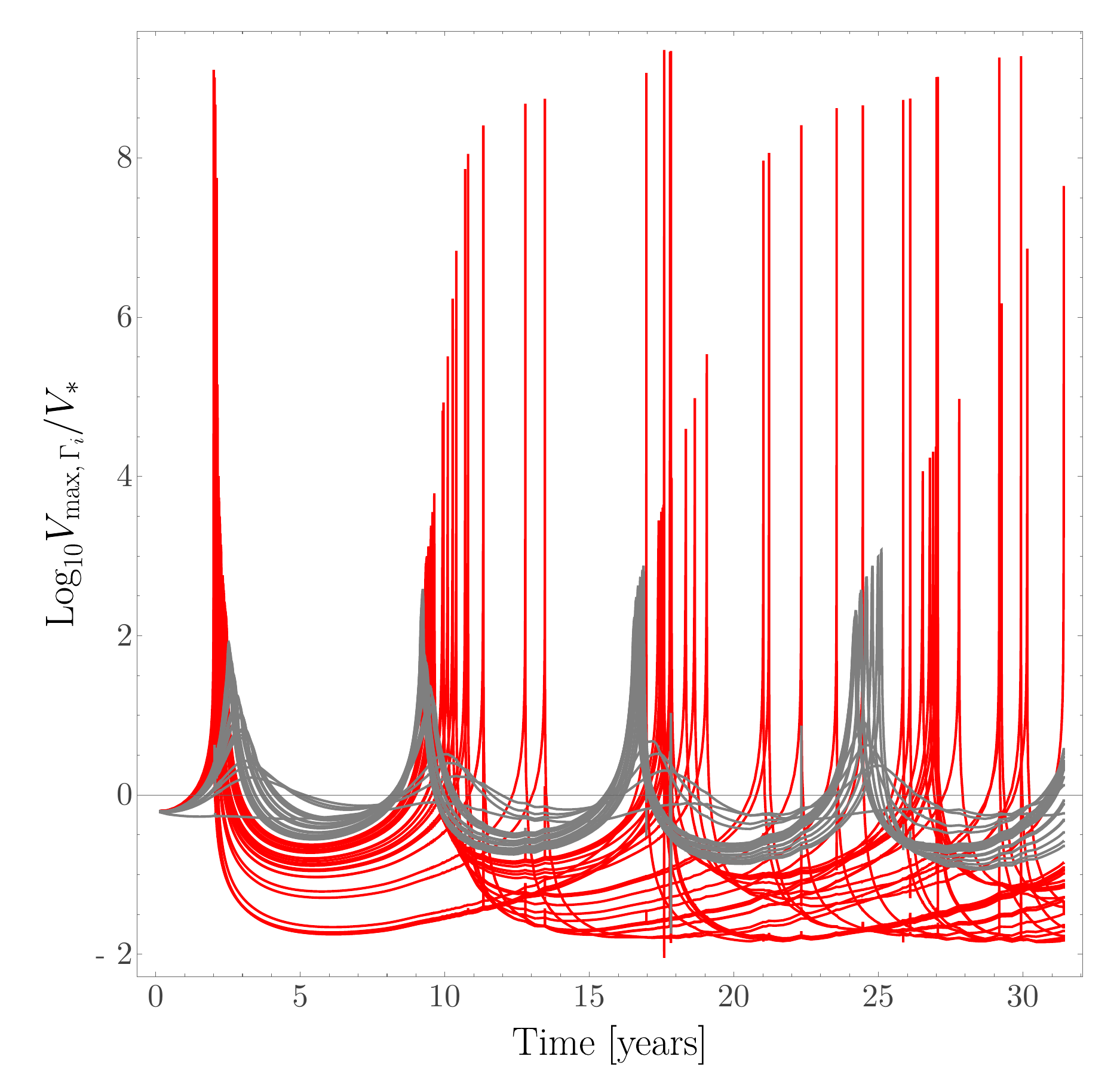} }}
\,\,
\subfloat[]{{\includegraphics[width=0.48\textwidth]{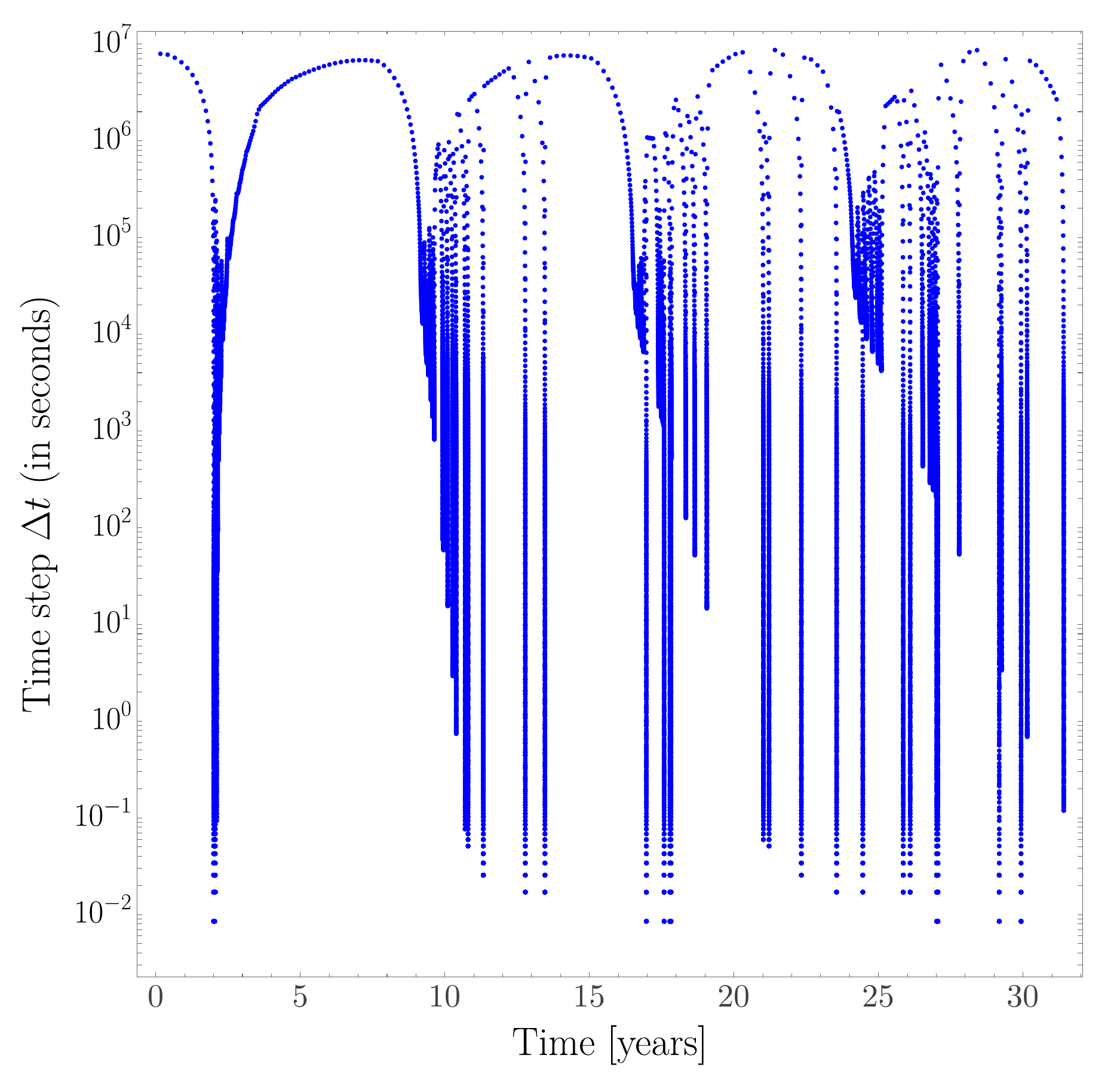} }}\\[0.5cm] 
\caption{Left: Logarithmic of maximum slip rate attained on each fault as a function of time. Faults classified as potentially seismogenic (red curves), i.e., faults whose total length exceeds the theoretical nucleation length ($L > L_c$; red labels in Fig.~\ref{fig:Golf_Corinth_problem}), are shown in red. Faults shorter than the nucleation length ($L < L_c$), which can only host aseismic transients, are shown in gray. Right: Evolution of the adaptive time step used during simulation. The time step spans nearly nine orders of magnitude between interseismic and coseismic phases: large time steps accelerate the simulation during slow deformation, whereas the time step collapses to $\Delta t_{\min} \simeq 0.0085$~s during dynamic rupture propagation.}
\label{fig:plot1Corinth}
\end{figure}

Figure~\ref{fig:plot1Corinth} illustrates the coupled evolution of fault-wise maximum slip rates and adaptive time stepping, highlighting the multi-scale nature of earthquake sequences and aseismic slip (SEAS) as well as the associated computational challenges.

The left panel reports the maximum slip rate attained on each fault as a function of time (in years). Faults classified as potentially seismogenic (red curves) undergo episodic dynamic slip instabilities, with scaled slip rates exceeding $\text{Log}_{10} V_{\max}/V_* = 9$ (i.e. $V_{\max}\gtrsim1$~m/s). These events are characterized by rapid coseismic slip acceleration driven by inertial effects and elastodynamic wave radiation. In contrast, faults shorter than the nucleation length (gray curves) exhibit only transient slip accelerations. Their slip remains aseismic and does not transition into fully dynamic rupture.
Although elastodynamic interactions between faults are clearly visible through the presence of slip-rate spikes on many segments (both red and gray curves), dynamic ruptures occurring on seismogenic faults do not trigger inertia-driven slip on non-seismogenic faults. Even when coseismic slip rates on seismogenic faults reach dynamic levels, the maximum slip rates attained on sub-nucleation faults remain well below dynamic instability threshold. In particular, the gray curves never exceed $\text{Log}_{10}(V_{\max}/V_*) \approx 3$ (i.e., $V \lesssim 10^{-6}$~m/s), which can be regarded as a threshold for slow-slip events. 
This indicates that, while stress perturbations are transmitted elastodynamically across the network, they remain insufficient to activate inertial instability on faults whose length is below the nucleation size. This behavior is consistent with theoretical predictions: under the Aging law formulation, slip is unconditionally stable to finite perturbations when the fault length satisfies $L < L_{\text{nuc}}$, as demonstrated by \citet{ciardo_nonlinear_2025}.

%\begin{figure}[t!]
%\centering
%\noindent\includegraphics[width=\textwidth]{Images/SnapshotsSlipRatesCorinth_compressed.pdf}
%\\[0.5cm]
%\caption{Spatial distribution of slip rate across the Gulf of Corinth fault network at three representative time snapshots centered around the nucleation and propagation of a dynamic rupture on Fault~7. Gray: snapshot taken immediately prior to rupture nucleation. Red: snapshot during dynamic rupture propagation.  Blue: snapshot taken shortly after rupture has traversed the entire fault length.}
%\label{fig:plot2Corinth}
%\end{figure}
\begin{figure}[t!]
  \centering
    \textcolor{gray}{$\left( t - t_{\text{nucl}} \right)\approx - 5.2411$ days;} \quad  \textcolor{blue}{$\left( t - t_{\text{nucl}} \right) \approx 9.79$ s;}  \quad \textcolor{red}{$\left(t - t_{\text{nucl}} \right) \approx 13.14$ s;}\\[0.1cm]
  % -------- Row 1 --------
  \includegraphics[width=0.76\textwidth]{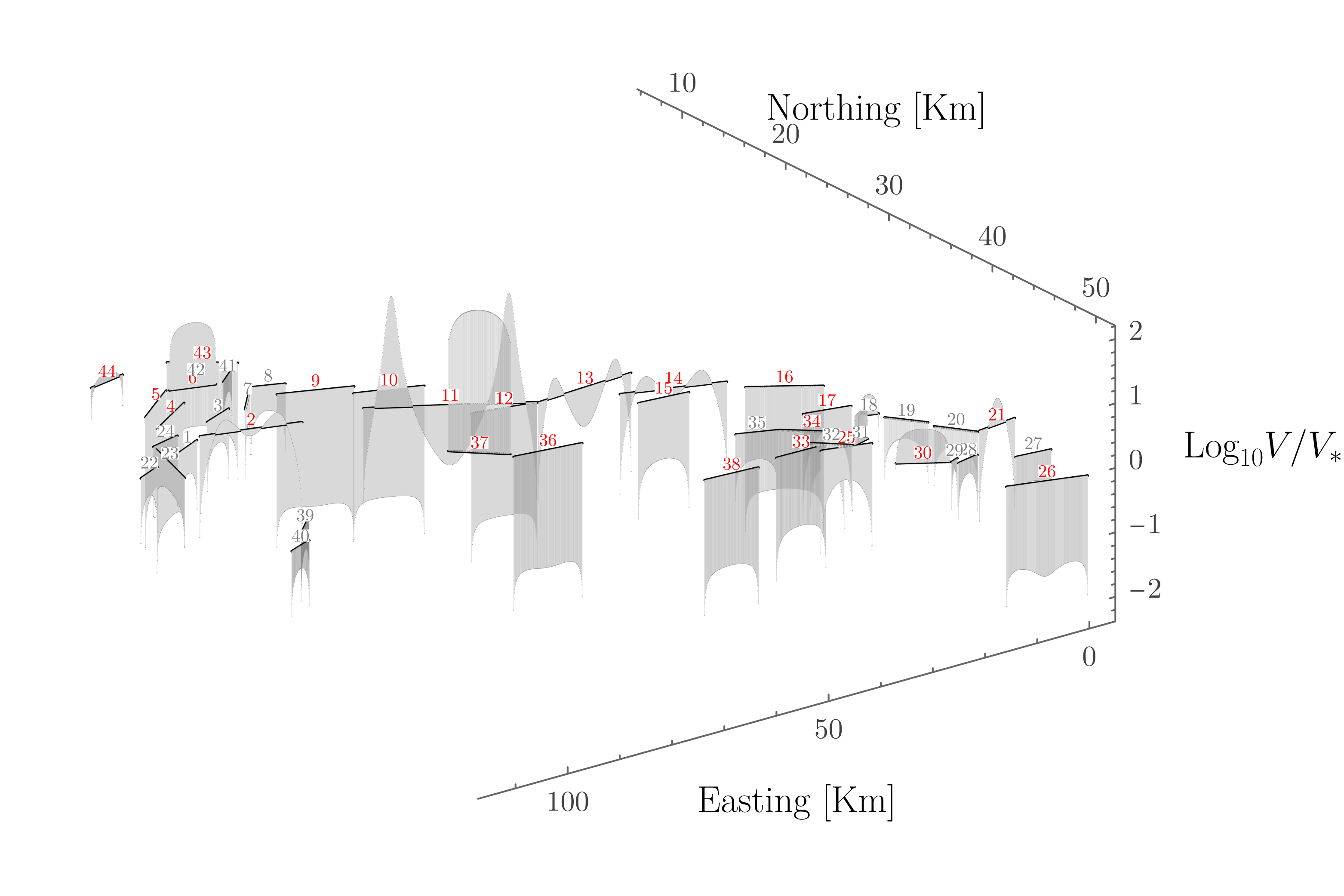}\\[-0.6cm]
  \includegraphics[width=0.76\textwidth]{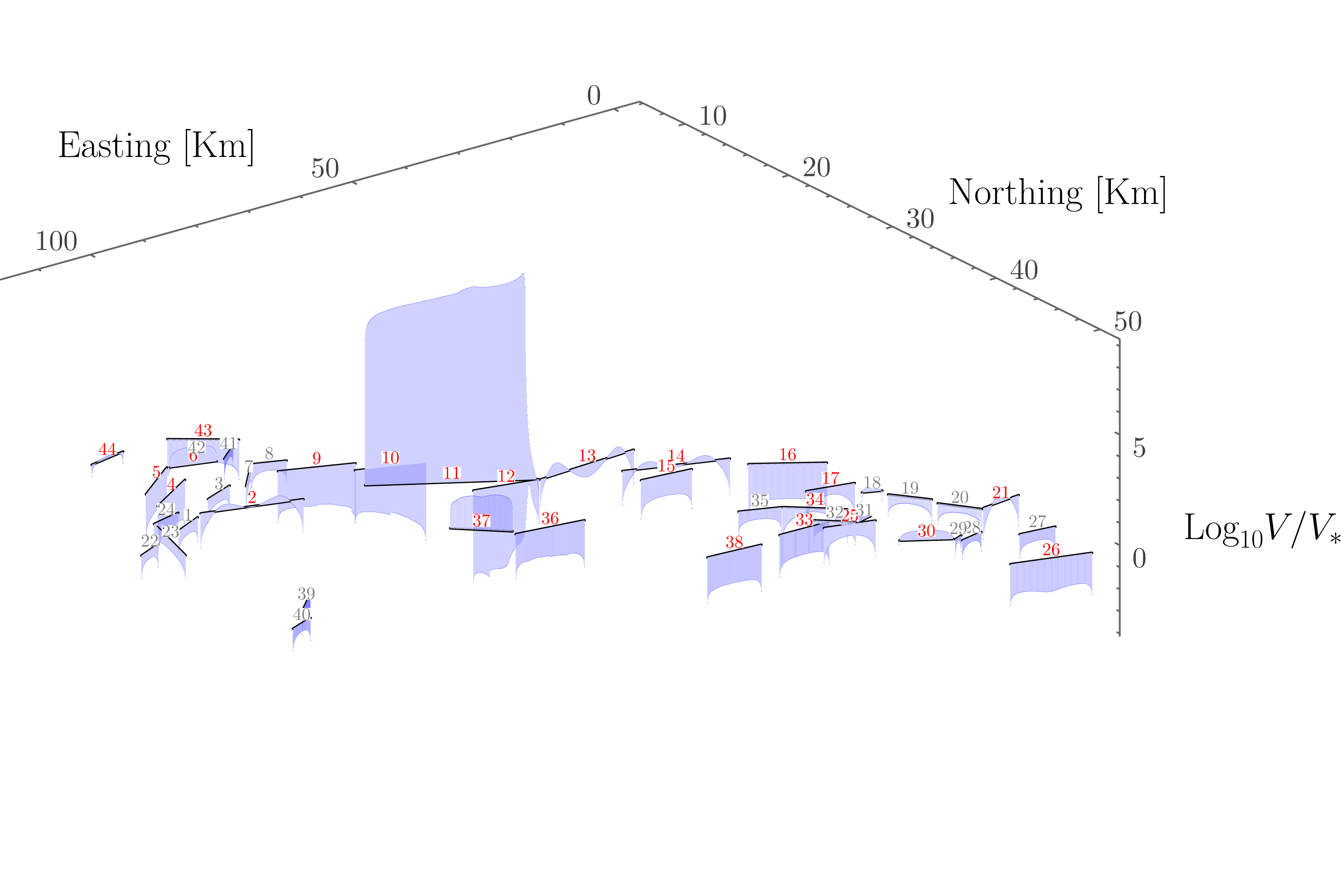}\\[-1.8cm]
    \includegraphics[width=0.76\textwidth]{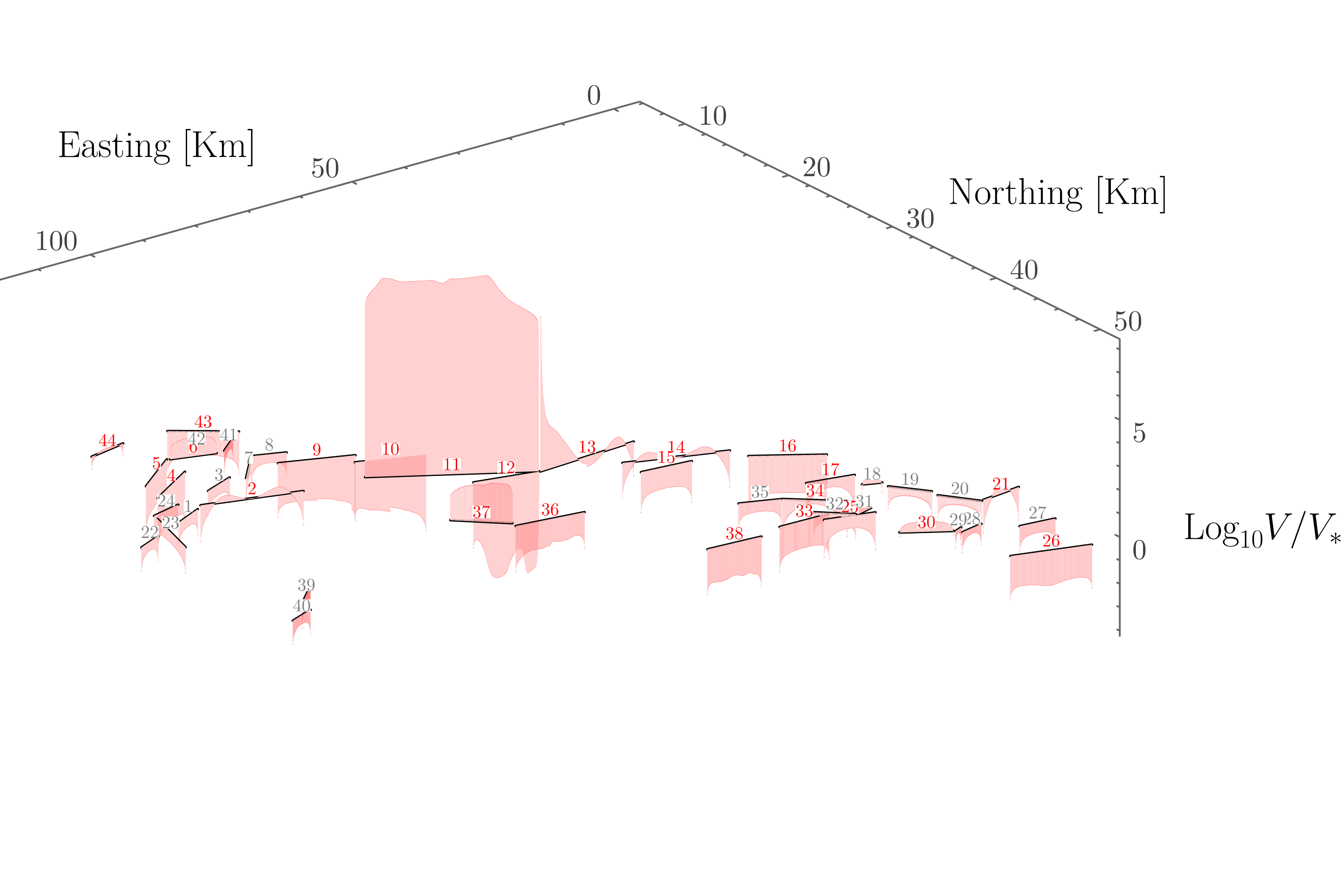}\\[-1.5 cm]
  \caption{Spatial distribution of slip rate across the Gulf of Corinth fault network at three representative time snapshots centered around the nucleation and propagation of a dynamic rupture on Fault~\ref{fig:plot3Corinth}. Gray: snapshot taken immediately prior to rupture nucleation. Red: snapshot during dynamic rupture propagation. Blue: snapshot taken shortly after rupture has traversed the entire fault length.}
  \label{fig:plot2Corinth}
\end{figure}

Taken together, the two panels illustrate the intrinsic complexity of modeling SEAS on large fault networks. Elastodynamic interactions generate strongly heterogeneous slip dynamics, with alternating periods of slow aseismic deformation and abrupt dynamic rupture. This multi-scale behavior imposes severe constraints on time integration.
The right panel of Fig. \ref{fig:plot1Corinth} shows the evolution of the adaptive time step used during numerical integration. The time step must decrease dramatically during dynamic rupture in order to resolve rapid slip acceleration and elastodynamic wave propagation, while it can increase substantially during interseismic and slow-slip phases to maintain computational efficiency. In the present simulation, the adaptive time step spans nearly nine orders of magnitude, collapsing to $\Delta t_{\min} \simeq 8.5\times10^{-3}$~s during dynamic events and reaching values on the order of years during quiescent phases.\\  %This extreme variability highlights both the numerical stiffness of the governing equations and the necessity of adaptive time stepping for tractable long-term simulations of earthquake sequences on complex fault systems.

%To visualize how rupture propagates and interacts across the network, Fig.~\ref{fig:plot2Corinth} shows the spatial distribution of slip rate on all faults at three representative time snapshots.
The temporal analysis discussed above provide insight into when and on which faults dynamic instabilities occur, but it does not reveal how rupture propagation and stress transfer manifest spatially across the fault network. To address this aspect, Fig.~\ref{fig:plot2Corinth} presents the spatial distribution of slip rate on all faults at three representative time snapshots taken to be close to the nucleation and propagation of a dynamic rupture on Fault~11, which occurs at $t_{\text{nucl}} \approx 17.58$ years $\approx6418.43$ days  $\approx 5.54552\cdot10^8$ seconds (see below for its definition).
The first snapshot (gray) is taken immediately prior to dynamic rupture nucleation, roughly $\sim 5$ days before it. At this stage, slip rates across the entire network remain below values typically associated with the onset of slow-slip events (i.e., $V \sim 10^{-6}\,\mathrm{m/s}$). The second snapshot (red) captures the dynamic rupture propagation phase, during which slip rates increase by several orders of magnitude along Fault~11. The third snapshot (blue) is taken shortly after rupture has traversed the entire fault length.
These maps reveal pronounced interaction effects across the surrounding fault system. Faults~10 and~9, which are sub-parallel to Fault~11 and share the same slip sign, experience a reduction in slip rate during rupture, as does Fault~12. In contrast, Fault~13—lying approximately along the geometric continuation of Fault~11—undergoes a marked increase in slip rate. 

\begin{figure}
\centering
\subfloat[Fault \textcolor{red}{11}]{{\includegraphics[width=0.48\textwidth]{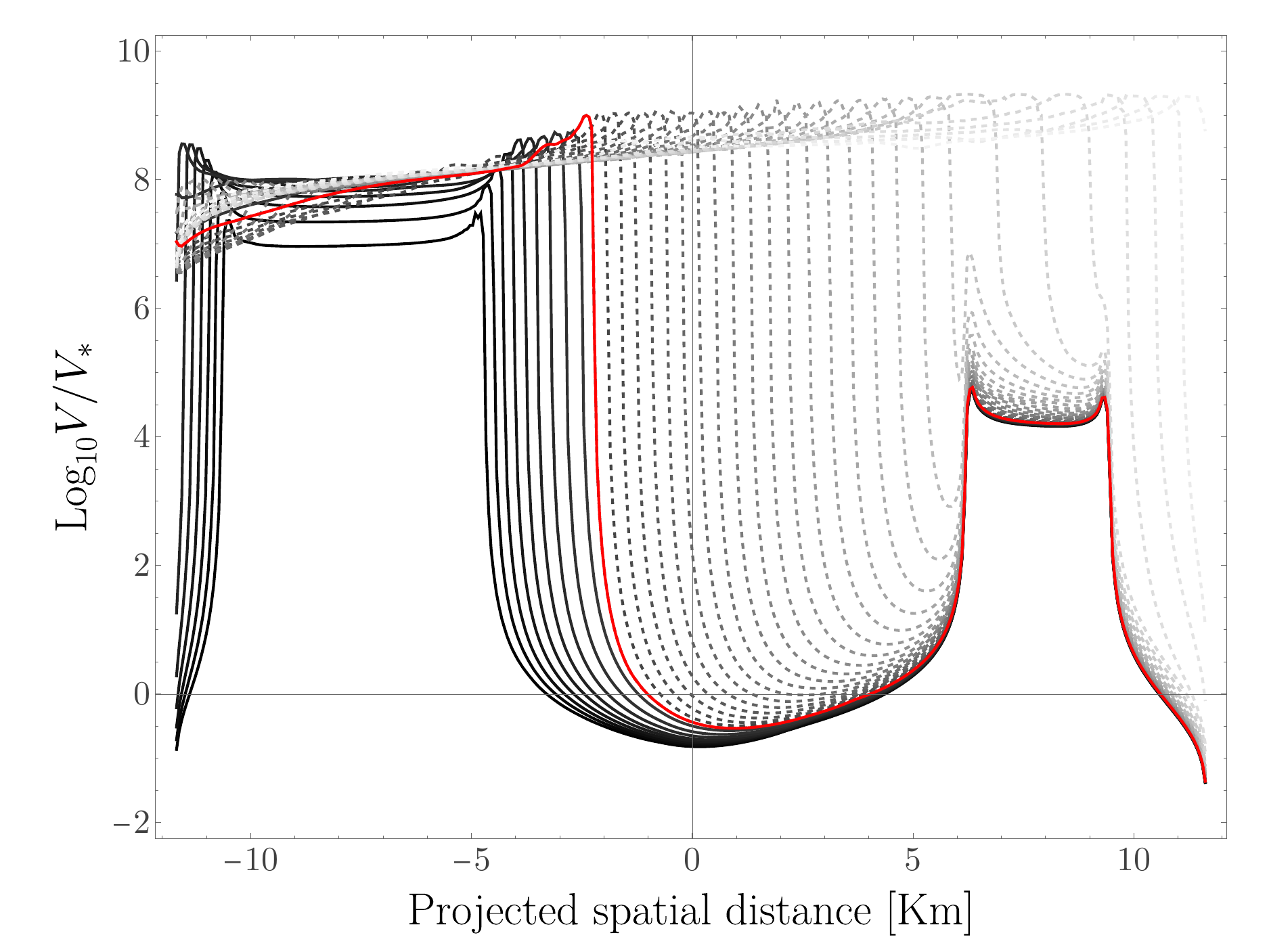} }}
\,\,
\subfloat[Fault \textcolor{red}{10}]{{\includegraphics[width=0.48\textwidth]{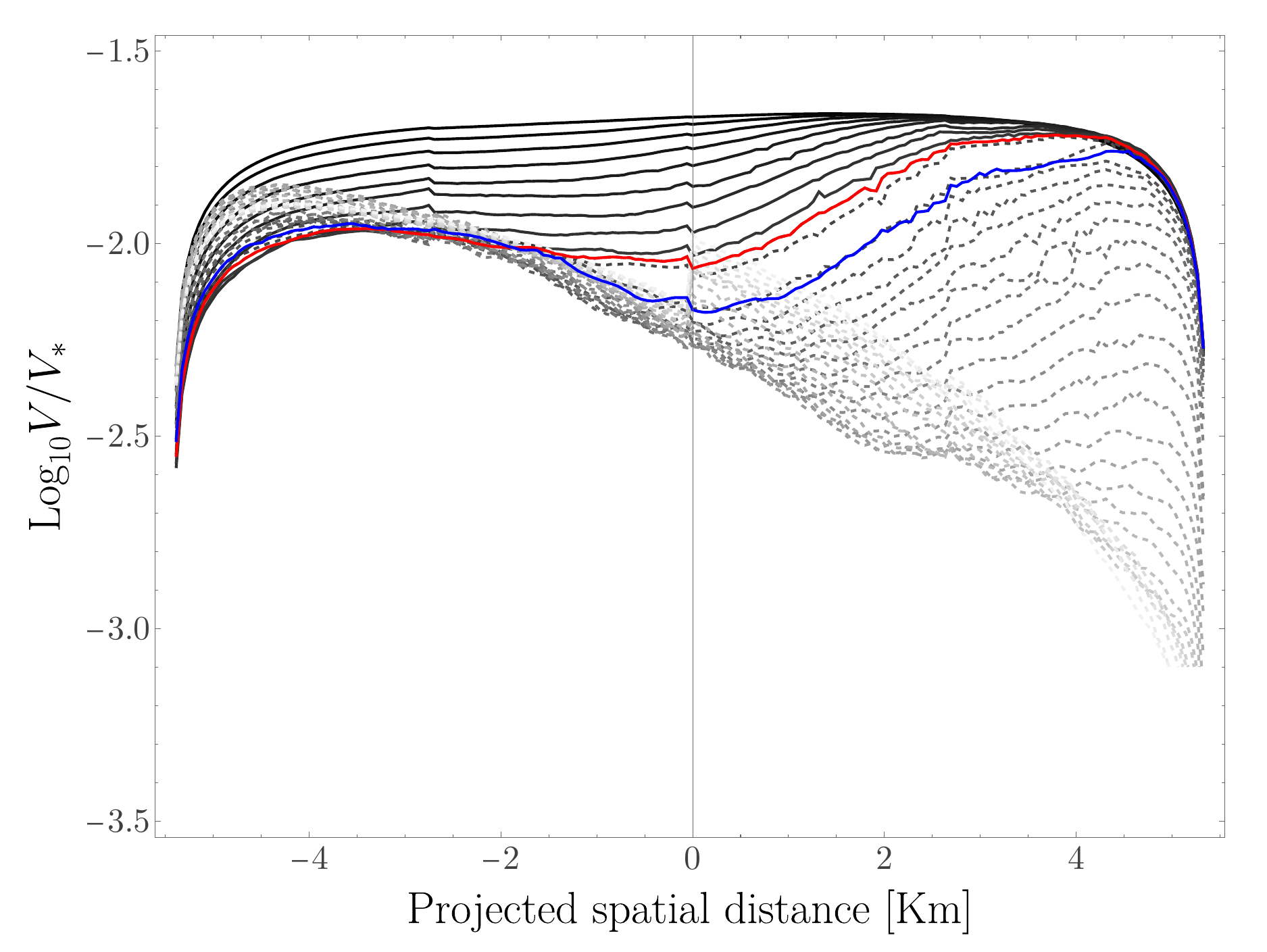} }}\\[0.5cm] 
\caption{Slip-rate profiles illustrating elastodynamic interaction between Fault~11 and Fault~10.
Left: Snapshots of slip-rate distribution along Fault~11. The red curve denotes the slip-rate distribution at the time of dynamic rupture nucleation, defined as the instant at which $\text{Log}_{10} V/V_* = 9$. The corresponding nucleation point is identified as the centroid node that first reaches this slip rate. Dashed curves denote subsequent coseismic
profiles plotted at constant time intervals equal to $50\,\Delta t_{\min}$, with gray shading that tracks rupture propagation along the fault.
Right: Slip-rate profiles along Fault~10. The blue curve corresponds to the slip-rate distribution evaluated at the theoretically predicted shear-wave arrival time (see Eq.~\ref{eq:predicted_time}). A pronounced negative jump in slip rate is observed. This abrupt variation constitutes a clear signature of arrival of the elastic shear wave radiated by the dynamic rupture on Fault~11.}
\label{fig:plot3Corinth}
\end{figure}

To further isolate elastodynamic interaction effects, Fig.~\ref{fig:plot3Corinth} reports slip-rate profiles along the source fault (Fault~11, left panel) and a representative receiver fault (Fault~10, right panel).

The red curve in Fig.~\ref{fig:plot3Corinth}a denotes the slip-rate distribution at the time of rupture nucleation. Nucleation is defined as the instant at which the slip rate first reaches $\text{Log}_{10}(V/V_*) = 9$. The corresponding nucleation point is identified as the centroid node along Fault~11 that first attains this slip rate.
Using this reference, the theoretical shear-wave arrival time on Fault~10 can be estimated as
\begin{equation}
t_{\text{arr}} = t_{\text{nucl}} + \min_{i \in \Gamma_7,\; j \in \Gamma_6} \left(\frac{r_{ij}}{c_s} \right),
\label{eq:predicted_time}
\end{equation}
where $r_{ij}$ denotes the Euclidean distance between source node $i$ located on Fault~11 (fault surface $\Gamma_7$) and receiver node $j$ located on Fault~10 (fault surface $\Gamma_6$), and $c_s$ is the shear-wave speed of the medium. The minimum operator selects the shortest source--receiver travel path, which controls the first arrival of the radiated shear wave on the receiver fault.

Coseismic slip-rate profiles are plotted at constant time intervals equal to $50\,\Delta t_{\min}$, with gray shading tracking rupture propagation along the source fault. In Fig.~\ref{fig:plot3Corinth}b, the blue profile corresponds to the slip-rate distribution along Fault~10 evaluated at the predicted arrival time $t_{\text{arr}}$. A pronounced negative jump in slip rate is observed between two successive coseismic profiles. This abrupt variation constitutes a clear signature of the arrival of the elastodynamic shear wave radiated by the dynamic rupture on Fault~11.
Importantly, this wave-induced perturbation can be unambiguously attributed to rupture propagation on Fault~11 alone. As shown by the pre-nucleation snapshot in Fig.~\ref{fig:plot2Corinth} (gray profiles), all faults in the network are undergoing slow aseismic slip prior to nucleation, with slip rates remaining below $10^{-6}$~m/s. Consequently, the sudden slip-rate drop observed on Fault~10 cannot result from background fault interactions or ongoing large aseismic transients, but instead reflects the direct arrival of the elastodynamic stress wave radiated by the ``source" fault 11.

\section{Discussions}

\subsection{Displacement versus velocity formulation for elastodynamic self-effects}

Spectral boundary integral formulations for elastodynamic fault slip with or without replication can be written either in terms of displacement or, after integration by parts in time, in terms of slip rate. The latter formulation, adopted for instance in the seminal work by \citet{lapusta_elastodynamic_2000}, offers several advantages. In particular, the quasi-static contribution to elastic traction is recovered exactly, and during slow (quasi-static) deformation the time-convolution term can be entirely avoided. As a result, algorithms that use the velocity formulation can naturally switch from a quasi-dynamic to a fully elastodynamic regime only when slip rates become sufficiently large, leading to computational savings when long interseismic phases dominate the simulation time.

In this contribution, however, we adopt the displacement formulation for the computation of elastodynamic self-effects (see Equations~(\ref{eq:functional_no_replication}--\ref{eq:non_replicant_kernel})): a temporal convolution is performed at any time regardless of the maximum slip rate achieved on pre-existing faults. This choice is motivated by a combination of numerical robustness and practical considerations that are specific to the non-replicating spectral formulation and to long-term simulations of earthquake sequences.

First, within the spectral formulation without spatial replication, the velocity formulation introduces an additional convolution kernel whose numerical evaluation is prohibitively expensive \citep{noda_dynamic_2021}. In contrast, the displacement formulation leads to convolution kernels that remain tractable and can be efficiently evaluated numerically for each spectral mode.

Second, recent community benchmark studies highlight an additional limitation of velocity-based formulations in long-term simulations when implemented within volumetric methods. In the fully dynamic SEAS benchmark BP1--FD, \citet{erickson_incorporating_2023} showed that several finite-difference and finite-element codes employing velocity formulations required explicit switching criteria to transition between quasi-dynamic and fully elastodynamic regimes. Different implementations relied on distinct thresholds—such as critical slip rates or nondimensional measures comparing quasi-static and dynamic contributions—which had to be carefully tuned to obtain quantitatively consistent results across codes. While good agreement could ultimately be achieved, the need for such user-defined switching parameters introduced additional implementation sensitivity. This behavior is partly rooted in the mixed elliptic--hyperbolic character of earthquake cycle simulations, which volumetric solvers must accommodate explicitly when transitioning between quasi-static and fully dynamic regimes.
It is important to emphasize that this requirement is not intrinsic to velocity formulations themselves, nor does it generally arise in boundary integral implementations, where elastodynamic radiation is treated analytically through convolution operators. The displacement formulation employed here therefore provides a conceptually simple framework in which elastodynamic effects are retained consistently throughout the entire seismic cycle without the introduction of ad hoc transition criteria. %This feature is particularly advantageous in earthquake sequence simulations, where slip rates span many orders of magnitude and repeated transitions between slow and fast deformation occur.

Third, the simulations presented here rely on a strongly adaptive time-stepping scheme, with time steps varying by up to nine orders of magnitude between slow interseismic deformation and fast coseismic slip (see for instance Fig. \ref{fig:plot1Corinth}-right). In practice, the overwhelming majority of time steps are taken during fast slip episodes, while slow deformation is resolved using a comparatively small number of large steps. As a consequence, although the displacement formulation requires a time convolution to be evaluated at every time step, the cumulative computational cost associated with slow deformation remains marginal relative to the total runtime. The overall cost of the convolution is instead dominated by fast-slip phases, for which velocity formulations would in any case require full elastodynamic evaluation based on slip-rate history.

Finally, in the non-replicating spectral BIEM used here, the elastodynamic convolution kernel decays rapidly and monotonically once the nondimensional time ratio $c_s t / L$ exceeds unity \citep{cochard_spectral_1997} (see also Fig. \ref{fig:Kernels_decay} in Appendix \ref{app:Appendix4}). This property enables efficient truncation of the elastodynamic history for each spectral mode, substantially reducing the effective convolution length and further mitigating the additional cost associated with the displacement formulation.

\subsection{Introduction of a traction-free surface}

The hybrid spectral / accelerated-space-time BIEM presented in this work is based on boundary integral equations derived for an unbounded elastic medium (see Appendix \ref{app:Appendix1}). As a result, traction-free boundaries are not included explicitly and must be introduced through an additional modeling strategy.

A flexible approach consists in representing the free surface as an auxiliary boundary on which a zero-traction condition is enforced \citep{ando_deep_2010,ando_dynamic_2017}. Within the BIEM framework, this can be achieved by introducing an additional surface carrying unknown displacement discontinuities, whose sole role is to generate elastic fields that cancel the traction induced by fault slip and external loading at the free surface. The traction-free condition can then be imposed by requiring that the total shear and normal tractions vanish at all collocation points on this auxiliary surface.

%In practice, this auxiliary surface must extend well beyond the region of interest to prevent ``edge effects" from contaminating the solution near the faults. Although this increases the total number of degrees of freedom, the associated computational cost can be efficiently controlled within the present framework. In particular, interactions involving the free surface are typically long-range and therefore strongly compressible, making them well suited for hierarchical matrix approximation. This allows the free surface to be incorporated without a prohibitive increase in memory usage or runtime, at least in two-dimensional settings.
In practice, this auxiliary surface must extend well beyond the region of interest to prevent ``edge effects'' from contaminating the solution near the faults. Although this increases the total number of degrees of freedom, the associated computational cost can be efficiently controlled within the present framework. Importantly, self-interactions along the (planar) free surface are evaluated using the spectral boundary integral formulation, which scales as $\mathcal{O}(n \log n)$ and therefore keeps the additional computational cost modest.
Interactions involving the free surface are typically long-range and therefore strongly compressible, making them well suited for hierarchical matrix approximation. This allows the free surface to be incorporated without a prohibitive increase in memory usage or runtime, at least in two-dimensional settings.

More elegant and accurate alternative strategies for introducing free-surface effects, such as the use of modified Green’s functions that satisfy traction-free boundary conditions or image-based methods, are well established for simple geometries \citep{feng_exact_2018}. However, their implementation is cumbersome. %their extension to non-replicating spectral formulations and complex fault networks is not straightforward.

\subsection{Improvement of the algorithm}

Several algorithmic developments could further improve the efficiency and scalability of the proposed hybrid approach.

First, the current hierarchical compression relies on a binary tree decomposition of each fault geometry. Alternative spatial partitioning strategies, such as quad-tree (in two dimensions) or octree (in three dimensions) decompositions, are commonly used to construct cluster trees for hierarchical matrix methods \citep{hackbusch_sparse_1999, boermGrasedyckHackbusch2003}. These geometry-driven decompositions may provide more flexible and adaptive clustering for complex fault networks, potentially reducing the number of near-field interactions and improving compression efficiency.

Second, the present implementation employs truncated singular value decomposition to compress admissible blocks. While robust, Singular Value Decomposition is computationally expensive and may become a bottleneck for large-scale or three-dimensional simulations. Algebraic low-rank approximation techniques such as adaptive cross approximation (ACA) offer an efficient alternative, constructing low-rank representations using only selected rows and columns of the matrix \citep{bebenendorf2000approximation}. ACA is widely used in boundary element methods and is particularly well suited for compressing far-field interaction kernels (see e.g.\ \citep{ciardo_fast_2020}).

Third, the truncation of the elastodynamic time window for fault--to--fault interactions is currently defined at the level of entire fault pairs. A natural extension would consist in defining the truncation time $\mathcal{T}_{\text{end}}^{\,i'j'}$ at the cluster level, consistently with the hierarchical block structure used for spatial compression. Because the maximum arrival time of elastodynamic waves depends on the geometric separation between interacting clusters, a cluster-dependent truncation criterion would allow shorter time windows for well-separated cluster pairs, thereby reducing the storage and evaluation of unnecessary convolution history. %Such a refinement could avoid the need to select an overly small global $\eta_{t,\min}$ for a given fault pair and may further reduce both memory usage and computational time in large-scale simulations.

Finally, the computational cost associated with fault--to--fault interactions could be further reduced by incorporating domain-partitioning strategies in the space--time formulation. In particular, the Fast Domain Partitioning Method proposed by \cite{sato_log-linear_2021} for elastodynamic boundary integral equations exploits the causal structure of wave propagation to decompose the space--time interaction domain into regions associated with distinct wavefront arrivals. When applied to fault--to--fault interactions and combined with hierarchical low-rank compression, this approach has been shown to achieve log--linear scaling of both memory usage and computational cost in time-marching BIEMs.
Integrating such space--time domain partitioning concepts within the present hybrid framework—specifically for the treatment of fault--to--fault interactions—represents a promising direction for further reducing the computational cost of large-scale simulations, while retaining the spectral treatment of fault self-effects with the same log--linear complexity.

\subsection{Extension to 3-Dimension}

Although the numerical examples presented in this work are restricted to anti-plane (mode III) deformation, the proposed hybrid spectral and space--time BIEM framework is not limited to two-dimensional settings. In principle, it can be extended to fully three-dimensional problems involving two-dimensional fault planes and mixed-mode (II--III) slip.

From a computational perspective, the extension to three dimensions is expected to further benefit from hierarchical matrix acceleration. In two-dimensional settings, both anti-plane (mode~III) and in-plane (mode~II) formulations lead to traction kernels whose far-field spatial decay behaves as $\sim r^{-2}$, once the static or late-time regime behind the shear-wave front is reached (e.g., \citealp{ando_efficient_2007}).
In three-dimensional elastodynamics, the spatial decay of space--time interaction kernels is generally stronger. Using the Fast Domain Partitioning Method, \citet{ando_fast_2016} showed that the decay rate depends on the region of the causality cone: kernel magnitudes scale as $\sim r^{-2}$ along wavefront phases, as $\sim r^{-3}$ in the static-equilibrium domain behind the wavefronts, and decay even more rapidly (up to $\sim r^{-5}$) in intermediate regions between the $P$- and $S$-wave arrivals. This overall stronger far-field decay in three dimensions is favorable for low-rank approximation and suggests that hierarchical compression of fault--to--fault interactions may be even more effective in fully three-dimensional settings. Moreover in three-dimensional elastodynamics the dynamic correction to the traction kernel vanishes identically once the shear wave has traversed the largest dimension of the rupture domain. In other words, the space--time convolution kernel has compact temporal support in three dimensions. This property is specific to 3-D problems; in two dimensions the dynamic kernel approaches the static limit only asymptotically as $t \to \infty$, so that time-window truncation can only be approximate. The finite temporal support of the 3-D kernel therefore enables exact truncation of the elastodynamic time window \citep{ando_fast_2016}.

The primary challenge lies in the development of a non-replicating spectral formulation for fault self-effects in three dimensions. While spectral BIEMs are well established for planar faults in 3-D (e.g. \citep{geubelle_spectral_1995, lapusta_three-dimensional_2009}), removing spatial replication requires additional numerical developments \citep{cochard_spectral_1997}. This aspect is currently the subject of ongoing work.

\section{Conclusion and summary}

We have presented a hybrid boundary integral framework for simulating fully dynamic sequences of earthquakes and aseismic slip on complex fault systems, including intersecting fault segments. The central idea is to evaluate the same elastodynamic boundary integral operator using different representations for elastodynamic self-effects and fault--to--fault interactions.

Elastodynamic self-effects are computed using a non-replicating spectral BIEM, which retains the efficiency of spectral diagonalization and FFT-based transforms while avoiding unwanted interactions associated with periodic rupture(s) replication. Fault--to--fault interactions instead are evaluated in the space--time domain and accelerated using hierarchical low-rank matrix techniques. %This combination allows fully dynamic effects to be retained throughout the earthquake cycle without introducing additional tuning parameters or a substantial enlargement of the computational domain.

For the two-dimensional anti-plane problems considered here, this hybrid strategy leads to a reduction of the computational cost by approximately three orders of magnitude and a reduction of memory usage by about one order of magnitude compared to traditional space--time formulations for problem sizes on the order of a few thousand degrees of freedom. 
Moreover, the proposed algorithm is inherently parallelizable, offering a clear pathway toward simulations at scales that have previously remained computationally inaccessible.
As a result, long-term, fully dynamic earthquake-cycle simulations can be carried out efficiently on a single shared-memory personal computer, without the need for large-scale high-performance computing resources.

The present study should therefore be viewed as a proof of concept for this hybrid strategy in the simplest fully dynamic elastodynamic setting. The same ideas naturally extend to three-dimensional fault geometries and mixed-mode ruptures, which are the focus of ongoing work.

%%%%%%%%%%%%%%%%%%%%%%%%%%%%%%%%%%%%%%%%%%%%%%%%%%%%%%%%%%%%%%%%%%%%
\section*{CrediT authorship contribution statement}
\textbf{Federico Ciardo}: Conceptualization, Formal analysis, Investigation, Software, Methodology, Visualization, Writing - original draft, Writing - review \& editing.\\
\textbf{Pierre Romanet}: Conceptualization, Methodology, Writing - review \& editing.

%%%%%%%%%%%%%%%%%%%%%%%%%%%%%%%%%%%%%%%%%%%%%%%%%%%%%%%%%%%%%%%%%%%%
\section*{Declaration of competing interest}
The authors declare that they have no known competing financial interests or personal relationships that could have appeared to influence the work reported in this paper. 

%%%%%%%%%%%%%%%%%%%%%%%%%%%%%%%%%%%%%%%%%%%%%%%%%%%%%%%%%%%%%%%%%%%%
\section*{Data availability}
No data were used for the research described in this article. 

\begin{appendices}
\section{Preliminaries on linear elastodynamics in fractured media and boundary integral equations}
\label{app:Appendix1}
%\comment{I think this should go to Appendix, and we can start directly from the BEM given in Aki and Richard}
%\textcolor{blue}{It could be an option. I've started from the basics because I'd like to show that our approach is generic.. in a sense that it can be applied to 3D fractures as well.. and here we restrict our attention only to anti-plane deformations. I don't want that a generic reader that would quickly read our manuscript thinks that this approach is farly "useless" as it is only applicable to anti-plane deformations..}\\
%\comment{What I think is a bit weird here is that you start using Helmholts decomposition, which is not strickly used in the derivation of the boundary integral method. It is used to find the 3D elastodynamic Green's function. We can just start from the 3D BIEM given in Aki, and then separated it into two integrals where one is the self interation and the others are the interaction of each other faults. I think if we reference it correctly (like where to find 2D and 3D Green's function, how to derive BIEM) it would be ok.}
%\textcolor{blue}{Ok, Pierre! I'll focus on the numerical example now and will go back to this part later.}
Consider a homogeneous, isotropic, linear elastic, unbounded medium in $\mathbb{R}^3$ with fourth-order Hooke tensor 
\begin{equation}
C_{ijkl} = \lambda \delta_{ij} \delta_{kl} + \mu \left( \delta_{ik} \delta_{jl} + \delta_{il} \delta_{jk}\right),
\label{eq:stiffness_tensor}
\end{equation}
where $\lambda$ and $\mu$ are the Lamé constants, expressed in terms of Young modulus $E$ and Poisson ratio $\nu$ as $\lambda = \dfrac{\nu E}{(1-2\nu) (1+\nu)}$ and $\mu = \dfrac{E}{2(1-\nu)}$. Combining Newton's second law, the kinematic relations $\varepsilon_{ij} = (u_{i,j},+u_{j,i})/2$ and the constitutive law $\sigma_{ij} = C_{ijkl} \varepsilon_{kl}$, and neglecting body forces, leads to the classical Lamé-Navier equation:
\begin{equation}
(2\mu + \lambda) \nabla (\nabla \cdot \textbf{u}) - \mu \nabla \times (\nabla \times \textbf{u}) = \rho \frac{\partial^2 \bf{u}}{\partial t^2},
\label{eq:NavierEquation}
\end{equation}
where $\textbf{u} = u_i e_i$ is the displacement field with $e_i$ defining the orthonormal global Cartesian reference frame $\mathcal{R} = \left( \mathcal{O},e_1,e_2,e_3\right)$.\footnote{We adopt standard notation: latin indices run over $1,2,3$, Einstein summation is used, and a comma denotes partial differentiation (e.g., the divergence operator reads $\nabla\cdot\mathbf{q}=q_{i,i}$).} 

Decomposing the displacement vector field into a longitudinal $\mathbf{u}_{L}$ and traversal part $\mathbf{u}_{T}$, i.e. $\textbf{u} = \mathbf{u}_{L} + \mathbf{u}_{T}$, such that 
\begin{equation}
\begin{split}
&\nabla \times \mathbf{u}_{L} = 0 \qquad \text{(curl-free, irrotational)} \\
& \nabla \cdot \mathbf{u}_{T} = 0 \qquad \text{(divergence-free, equivoluminal)}
\end{split}
\end{equation}
yields two uncoupled (Helmholtz) wave equations:
\begin{equation}
c_d^2 \nabla^2 \textbf{u}_{L} - \frac{\partial ^2 \textbf{u}_{L}}{\partial t^2} = 0, \quad     c_s^2 \nabla^2  \textbf{u}_{T} - \frac{\partial ^2 \textbf{u}_{T}}{\partial t^2} = 0,   
\end{equation}
where $c_d = \sqrt{(\lambda + 2 \mu)/\rho}$ and $c_s = \sqrt{\mu/\rho}$ are, respectively, the compressive (or longitudinal) and shear (or transverse) wave speeds.

Now assume that the medium contains a set of pre-existing faults or fractures with arbitrary orientations (see e.g. Figure \ref{fig:fault_network}). These faults form a lower dimensional boundary region $\Gamma = \bigcup_{i=1}^N \Gamma_i$ across which displacement discontinuities occur. The displacement jump at point $\mathbf{y} \in \Gamma$ is defined as $\Delta \mathbf{u}(\mathbf{y},t) = \mathbf{u}^+(\mathbf{y},t) - \mathbf{u}^-(\mathbf{y},t)$, where the superscript $^+$ and $^-$ denote the two sides of the fault at the same point. 

The exact solution of Equation (\ref{eq:NavierEquation}) in this fractured medium can be expressed in the form of a boundary integral equation that relates the displacement at a given point in the medium with the history of displacement discontinuities at $\Gamma$. Using Betti’s reciprocity theorem and assuming (i) zero initial conditions for $t\le 0$, (ii) absence of body forces, and (iii) continuity of traction across each fault plane $\Gamma_i$, one obtains \citep{bonnet_boundary_1999}
\begin{equation}
    u_k (\mathbf{x}, t) = - \int_{\Gamma} \Delta u_i (\mathbf{y},t) * n_j(\mathbf{y}) \Sigma_{ij}^k \left( \mathbf{x}, t, \mathbf{y}\right) \, \text{d} S,
    \label{eq:representation1}
\end{equation}
where $*$ denotes the convolution product in time of two vector fields, i.e. 
\begin{equation}
\mathbf{a}(t) * \mathbf{b}(t) = \int_{0}^t \mathbf{a}(t-\Theta) \cdot \mathbf{b}(\Theta) \, \text{d}\Theta = \int_{0}^t \mathbf{a}(t) \cdot \mathbf{b}(t-\Theta) \, \text{d}\Theta,
\label{eq:convolution_product}
\end{equation}
$u_k (\mathbf{x}, t)$ is the displacement in the $k$th direction at position $\mathbf{x}$ and time $t$, $\Delta u_i (\mathbf{y},t)$ is the displacement discontinuity in the $i$th direction lying at point $\mathbf{y}$ on $\Gamma$, and $\Sigma_{ij}^k \left( \mathbf{x}, t, \mathbf{y}\right)$ is the stress Green's function denoting that $ij$-component of the stress tensor observed at position $\mathbf{x}$ and time $t$ due to a unit force in the $k$th direction applied at position $\mathbf{y}$ and time 0. Using Hooke’s law, this representation can be rewritten in terms of the displacement Green’s function $U_a^k$:
\begin{equation}
    u_k (\mathbf{x}, t) = - \int_{\Gamma} \Delta u_i (\mathbf{y},t) * n_j(\mathbf{y}) C_{ijab} \frac{\partial }{\partial x_b}U_{a}^k \left( \mathbf{x}, t, \mathbf{y}\right) \, \text{d} S,
    \label{eq:representation2}
\end{equation}

Hereunder we specialize these elastodynamic relations to anti-plane (mode III) deformations, which represent the theoretical foundations of this work. The same approach can be extended straightforwardly to in-plane (mode II) deformations, or mode I (opening) deformations.

%\subsubsection{Anti-plane deformations}
%\comment{If we keep the previous section in the paper, maybe we can just merge this section with the previous one? What do you think?}
%\textcolor{blue}{Yes, I agree with you.}\\
%We now restrict the attention to anti-plane (mode III) deformations.
\paragraph{Anti-plane deformations}
Assuming translation invariance along $e_3$-direction, the displacement field takes the form $u(\mathbf{x},t) = u(x_1,x_2,t) e_3$, so that the only non-zero stress components are $\sigma_{3i} = \mu \, 
u_{,i}$. Under this assumption, the irrotational (longitudinal) part of the displacement vanishes, and the Lamé–Navier equation (\ref{eq:NavierEquation}) reduces to a single scalar equation. Since $\nabla \times (\nabla \times u) = \nabla (\nabla \cdot u) - \nabla^2 u$ and $\nabla \cdot u = 0$, the governing equation of motion simplifies to
\begin{equation}
c_s^2 \nabla^2 u(\mathbf{x},t) = \frac{\partial^2 u(\mathbf{x},t)}{\partial t^2}
\end{equation}

The Betti representation theorem (\ref{eq:representation2}) in anti-plane settings with the change of notation $\Delta u = \delta$ reduces to the following scalar boundary integral equation
\begin{equation}
    u(\mathbf{x}, t) = -\mu \int_{\Gamma} \delta (\mathbf{y}(\xi),t) * n_j(\mathbf{y}(\xi))\frac{\partial}{x_j} U_3^3 \left( \mathbf{x}(\xi), t, \mathbf{y}(\xi)\right) \, \text{d} \xi,
    \label{eq:representation_antiplane_displ}
\end{equation}
where the displacement Green's function reads
\begin{equation}
    U_3^3(\mathbf{x}, t, \mathbf{y}) = \frac{1}{2 \pi \mu} \frac{H(t-t_s)}{\sqrt{t^2 - t_s^2}}
    \label{eq:GreenFunction}
\end{equation}  
with $t_s = r/c_s$ and $r = \left\Vert \mathbf{x} - \mathbf{y}\right\Vert$. To arrive at \eqref{eq:representation_antiplane_displ}, we observe that the only non-zero components of the stiffness tensor (\ref{eq:stiffness_tensor}) involved are $C_{3,j,3,l} = \lambda \delta_{3j} \delta_{3l} + \mu (\delta_{jl} + \delta_{3l} \delta_{j3})$, and noting that the unit normal $\mathbf{n}$ lies in the $(x_1,x_2)$-plane eliminates terms with $j=3$, leaving only $\mu \, \delta_{jl}$. Contraction over the dummy indices yields the scalar integral equation above. %The traction component $\tau$ along direction $e_3$, at point $\mathbf{x}$ and time $t$ can be calculated using the displacement representation (\ref{eq:representation_antiplane_displ}) and invoking Hooke's law: 
%as $\tau(\mathbf{x},t) = n_{l}(\mathbf{x}) \sigma_{3l} (\mathbf{x},t)$, with the stress component given by
Using Hooke's law and projecting the associated stress tensor onto a plane passing at $\mathbf{x}$ with normal vector $\mathbf{n(x)}$ then yields

\begin{equation}
\begin{split}
    \tau(\mathbf{x},t) &= -\mu^2 n_{l}(\mathbf{x})\int_{\Gamma} \delta (\mathbf{y}(\xi),t) * n_j(\mathbf{y}(\xi))\frac{\partial}{x_l}\frac{\partial}{x_j} U_3^3 \left( \mathbf{x}(\xi), t, \mathbf{y}(\xi)\right) \, \text{d} \xi = \\
    &= -\mu^2 n_{l} (\mathbf{x})\sum_{i=1}^N\int_{\Gamma_i} \delta (\mathbf{y}(\xi),t) * n_j(\mathbf{y}(\xi))\frac{\partial}{x_l}\frac{\partial}{x_j} U_3^3 \left( \mathbf{x}(\xi), t, \mathbf{y}(\xi)\right) \, \text{d} \xi
    \label{eq:representation_antiplane_stress}
\end{split}
\end{equation}
Note that Equation (\ref{eq:representation_antiplane_stress}) contains a hyper-singular kernel due to the first-order derivatives of the Green's function (\ref{eq:GreenFunction}). Specific integration technics, or regularization is therefore required to remove such strong singularities at the wavefront $t-\Theta - r/c_s = 0$ (see e.g., \cite{cochard_dynamic_1994, tada_non-hypersingular_1997}). 
%After regularization, the projection of (\ref{eq:representation_antiplane_stress}) provides the change of shear traction at any point on the fractured medium due to the slip history on $\Gamma_f$. Indeed, in this contribution, traction are treated as \textit{effective} traction, defined as the difference between i) traction produced at $\mathbf{x}$ by history of fault slip $\delta = \Delta u$ on $\Gamma_f$, and traction that would occur at the same point by external loading if the medium were constrained against slip. Thus:
%Equation \eqref{eq:representation_antiplane_stress} provides the change in shear traction at a point $\mathbf{x}$ resulting from the slip history on each fracture plan $\Gamma_i$. Because this quantity represents only the slip-induced component of traction, and not the absolute shear traction, we treat it as an `effective' traction.
%That is, the elastodynamic expression gives the increment of shear traction relative to the traction that would exist under the same external loading if the medium were constrained against slip.

\section{Verification of spectral BIEM without replication: single-fault problem}
\label{app:Appendix2}

\begin{figure}[ht!]
  \centering

  % -------- Row 1 --------
  \subfloat[$z=0$ m]{
    \includegraphics[width=0.4\textwidth]{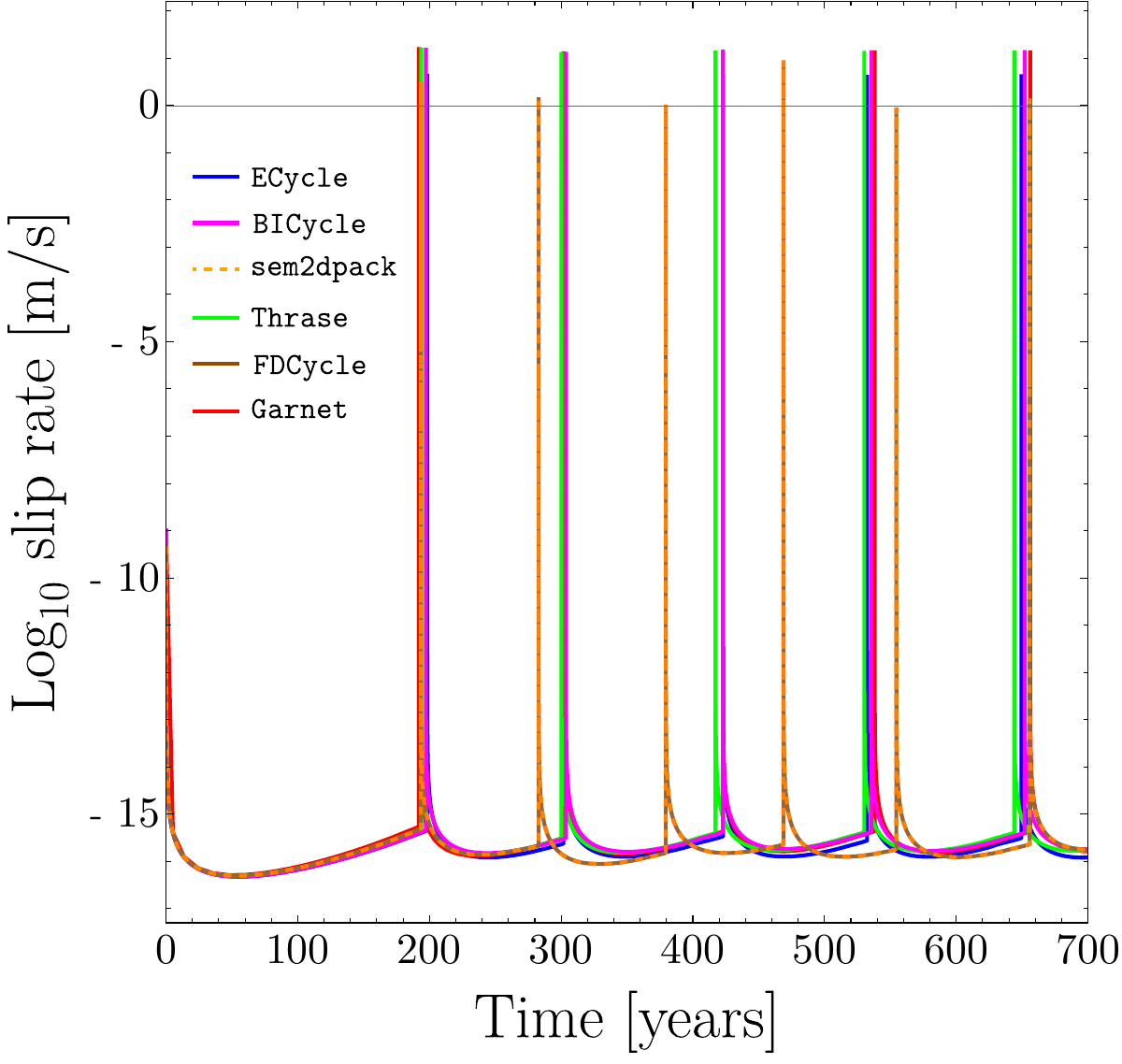}
  }\,\,\,
  \subfloat[$z=0$ m]{
    \includegraphics[width=0.39\textwidth]{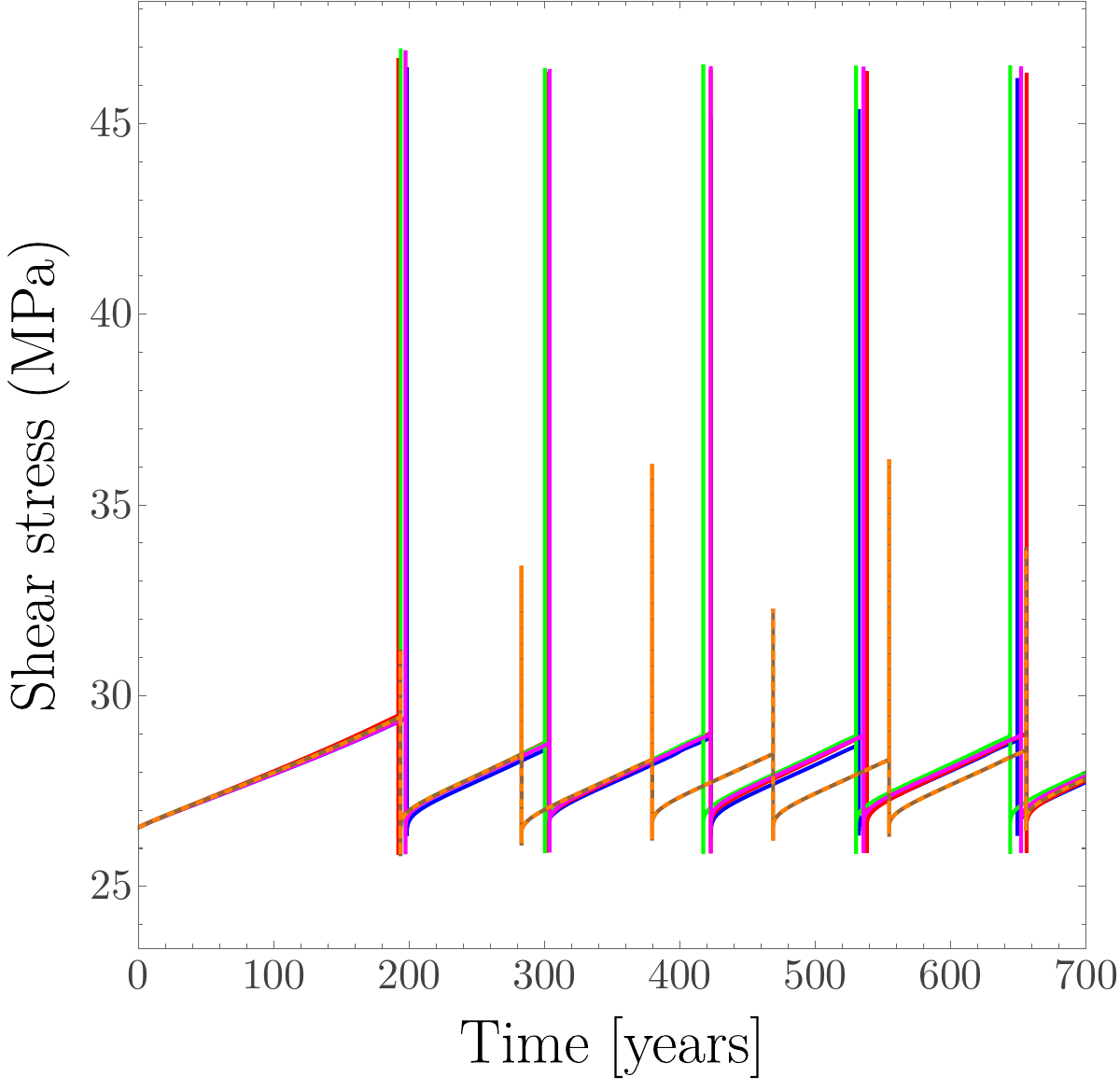}
  }\\[6pt]

  % -------- Row 2 --------
  \subfloat[$z=15$ m]{
    \includegraphics[width=0.4\textwidth]{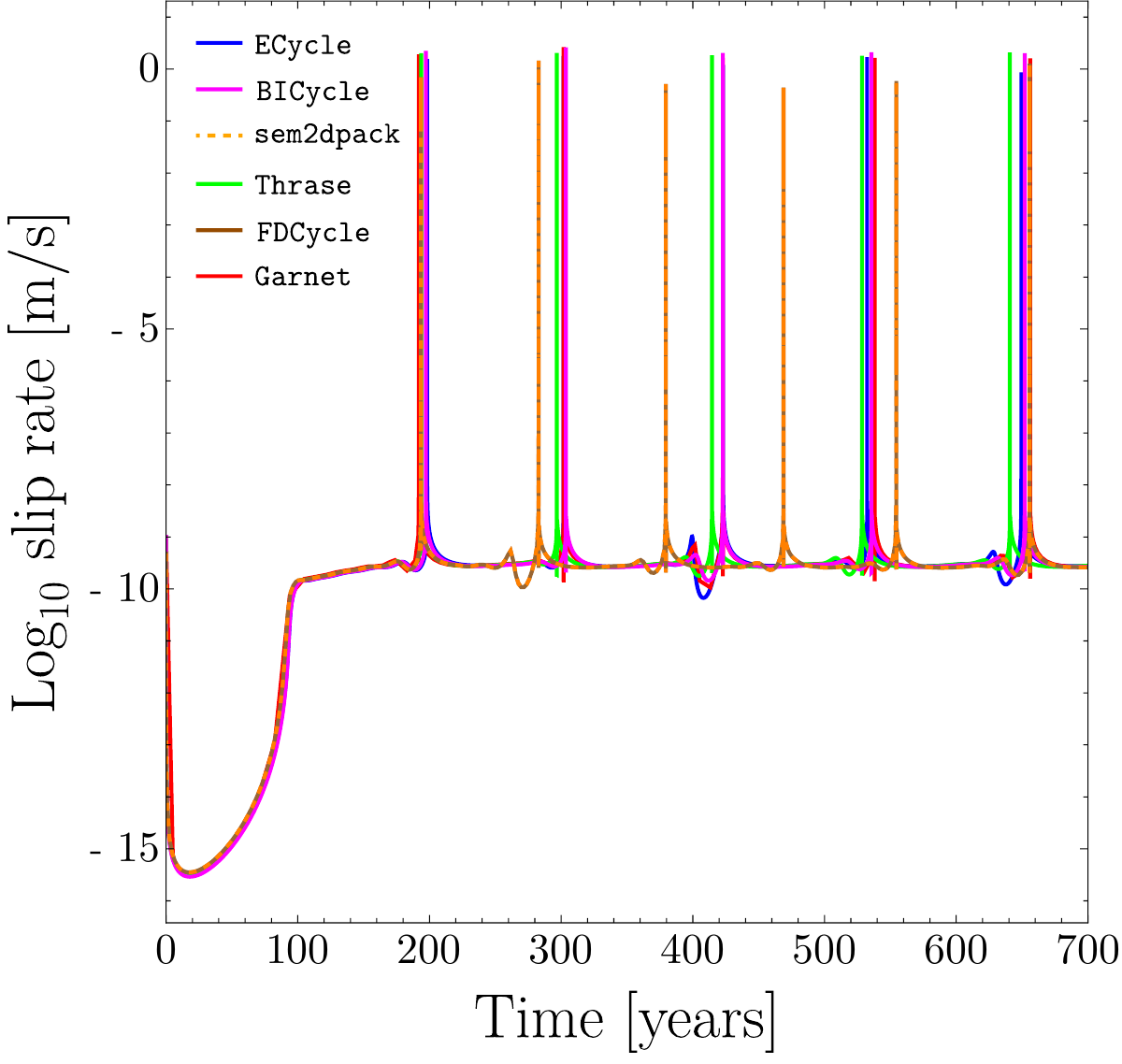}
  }\,\,\,
  \subfloat[$z=15$ m]{
    \includegraphics[width=0.39\textwidth]{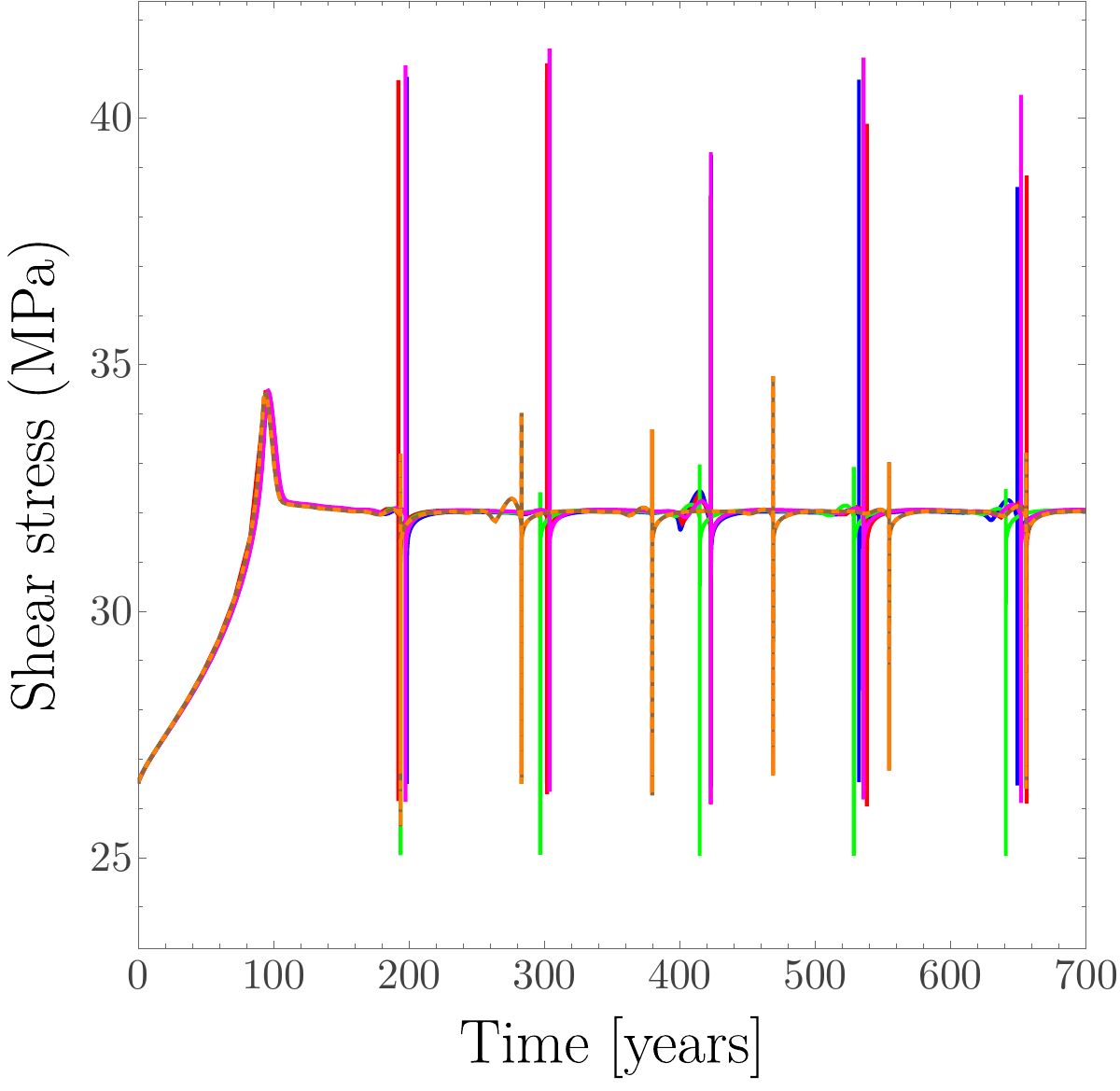}
  }\\[6pt]

  % -------- Row 3 --------
  \subfloat[$z=30$ m]{
    \includegraphics[width=0.4\textwidth]{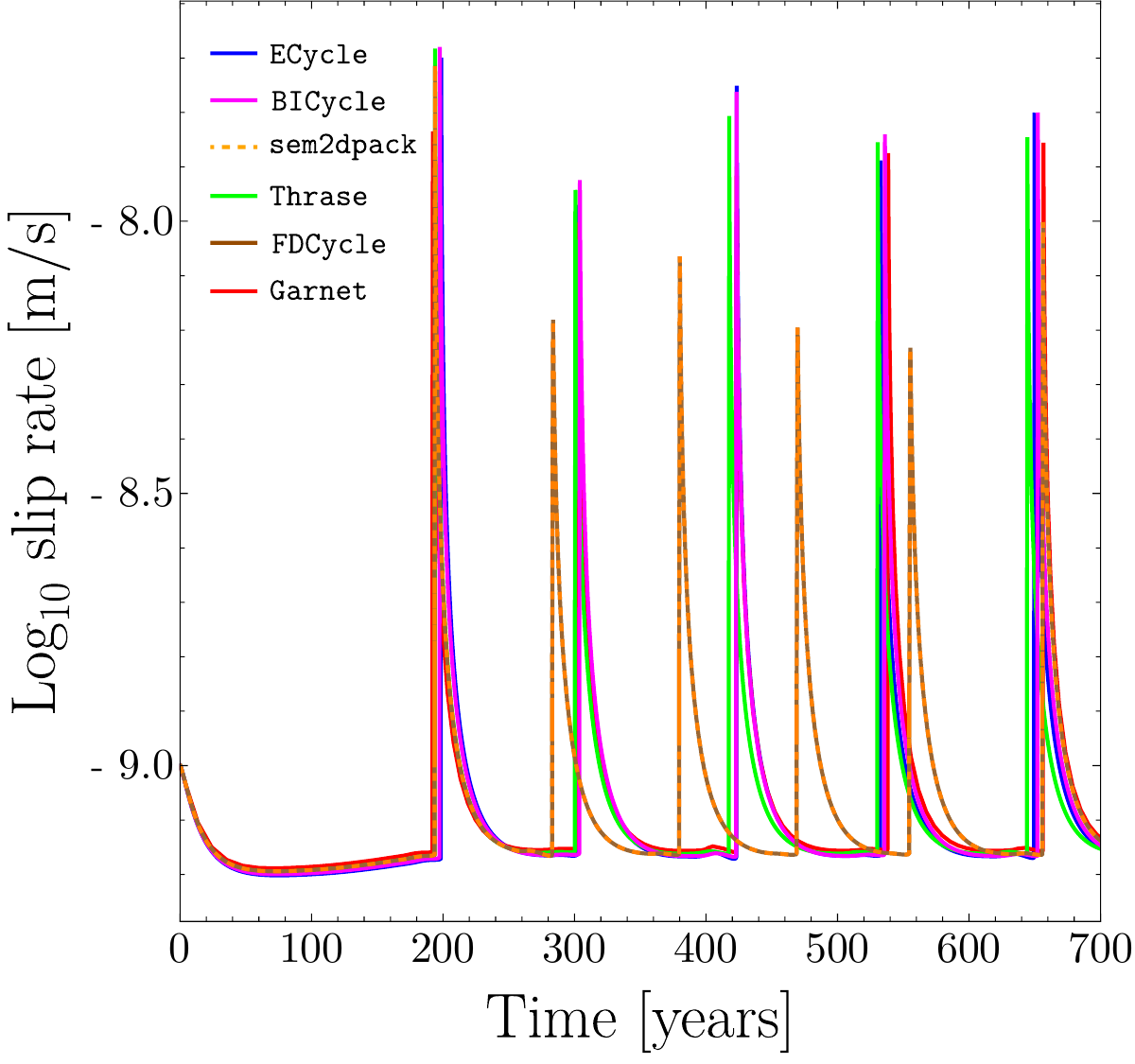}
  }\,\,\,
  \subfloat[$z=30$ m]{
    \includegraphics[width=0.4\textwidth]{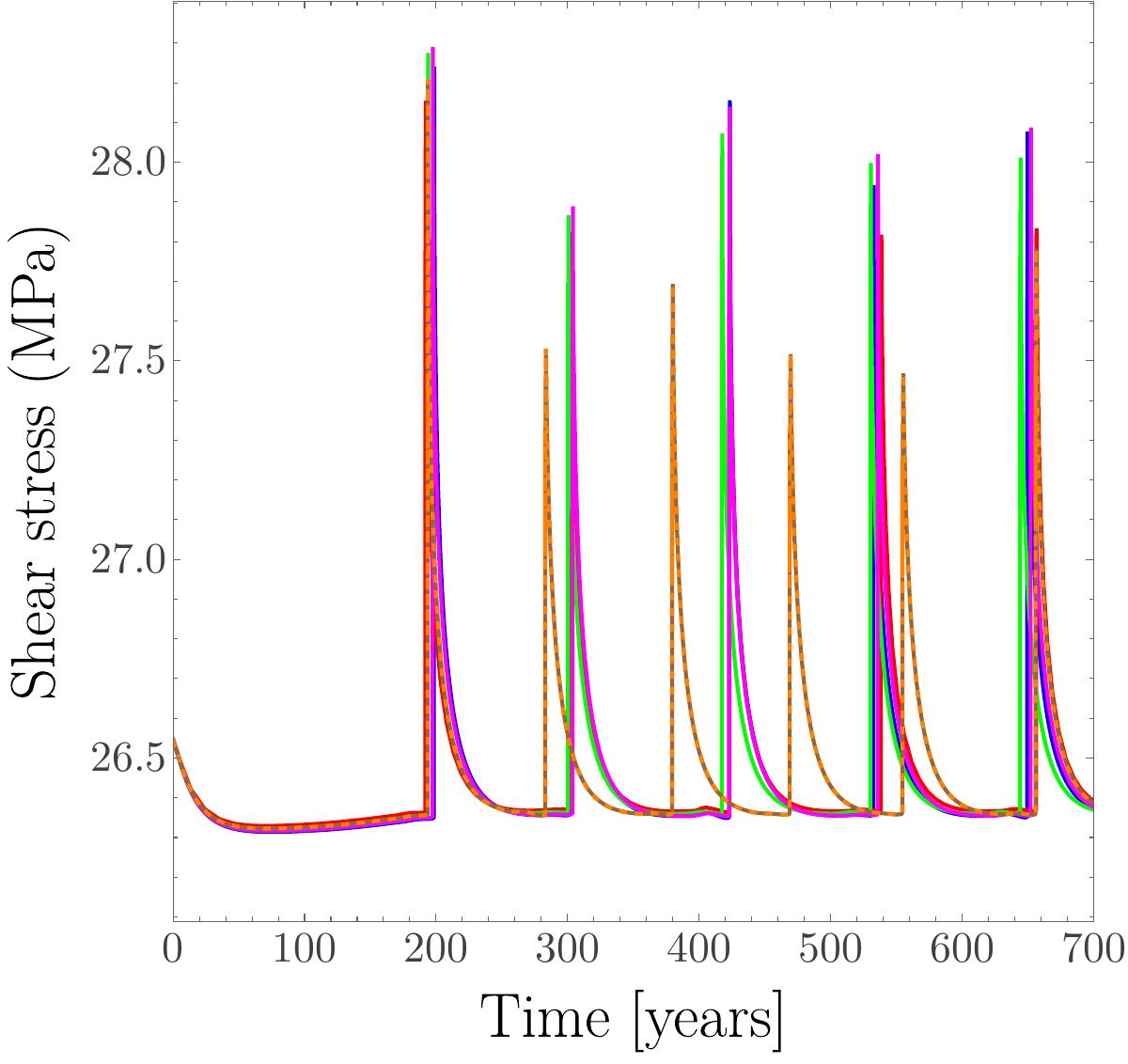}
  }
  \\[0.5 cm]
  \caption{Comparison of time histories of slip rate and shear stress at three representative depths for the BP1--FD benchmark. Blue: this new hybrid approach (\texttt{ECycle} code); Red: \texttt{GARNET} code; Green= \texttt{Thrase} code; Magenta: \texttt{BICycle} code; Brown: \texttt{FDCycle} code; Dashed-Orange: \texttt{sem2dpack} code.}
  \label{fig:bp1fd}
\end{figure}

Because no analytical solution exists for time-dependent elastodynamic fault slip problems with rate-and-state friction, verification of numerical solvers rely on well-established community benchmarks. Here we verify the implementation of the spectral BIEM without spatial replication against the first fully dynamic benchmark of the Statewide California Earthquake Center initiative, namely BP1--FD \citep{erickson_community_2020}. This benchmark involves a single, planar, vertical strike-slip fault embedded in a homogeneous elastic half-space and therefore allows us to directly assess the accuracy of the spectral BIEM formulation without replication of the rupture domain.

To model the elastic half-space with a free surface, we employ the method of images and extend the computational domain symmetrically above and below the free surface. This approach is exact for the anti-plane (mode~III) deformation considered in BP1--FD, for which the problem is scalar and the free-surface boundary condition of zero traction can be enforced by antisymmetric continuation of slip across the surface.

For brevity, we do not report here the full set of input parameters, initial conditions, constitutive properties, and geometrical specifications defining the BP1--FD benchmark. These are identical to those reported in the original benchmark description and can be found in \citet{erickson_community_2020}. Instead, we focus on comparing the results obtained with the present spectral BIEM implementation to the reference numerical solutions available at \url{https://strike.scec.org/cvws/seas/}.\\

In our numerical model, the total fault rupture length is thus $80~\mathrm{km}$. To avoid spatial replication, padding regions are introduced that extend $40~\mathrm{km}$ to the left and $40~\mathrm{km}$ to the right of the original fault, effectively doubling the computational domain. The fault is discretized using $3200$ constant-size elements, corresponding to a uniform mesh spacing of $25~\mathrm{m}$, which is sufficient to resolve the cohesive-zone length scale $\Lambda$ with more than 12 elements ($\Lambda/\Delta x = (9 \pi / 32) (L_b/\Delta x) \sim 12.1$). The padding regions are discretized using the same element size.
The mode-dependent elastodynamic time window for self-effects is truncated using the parameters $\eta_w = 2$ and $q_w = N_{\text{elts}}/2$.

\begin{figure}[t!]
\centering
\noindent\includegraphics[width=0.9\textwidth]{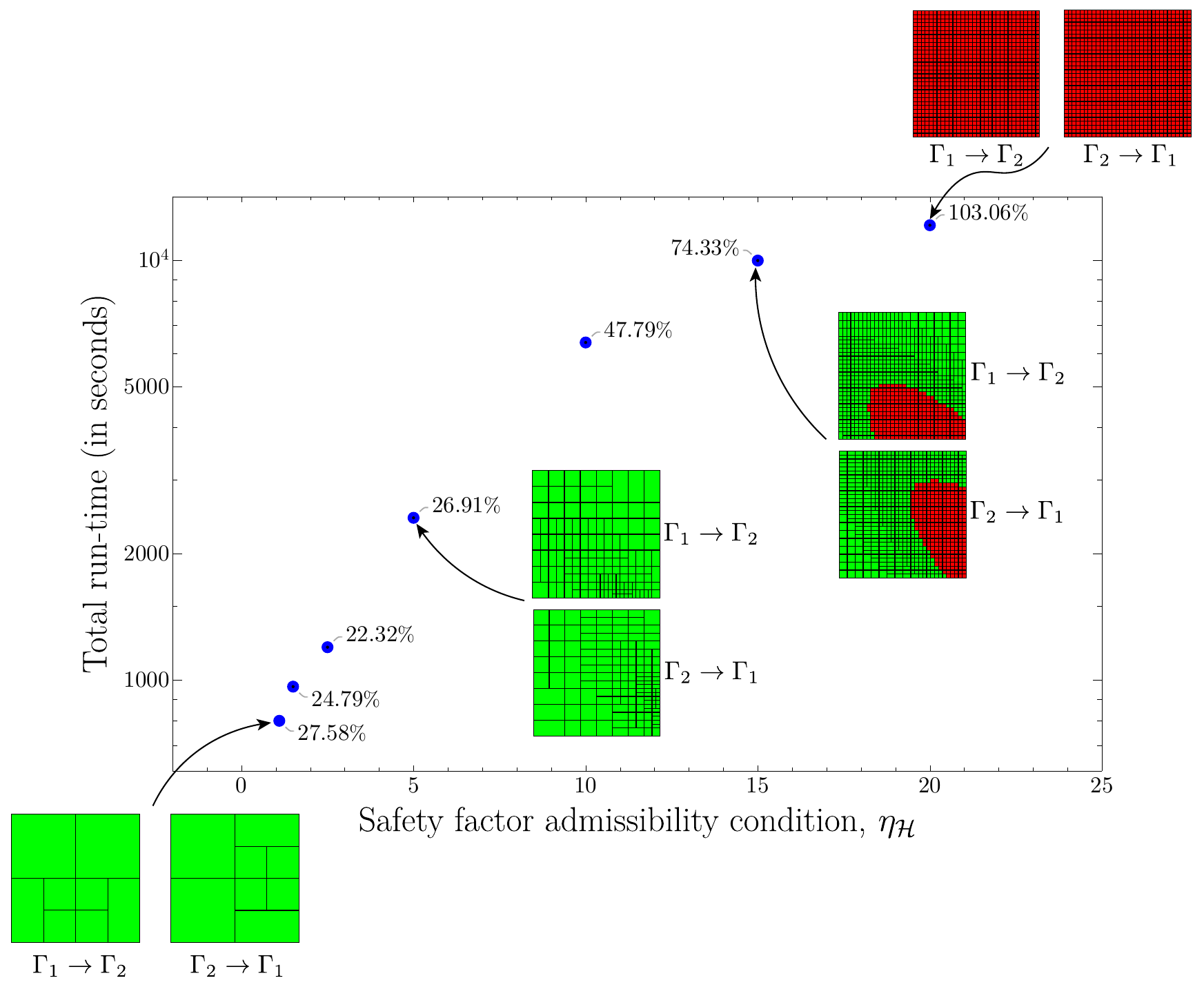}
\\[0.5cm]
\caption{Total runtime as a function of the $\mathcal{H}$-matrix admissibility parameter $\eta_H$ for the two--fault benchmark problem. Percentages reported next to the symbols indicate the overall compression ratio associated with each value of $\eta_H$.}
\label{fig:TotalRuntime_Verification1}
\end{figure}

Figure~\ref{fig:bp1fd} compares the time histories of slip rate (in m/s) and shear stress (in MPa) at three representative depths along the fault, specifically $z=0$ m (free surface), $x = 15$ m and $z=30$ m. Each row corresponds to a different depth, with slip rate and shear stress shown side by side. Overall, our results obtained with the present spectral BIEM without replication (blue) show a good agreement with the reference solutions produced by \texttt{BICycle}, \texttt{GARNET} (red), and \texttt{Thrase} (green). 
In contrast, the solutions obtained with \texttt{sem2dpack} (dashed-orange) and \texttt{FDCycle} (brown) exhibit a systematic temporal lag relative to the other results, which is consistently observed at all depths considered. All reference solutions were retrieved from the public repository (\url{https://strike.scec.org/cvws/cgi-bin/seas.cgi}), and for each code the most refined simulation available in the archive was used in the comparison.

The particularly close agreement with \texttt{BICycle} is noteworthy, as this code is based on the traditional spectral BIEM with spatial replication of the rupture domain. Achieving such agreement implies that the replication distance adopted in that simulation must be at least four times larger than the effective rupture domain in order to avoid spurious interactions \citep{noda_dynamic_2021}. Consequently, attaining comparable accuracy within a replicating spectral BIEM framework requires a substantially larger computational domain and incurs at least a twofold increase in computational cost compared to the non-replicating formulation employed here.

\section{Influence of the admissibility parameter on performance and accuracy}
\label{app:Appendix3}

\begin{figure}[t!]
\centering
\noindent\includegraphics[width=0.85\textwidth]{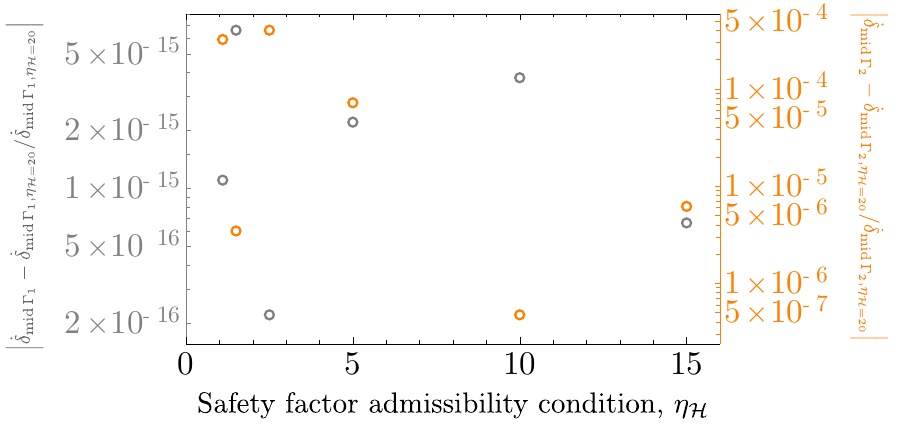}
\\[0.5cm]
\caption{Relative error in slip rate at the midpoint of each fault at the end of the simulation, $t = T_{\mathrm{end}} = 2\,\mathrm{s}$, as a function of the admissibility parameter $\eta_H$. Errors are computed with respect to the reference solution obtained for $\eta_H = 20$ (uncompressed case).}
\label{fig:RelativeErrors_Verification1}
\end{figure}

Here we investigate the influence of the $\mathcal{H}$-matrix admissibility parameter $\eta_{\mathcal{H}}$ on computational performance and solution accuracy for the two--fault benchmark problem discussed in Section~\ref{subsec:two_fault_problem}. The admissibility parameter enters the geometric criterion used to identify well-separated source--receiver cluster pairs and therefore directly controls both the number and the size of interaction blocks that are approximated in low-rank form.

Figure~\ref{fig:TotalRuntime_Verification1} reports the total runtime (in seconds) as a function of $\eta_{\mathcal{H}}$. As expected, the total runtime increases monotonically with increasing $\eta_{\mathcal{H}}$. Larger values of $\eta_H$ correspond to a more restrictive admissibility condition \eqref{eq:admissibility_condition}, which reduces the number of admissible blocks per convolution time step and increases the number of near-field blocks stored as full, dense matrices (see the matrix decompositions shown in Fig.~\ref{fig:TotalRuntime_Verification1} for some values of $\eta_{\mathcal{H}}$, which correspond to the first convolution time step). As a result, fewer interactions benefit from low-rank approximation, leading to a higher overall computational cost.

In the opposite regime, decreasing $\eta_{\mathcal{H}}$ relaxes the admissibility condition and increases the fraction of interactions that are approximated in low rank. This produces a systematic reduction of the total runtime, with the lowest runtimes observed for the smallest values of $\eta_{\mathcal{H}}$ considered. Interestingly, this regime does not necessarily coincide with the strongest overall compression, as indicated by the \textit{total} compression ratios reported next to the symbols. For small $\eta_{\mathcal{H}}$, the hierarchical partitioning becomes coarser, yielding a smaller number of admissible blocks of larger size. Although the resulting compression ratio is not minimal, the reduced number of low-rank blocks lowers memory-access and traversal overhead, ultimately leading to a smaller total runtime.

Figure~\ref{fig:RelativeErrors_Verification1} shows the relative error in slip rate evaluated at the centers of the two faults at the end of the simulation, $t = T_{\mathrm{end}} = 2\,\mathrm{s}$. The relative error is computed with respect to a reference solution obtained using the largest admissibility parameter considered, $\eta_{\mathcal{H}} = 20$, for which the admissibility criterion is the most restrictive and the interaction kernels are stored entirely in dense form (see red interaction matrices in the top-right corner). Results are shown separately for fault~$\Gamma_1$ (gray symbols) and fault~$\Gamma_2$ (orange symbols).

For all values of $\eta_H$ explored, the relative error remains very small on both faults, with noticeably smaller errors on fault~$\Gamma_1$, where dynamic rupture propagates, than on fault~$\Gamma_2$. This indicates that the hierarchical approximation does not significantly affect solution accuracy, although it introduces slightly larger errors than the spectral--time representation used for elastodynamic self-effects. In particular, for small and moderate values of $\eta_H$, which correspond to the most efficient compression and the lowest runtimes (Figure~\ref{fig:TotalRuntime_Verification1}), the slip-rate error remains very small. As $\eta_H$ increases and the admissibility criterion becomes more restrictive, the hierarchical approximation approaches the reference solution, and the relative error decreases accordingly.

Overall, these results demonstrate that the computational speedups achieved by relaxing the admissibility criterion do not come at the expense of accuracy.

\section{Comparison between replicant and non-replicant convolution kernels for elastodynamic self-effects}
\label{app:Appendix4}

\begin{figure}[h!]
\centering
\noindent\includegraphics[width=0.55\textwidth]{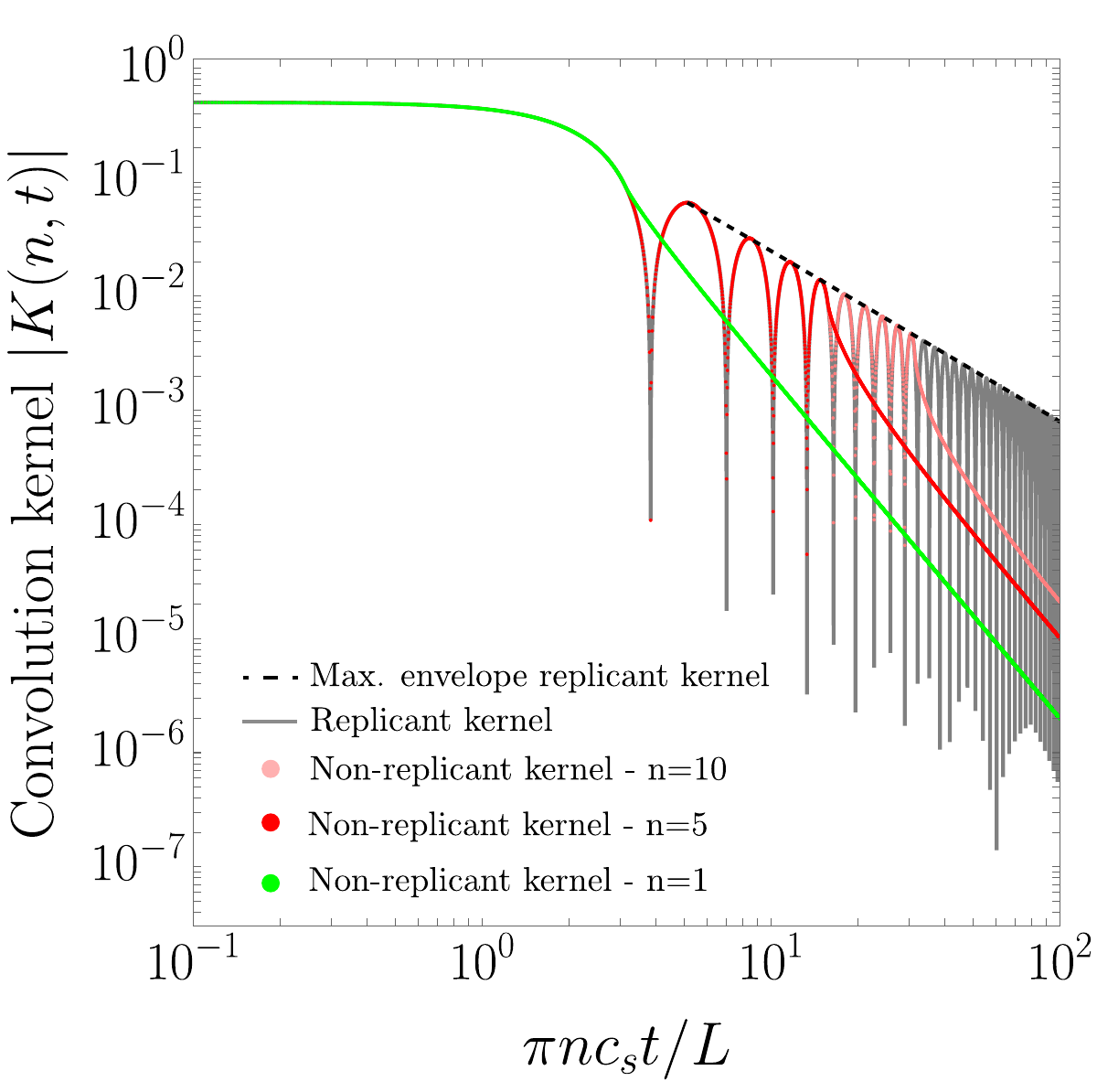}
\\[0.5cm]
\caption{Comparison between replicating and non-replicating spectral convolution kernels in log-log plot. The figure shows the temporal evolution of the absolute value of kernel $K(n,t)$ for selected Fourier modes. While both formulations coincide at early times, the non-replicating kernel exhibits a faster and monotonic decay at large convolution times. In contrast, the replicating formulation retains an oscillatory behavior. This difference in decay behavior enables a more efficient, mode-dependent truncation of the elastodynamic convolution window in the non-replicating formulation.}
\label{fig:Kernels_decay}
\end{figure}

\end{appendices}

% Bibliography
\bibliographystyle{apalike}
\bibliography{bibliography.bib}
\end{document}